\documentclass[11pt,a4paper]{article}
\usepackage{jheppub}
\usepackage{amsmath}
\usepackage{amssymb}
\usepackage{subcaption}
\usepackage{graphicx}
\usepackage{float}
\usepackage{physics}
\usepackage{tikz}
\usepackage{tkz-euclide}
\usepackage{siunitx}
\usepackage{xcolor}
\newcommand{\red}[1]{{\color{red}{#1}}}
\newcommand{\blue}[1]{{\color{blue}{#1}}}
\newcommand{\violet}[1]{{\color{violet}{#1}}}

\let\vec\boldsymbol
\newcommand{\orange}[1]{#1}

\tikzset{every picture/.style={line width=0.3mm}}
\definecolor{pred}{RGB}{238,28,37}
\definecolor{pblue}{RGB}{48,49,146}
\definecolor{pgreen}{RGB}{00,163,80}
\usetikzlibrary{decorations.pathmorphing}
\tikzset{snake it/.style={decorate, decoration=snake}}

\title{Dynamic critical behavior of the chiral phase transition from the real-time functional renormalization group}
\author[a]{Johannes V.~Roth,}
\author[b]{Yunxin Ye,}
\author[b]{S\"oren Schlichting,}
\author[a,c]{and Lorenz von Smekal}

\affiliation[a]{Institut f\"ur Theoretische Physik,\\
Justus-Liebig-Universit\"at, Heinrich-Buff-Ring 16, 35392 Gie{\ss}en, Germany}
\affiliation[b]{Fakult\"at f\"ur Physik,\\
Universit\"at Bielefeld, D-33615 Bielefeld, Germany}
\affiliation[c]{Helmholtz Research Academy Hesse for FAIR (HFHF),\\
Campus Gießen, 35392 Gie{\ss}en, Germany}

\emailAdd{johannes.v.roth@physik.uni-giessen.de}
\emailAdd{ye@physik.uni-bielefeld.de}
\abstract{In the chiral limit the complicated many-body dynamics around the second-order chiral phase transition of two-flavor QCD can be understood by appealing to universality. We present a novel formulation of the real-time functional renormalization group that describes the stochastic hydrodynamic equations of motion for systems in the same dynamic universality class, which corresponds to Model G in the Halperin-Hohenberg classification. Our approach preserves all relevant symmetries of such systems with reversible mode couplings. We show that the calculations indeed produce the non-trivial value $z=d/2$ for the dynamic critical exponent, where $d$ is the number of spatial dimensions. From the momentum and temperature dependence of the diffusion coefficient of the conserved charge densities, we extract the dimensionless universal scaling function.}
\begin{document}
\maketitle

\section{Introduction}

The chiral phase transition of QCD in the limit of two  massless quark flavors is generally believed to be of second order and to fall into the $O(4)$ universality class~\cite{Pisarski:1983ms}. This is because QCD is then invariant under independent $SU(2)$ rotations of left and right-handed quarks, giving rise to  the $SU(2)_L \times SU(2)_R$ chiral symmetry, the double cover of the identity component of $O(4)$, which in the vacuum is spontaneously broken down to the two-flavor $SU(2)_V$ isospin symmetry, that of $SO(3)$.\footnote{The argument requires the anomalous breaking of the axial $U(1)_A$ symmetry to be sufficiently strong at the transition. Otherwise, if it is effectively restored at the chiral phase transition, the symmetry breaking pattern would be $U(2)_L \times U(2)_R \to U(2)_V$. While no infrared stable fixed point was found at first-order in the $\varepsilon$-expansion~\cite{Pisarski:1983ms}, beyond this perturbative expansion around the upper-critical dimension of four, it is now widely accepted that such a fixed point in three dimensions exists~\cite{Pelissetto:2013hqa,Grahl:2014fna}. So the fate of the axial $U_A(1)$ merely affects the universality class of the two-flavor chiral phase transition, but not its order.}
Lattice QCD simulations performed close to the two-flavor chiral limit for lighter than physical pion masses provide evidence of a second-order chiral phase transition \cite{HotQCD:2019xnw,Cuteri:2021ikv}. Although the universality class cannot be determined with confidence yet, the results from the HotQCD collaboration at least indicate that the quark-mass and temperature dependence of chiral condensate and susceptibility are consistent  with $O(4)$ universality, as described by the corresponding universal scaling functions  \cite{HotQCD:2019xnw,Kaczmarek:2020sif}.
While the discriminating power of available lattice results does not allow to nail the universality class, the window for a description in terms of universal scaling functions appears to extend up to the physical pion mass \cite{Kotov:2021rah}. This window is not necessarily equal to the true scaling window, however. Results from functional methods for effective low-energy models and QCD have long suggested that the actual scaling region is restricted to much smaller pion masses~\cite{Braun:2010vd,Braun:2020ada,Chen:2021iuo,Gao:2021vsf}, with the most recent QCD estimate limiting the critical region for $O(4)$ scaling to pion masses below a few MeV, and to a similarly small range of temperatures around the transition~\cite{Braun:2023qak}. 

The \emph{dynamic} universality class of the two-flavor chiral phase transition is believed to be an extension of the original Model G  by Halperin and Hohenberg \cite{RevModPhys.49.435} to an $O(4)$ order parameter \cite{Rajagopal:1992qz}.
This extension is frequently referred to also simply as `Model G', although strictly speaking `Model~G' specifically referred to the antiferromagnetic  $O(3)$ Heisenberg case in the original classification.
The phenomenological consequences for heavy-ion collisions and the critical dynamics of Model~G were previously studied in Refs.~\cite{Grossi:2020ezz,Grossi:2021gqi,Florio:2021jlx,Florio:2023kmy}. In fact, it was suggested that an observed excess of soft pions in heavy-ion collisions \cite{Devetak:2019lsk} might be attributed to remnants of the second-order $O(4)$ transition \cite{Florio:2023kmy}.
With its critical dynamics thus being potentially relevant for the phenomenology of heavy-ion collisions, here we therefore further pursue the study of this model which has historically also been known as the S\'asvari-Schwabl-Sz\'epfalusy (SSS) model \cite{SASVARI1975108,taeuber_2014} in the literature. Luckily, to observe \emph{strong dynamic scaling}, where  the order parameter and the conserved charges are expected to relax with the same dynamic critical exponent $z=d/2$ in $d$ spatial dimensions, it is not necessarily required to be strictly inside the potentially small static $O(4)$ scaling region.

In this work, we focus on studying the $O(4)$ Model~G dynamics within a \emph{real-time formulation} of the functional renormalization group (FRG). 
This extends previous work in several ways:
For example, a real-time FRG formulation of classical Langevin dynamics was used to calculate the dynamic critical exponent $z$ of the purely dissipative Model~A in \cite{Canet:2006xu,Canet:2011wf,Duclut:2016jct}
and of Model~C in \cite{Mesterhazy:2013naa}. Quantum fluctuations can be incorporated by formulating the FRG on the Schwinger-Keldysh closed-time path (CTP). Such real-time formulations of the FRG on the CTP were used to study critical dynamics and spectral functions of the relativistic $O(4)$ model in Refs.~\cite{Mesterhazy:2015uja,Tan:2021zid}. Without including the reversible mode couplings in presence of the conserved charges in these references this led to the dissipative dynamics of Model~A, however, where the charges are not conserved.
The truncation of Ref.~\cite{Tan:2021zid} had thereby been adapted from Ref.~\cite{Huelsmann:2020xcy} were it had been used to study spectral functions of the quartic-anharmonic oscillator in quantum mechanics.
A combined vertex and loop expansion for the FRG on the CTP was formulated in \cite{Roth:2021nrd} where the same quantum-mechanical system was studied with various different real-time methods for spectral functions. 
A field-theory extension of this real-time FRG formulation was developed in \cite{Roth:2023wbp} and used to compute the critical spectral functions of the relaxational Models A, B, and~C. 
In a recent update, a derivative expansion similar to the one used in Ref.~\cite{Canet:2006xu} was employed to resolve the field dependence of the kinetic coefficient of Model~A \cite{Batini:2023nan}.
In our present paper we take the real-time FRG to the next level by including the reversible mode couplings needed to study Model~G dynamics in this framework. The subtleties associated with these reversible mode couplings will also be relevant for Model H, the dynamic universality class of the liquid-gas transition in a pure fluid and the conjectured one of the QCD critical point, in the future. 

Other genuine real-time methods are provided by classical-statistical simulations \cite{Aarts:2001yx,Berges:2009jz,Schlichting:2019tbr,Schweitzer:2020noq,Schweitzer:2021iqk,Roth:2021nrd}, and extensions including the leading quantum corrections in a Gaussian-state approximation \cite{Buividovich:2017kfk,Buividovich:2018scl,Roth:2021nrd}. The relaxational Models A and C were simulated in a single-component classical field theory in \cite{Berges:2009jz,Schweitzer:2020noq}. These classical-statistical simulations were later extended to the diffusive dynamics of a conserved order parameter in a microscopic model with Model B dynamics and dissipation, including a new non-dissipative limit with the same set of conserved quantities as in Model D but without a diffusive mode \cite{Schweitzer:2021iqk}. Particularly relevant for the present work are Refs.\ \cite{Schlichting:2019tbr,Florio:2021jlx,Florio:2023kmy} in which the $O(4)$ model close to the critical point was studied with classical-statistical simulations.
The simulations in \cite{Schlichting:2019tbr} were performed using the relativistic microscopic $O(4)$ model, but without explicitly warranting the conservation of the charges, and therefore remained inconclusive on the precise value of the resulting dynamic critical exponent.
In \cite{Florio:2021jlx} the dynamic universality class of Model G was forced by an explicit coupling to the conserved isovector and isoaxial charge densities using the Poisson-bracket technique.
They obtained a numerical value of $z=1.47 \pm 0.01\text{(stat)}$ for the dynamic critical exponent which is consistent with the exact prediction of $z=d/2$ from the $\varepsilon$-expansion in $d=3$ spatial dimensions \cite{SASVARI1975108,janssen_renormalized_1977,PhysRevLett.38.505,PhysRevB.18.353,Rajagopal:1992qz,PhysRevE.55.4120,taeuber_2014}.
The real-time FRG framework needed to study this dynamic universality near the associated strong-scaling fixed point is developed in our present work. 

This paper is organized as follows.  In Sec.~\ref{sec:model_G} we introduce the basics of the dynamic universality class of Model~G and construct a corresponding generating functional using the Martin-Siggia-Rose-Janssen-De Dominicis (MSRJD or often simply MSR) formalism \cite{Martin:1973zz,Dominicis:1976,Janssen:1976}. In the same section, we also discuss the symmetries of the MSR action of Model~G (with particular focus on the symmetry that arises due to the reversible mode couplings), which will serve as a guiding principle for finding a suitable truncation  within the FRG. In Sec.~\ref{sec:FRG_rev_model_coup}, we formulate the FRG for systems with reversible mode couplings and discuss its regularisation. We show that the presence of the dynamic in reversible mode coupling systems should not change the static flow. In Sec.~\ref{sec:truncation} we formulate a truncation of  effective average action and derive corresponding flow equations for the  static and dynamic couplings. In Sec.~\ref{sec:numeric}, we  introduce the numerical methods we use for solving the flow equations. In Sec.~\ref{sec:result}, we discuss the critical behavior of Model~G obtained from our real-time FRG study, including the critical exponents and dynamic universal scaling function for the diffusion coefficient of the charge densities.

Several appendices  are added where we provide additional explanations. In App.~\ref{sec:detailsOnMSRPathIntegral} we discuss technical subtleties which arise in the MSR path-integral formulation of Model~G, like a proper continuum formulation of the underlying Ito discretization, and the presence of Faddeev-Popov ghosts. In App.~\ref{appendix:fdr} we derive generalized fluctuation-dissipation relations for $n$-point vertices that hold within our truncation. In App.~\ref{sec:detailsOnFRG} we derive the FRG flow equation for systems with reversible mode couplings and show that even though the regulator necessarily couples to composite fields, the effective average action converges to the bare action in the ultraviolet (UV) limit $k \to \Lambda$. In particular, in App.~\ref{sct:CorrespondenceRealTimeAndEuclFlows} we provide an explicit demonstration that the flow equation of the static free energy decouples from the rest and satisfies a closed flow equation on its own, which is the same as the standard Euclidean flow equation for a dimensionally reduced $d$-dimensional system. Technical details about the extraction of the dynamic critical exponent are given in App.~\ref{extraction_app}.

\section{Model G}\label{sec:model_G}

In its original formulation by Halperin and Hohenberg, Model G describes the dynamic universality class of an isotropic Heisenberg antiferromagnet \cite{RevModPhys.49.435}.
The order parameter is a three-component vector and is interpreted as the staggered magnetization of the model which is not conserved.
It  couples dynamically to the conserved three-component magnetization through reversible mode couplings.
This leads to the non-trivial value $z=d/2$ of the dynamic critical exponent (where $d$ is the number of spatial dimensions).
In this work, we consider a generalized formulation, where the order parameter is an $N$-component scalar field $\phi_a(x)$ which couples dynamically to an antisymmetric matrix of $N(N-1)/2$ charge densities $n_{ab}(x) = -n_{ba}(x)$.
This generalization of Model~G is also known as the S\'asvari-Schwabl-Sz\'epfalusy (SSS) model \cite{SASVARI1975108,taeuber_2014}.
The SSS model was previously inspected using field-theoretic methods in (e.g.)\ Refs.\ \cite{janssen_renormalized_1977,PhysRevB.18.353,PhysRevLett.38.505,PhysRevE.55.4120,taeuber_2014}. For $N=2$ one recovers Model~E, which describes the dynamic universality class of the critical point in the phase diagram of planar antiferromagnets. For $N=3$ one obtains the standard formulation of Model~G which describes the dynamic universality class of the Heisenberg antiferromagnet. See \cite{Yao:2022fwm} for a recent study thereof.
For $N=4$ one obtains the anticipated extension of Model G which describes the dynamic universality class of the chiral phase transition in two-flavor QCD \cite{Rajagopal:1992qz}.
This is the relevant case for this work, although we will keep the number of field components $N$ general in the derivation of the flow equations.
Note that larger values of $N > 3$ do not correspond to any of the standard Halperin-Hohenberg dynamic universality classes \cite{RevModPhys.49.435}.
However, previous analyses showed  that $z=d/2$  as in the standard formulation of Model~G  still holds \cite{SASVARI1975108,janssen_renormalized_1977,PhysRevLett.38.505,PhysRevB.18.353,Rajagopal:1992qz,PhysRevE.55.4120,Nakano:2011re,taeuber_2014,Florio:2021jlx}.
We thus use the term `Model G' in the general sense also for $N \geq 3$, as frequently done especially in the literature on the SSS model. 
In the context of QCD  with two (nearly) massless quark flavors, the reasoning by Rajagopal and Wilczek \cite{Wilczek:1992sf,Rajagopal:1992qz} is as follows: The order parameter for the chiral phase transition is
the expectation value of the quark bilinear $M^{i}{}_{j} = \langle\bar{q}_L^i q_{Rj}\rangle$, which contains the left and right-handed components $q_{R/L} = \tfrac{1}{2} (1\pm \gamma^5)\,q$ of the quark Dirac spinors.
Under independent left and right-handed isospin $(U_L,U_R) \in SU(2)_L \times SU(2)_R$ transformations the order parameter transforms as 
\begin{align}
M \to U_L^\dagger M U_R \,. \label{eq:trafoOfM}
\end{align}
Hence, a non-vanishing value for $M$ signals a (spontaneous) breaking of chiral symmetry.
In the special case of $N_f=2$ quark flavors, the $2 \times 2$ matrix $M$ can be parameterized by four real numbers $\phi = (\sigma,\vec{\pi})$ as  
\begin{align}
    M = \sigma + i\vec{\pi}\cdot\vec{\tau} \,,
\end{align}
where $\vec{\tau} = (\tau^{i}) = (\tau_{i})$ denotes the vector of Pauli matrices.
Hence, in the context of QCD, the order parameter $\phi$ from $O(4)$ Model~G can be interpreted as a proxy for the quark bilinear $M$ which itself is not conserved but reversibly coupled to the generators of the
chiral transformations in \eqref{eq:trafoOfM}, i.e.~the $O(4)$ rotations of $\phi$. In particular, Noether's theorem implies six conserved currents for the chiral $SU(2)_L \times SU(2)_R$ symmetry, namely
\begin{align}
	L_{\mu}^i \equiv \bar{q}_L \gamma_{\mu} \frac{\tau^i}{2} q_L \,, \hspace{0.5cm}
	R_{\mu}^i \equiv \bar{q}_R \gamma_{\mu} \frac{\tau^i}{2} q_R \,.
\end{align}
These are equivalent to the isovector and isoaxial-vector currents
\begin{subequations}
\begin{align}
    V_{\mu}^i &\equiv L_{\mu}^i + R_{\mu}^i = \bar{q} \gamma_{\mu} \frac{\tau^i}{2} q \,, \\
    A_{\mu}^i &\equiv L_{\mu}^i - R_{\mu}^i = \bar{q} \gamma_{\mu} \frac{\tau^i}{2} \gamma^5  q \,,
\end{align} \label{eq:VACurrents}%
\end{subequations}
which can be combined into one antisymmetric $4\times 4$ matrix $(n_{ab})$ by setting
$n_{0i} = A^{0}_{i}$ and $n_{ij} = \varepsilon_{ijk} V_{k}^{0}$.
From the commutation relations of the currents \eqref{eq:VACurrents} one can verify that the $n_{ab}$'s satisfy the $O(4)$ Lie algebra relations 
\begin{equation}
    [n_{ab},n_{cd}] = i \,\left( \delta_{ac}n_{bd}+\delta_{bd}n_{ac}-\delta_{ad}n_{bc}-\delta_{bc}n_{ad} \right) \, . \label{eq:nO4Algebra}
\end{equation}
By Noether's theorem again, the $n_{ab}$'s are the densities of the conserved charges that generate the $O(4)$ transformations.  They thus represent the six charge densities needed for Model~G with $N=4$. In particular, one obtains the commutation relation
\begin{equation}
    [\phi_a,n_{bc}] = i\left(\delta_{ac}\phi_b - \delta_{ab}\phi_c \right) \label{eq:phiNCommutator}
\end{equation}
which describes how the order parameter $\phi$ (as an $O(4)$ vector) transforms under infinitesimal $O(4)$ transformations.

Static universal quantities of the $O(N)$ scalar field theory in $d$-dimensional Euclidean space are encoded in the Landau-Ginzburg-Wilson (LGW) free energy
\begin{equation}
    F=\int_{\vec{x}} \left\{ \frac{1}{2}(\partial^i \phi_a)(\partial^i \phi_a)+\frac{m^2}{2} \phi_a \phi_a + \frac{\lambda}{4!N}(\phi_a\phi_a)^2 +\frac{1}{\orange{4}\chi}n_{ab}n_{ab} \right\} \label{freeEnergy}
\end{equation}
with the shorthand notation
\begin{equation}
    \int_{\vec{x}} \equiv \int d^d x
\end{equation}
for spatial integrals,
and with $m^2<0$ to spontaneously break the $O(N)$ symmetry at low temperatures, and $\lambda > 0$ for stability at large field values.
Thereby $\chi$ denotes the (static) susceptibility of the charge densities $n_{ab}$, which would be microscopically, by Noether's theorem in the corresponding  $\lambda\phi^4$ theory, given by
\begin{equation}
    n_{ab}=\phi_a \frac{\partial}{\partial t}\phi_b-\phi_b \frac{\partial}{\partial t}\phi_a \, .\label{current1}
\end{equation}
Using this definition of the Noether charge densities one readily derives the following explicit expressions for their Poisson brackets,
\begin{equation}
    \{\phi_a, n_{bc}\} =\delta_{ac}\phi_b-\delta_{ab}\phi_c  \label{current2}
\end{equation}
and
\begin{equation}
    \{n_{ab}, n_{cd}\} =\delta_{ac}n_{bd}+\delta_{bd}n_{ac}-\delta_{ad}n_{bc}-\delta_{bc}n_{ad} \, . \label{current3}
\end{equation}
In particular, these correspond to the microscopic commutator expressions in Eqs.~\eqref{eq:nO4Algebra}, \eqref{eq:phiNCommutator} derived from the isovector and isoaxial-vector currents in QCD.\footnote{Poisson brackets and quantum-mechanical commutators are all defined at equal times and implicitly contain spatial $\delta $-functions. They are related via the correspondence principle $\{\cdot,\cdot\} \to - {i}[\cdot,\cdot]$.}

In the following, however, we consider the charge densities $n_{ab}$ as separate degrees of freedom, independent the $\phi$'s. Rather, they can be considered as Hubbard fields to explicitly ensure the conservation of the corresponding Noether currents, with requiring the Poisson-bracket relations in Eqs.~\eqref{current2}, \eqref{current3}.

Close to a second-order phase transition the dynamics is dominated by long-wavelength infrared modes.
As it stands, however, the free energy \eqref{freeEnergy} only encodes information on the static equilibrium distribution and hence only on the `static' universality class of the system. For the `dynamic' universality class we need the complete set of equations of motion, in addition.
The equations of motion for the $O(4)$ Model~G were derived by Rajagopal and Wilczek~\cite{Rajagopal:1992qz}. They are of the same form for any $N$, as described by the SSS Model with $N$ non-conserved order-parameter components reversibly coupled to the conserved $O(N)$ generators  \cite{SASVARI1975108,taeuber_2014},
\begin{subequations}
\begin{align}
    \frac{\partial \phi_a}{\partial t} &= -\Gamma_{0} \frac{\delta F}{\delta \phi_a}+\orange{\frac{g}{2}}\,\{\phi_a,n_{bc}\} \, \frac{\delta F}{\delta n_{bc}}+\theta_a \,, \label{eomop} \\
    \frac{\partial n_{ab}}{\partial t} &= \gamma \vec{\nabla}^2 \frac{\delta F}{\delta n_{ab}}+g\,\{n_{ab},\phi_c\} \, \frac{\delta F}{\delta \phi_c}+\orange{\frac{g}{2}}\,\{n_{ab},n_{cd}\} \, \frac{\delta F}{\delta n_{cd}}+\vec{\nabla} \cdot \vec{\zeta}_{ab} \, . \label{eomcc}
\end{align} \label{eq:eoms}%
\end{subequations}
Here, $\Gamma_{0}$ and $-\gamma\vec{\nabla}^2$ are the kinetic (Onsager) coefficients for the dynamics of the (non-conserved) order parameter and the (conserved) charge densities, respectively. The implicit $\delta$-functions in the equal-time Poisson brackets cancel the spatial integrations over the functional derivatives of the LGW free energy in the reversible force terms with mode coupling constant $g$. The terms $\theta,\vec{\zeta}$ on the right-hand sides generate thermal fluctuations and are implemented as Gaussian white noises with vanishing expectation values,
\begin{align}
    \langle \theta_a(t,\vec{x}) \rangle = 0 \,, \hspace{0.5cm}
    \langle \zeta^{i}_{ab}(t,\vec{x}) \rangle = 0 \,,
\end{align}
and variances 
\begin{subequations}
\begin{align}
    \langle \theta_a(t,\vec{x})\theta_b(t',\vec{x}') \rangle &= 2 \Gamma_{0} T \delta_{ab} \delta(t-t') \delta(\vec{x}-\vec{x}') \,, \\
    \langle \zeta^{i}_{ab}(t,\vec{x}) \zeta^{j}_{cd}(t',\vec{x}') \rangle &= 2\gamma T (\delta_{ac}\delta_{bd}-\delta_{ad}\delta_{bc}) \delta^{ij} \delta(t-t') \delta(\vec{x}-\vec{x}') \,,
\end{align} \label{eq:whiteNoise}%
\end{subequations}
proportional to temperature $T$ from Einstein relations to guarantee 
classical fluctuation-dissipation relation (FDR).
On a structural level, the equations of motion \eqref{eq:eoms} contain $(i)$ irreversible dissipative or diffusive parts,  as well as $(ii)$ reversible conservative parts:

$(i)$ The irreversible terms are the dissipative and diffusive terms with the kinetic coefficients $\Gamma_{0}$ and $-\gamma\vec{\nabla}^2$ but no Poisson brackets, and the thermal noises $\theta$ and $\zeta$.
The irreversible forces are diagonal here, i.e.~only the gradient $\delta F/\delta \phi_a$ of the free energy with respect to $\phi_a$ enters the equation of motion (eom) of $\phi_a$, and only $\delta F/\delta n_{ab}$ enters that of $n_{ab}$. The irreversible terms do not conserve the LGW free energy $F$ and are responsible for driving the system towards the equilibrium (Boltzmann) distribution
\begin{align}
    P_{eq}[\phi,n] &= \frac{1}{Z_{eq}} e^{-F[\phi,n]/T} \label{eq:boltzmannDist}
\end{align}
which also defines the associated equilibrium partition function\footnote{Path integrals over antisymmetric matrices $n_{ab}$ are here defined to run only over the $N(N-1)/2$ independent components, i.e. $$ \int \mathcal{D}n \equiv \int \prod_{a<b} \mathcal{D}n_{ab} $$ together with identifying $n_{ab} \equiv -n_{ba}$ in the integrand for $a > b$.}
\begin{align}
    Z_{eq} = \int \mathcal{D}\phi\,\mathcal{D}n\,e^{-F[\phi,n]/T} \,. \label{eq:partFnc}
\end{align}
In the context of linearized hydrodynamics, the dissipative terms can be generally derived by expansion in conjugate variables (i.e.\ in gradients of the free energy) and requiring entropy production to be positive~\cite{Grossi:2021gqi}.

$(ii)$ The reversible parts are all those that do contain a Poisson bracket. 
These are the conservative but off-diagonal forces and hence traditionally called `reversible mode couplings'.
They arise from the conservative dynamics with the underlying Lie algebra relations of charge densities and order parameter, cf.~Eqs.~\eqref{eq:nO4Algebra} and \eqref{eq:phiNCommutator} or \eqref{current2} and \eqref{current3}.
They conserve the free energy $F$ exactly and are generally present whenever there are non-linear couplings between hydrodynamic modes.
One can readily verify that they do not generate a current in the probability distribution, i.e.~the `streaming velocities'
\begin{align}
    V^{\phi}_{a} &\equiv \orange{\frac{g}{2}}\,\{\phi_a,n_{bc}\} \, \frac{\delta F}{\delta n_{bc}} \,, \hspace{0.5cm} V^{n}_{ab} \equiv g\,
\{n_{ab},\phi_c\} \frac{\delta F}{\delta \phi_c} + \orange{\frac{g}{2}}\,\{n_{ab},n_{cd}\} \, \frac{\delta F}{\delta n_{cd}} \,, \label{eq:streamingVel}
\end{align}
represent a vector field in the $(\phi,n)$ field space that is divergence-free with respect to the equilibrium Boltzmann distribution \eqref{eq:boltzmannDist},
\begin{equation}
    \int_{\vec{x}} \left[ \frac{\delta}{\delta \phi_a} (V^{\phi}_{a} P_{eq} ) + \orange{\frac{1}{2}} \frac{\delta}{\delta n_{ab}} ( V^{n}_{ab} P_{eq} ) \right] = 0 \, . \label{eq:revModeCouplAreDivFree}
\end{equation}
Hence the Boltzmann distribution remains unchanged in the presence of reversible mode couplings.
This reflects the fact that knowing the equilibrium distribution \eqref{eq:boltzmannDist}, which encodes all \emph{static} critical properties, is in general not enough to also know the \emph{dynamic} universality class, because Eqs.~\eqref{eq:streamingVel} do not represent a unique solution to \eqref{eq:revModeCouplAreDivFree} but there are in general multiple others as well \cite{Zinn-Justin:2002ecy}.

The physical interpretation of the reversible mode couplings $(ii)$ in Model~G is rather intuitive: the conserved Noether charges are the generators of $O(N)$ rotations.
Therefore, a non-vanishing field value for some $n$ induces a time-dependent $O(N)$ rotation of the $\phi$'s, as expressed by the Poisson-bracket term in \eqref{eomop}.
Reversibility then requires the presence of a corresponding term in the equation of motion for the charge densities, as represented by the first Poisson bracket in \eqref{eomcc}.
The second Poisson-bracket term in \eqref{eomcc}, which is present for $N>2$, reflects the non-Abelian nature of the $O(N)$ symmetry group with non-vanishing Poisson brackets $\{n_{ab},n_{cd}\} \neq 0$.
It implies that a time-dependent rotation of the $\phi$'s caused by some $n$ automatically comes along with a time-dependent rotation of other components of the conserved charges.
These Poisson brackets  lead to  self-interactions among the conserved charges when the underlying symmetry group that generates the reversible mode couplings is non-Abelian.\footnote{For $N=2$ there is only one independent conserved charge $n \equiv n_{12} = - n_{21}$ and hence no non-vanishing Poisson bracket. The underlying $O(2)$ symmetry is Abelian,  and there are no self-interactions of charge density. In this case, the dynamic universality class is that of Model~E according to Hohenberg and Halperin.}

Ultimately, we are interested in the universal properties of the system defined by Eqs.~\eqref{eq:eoms} and \eqref{freeEnergy} near a critical point, i.e.\ near a second-order phase transition.
These are generally identified with fixed points of the (functional) RG flow. In general, there can be multiple fixed points, corresponding to the possible macroscopic (thermodynamic) phases the system can find itself in. They may be stable or unstable upon flowing towards the infrared.
Linearizing the flow around a given fixed point reveals the set of associated critical exponents. The SSS model admits five different fixed points of the kinetic coefficients, see e.g.\ Ref.\ \cite{PhysRevE.55.4120} and references therein.
These are the usual Gaussian fixed point, the Model A fixed point (where the order parameter decouples from the conserved charges, which leads to purely dissipative dynamics), two `weak-scaling' fixed points where the autocorrelation times of the order parameter and the conserved charges diverge with different (dynamic) critical exponents $z_{\phi}$ and $z_n$, respectively, and of course the `strong-scaling' fixed point where the dynamic critical exponents of both sectors are locked, $z=z_{\phi}=z_n$, and thus the autocorrelation times for both the order parameter and the charge densities diverge at the same rate, $\xi_t \sim \xi^z$. A stability analysis to one-loop order reveals that of the latter three fixed points only the strong-scaling fixed point  is stable  \cite{PhysRevE.55.4120}, so we expect to see strong-scaling behavior in our results below. 

\subsection{Martin-Siggia-Rose path-integral formulation}
\label{sct:MSRpi}

By adding external sources to the free energy,
\begin{equation}
    F\to F -\int_{\vec x}\big(H_a\phi_a+\orange{\frac{1}{2}}\mathcal{H}_{ab}n_{ab}\big) \label{eq:freeEnergyReplacementWithSources}
\end{equation}
we can promote the partition function \eqref{eq:partFnc} to a generating functional for static (equal-time) correlation functions in thermal equilibrium,
\begin{align}
    Z_{eq}[H,\mathcal{H}] = \int \mathcal{D}\phi\,\mathcal{D}n\,\exp\left\{-\beta F[\phi,n] +\beta \int_{\vec{x}}\Big(H_a(\vec{x}) \phi_a(\vec{x})+\orange{\frac{1}{2}}\mathcal{H}_{ab}(\vec{x}) n_{ab}(\vec{x}) \Big) \right\} \label{eq:equilibriumGenFunc}
\end{align}
From its behaviour around the critical point all static universal properties like the critical exponents $\nu$ and $\eta$ may be derived.

To study \emph{dynamic} critical phenomena we need \emph{unequal-time} correlation functions.
A corresponding generating functional can be constructed from 
the Martin-Siggia-Rose (MSR) path-integral formulation \cite{Martin:1973zz,Dominicis:1976,Janssen:1976,Hertz_2017}.
In this approach, one reformulates a thermal expectation value $\langle O(\phi,n) \rangle$ of an operator $O(\phi,n)$ which may involve products of fields at different times as a real-time path integral. As in the Keldysh formalism \cite{Keldysh:1964ud}, this is achieved by doubling the number of fields. In the MSR formalism one introduces auxiliary `response' fields $\tilde{\phi}$ and $\tilde{n}$, and constructs a corresponding MSR action
\begin{align}
    S = \int_x \bigg[ &-\tilde{\phi}_a \left( \frac{\partial \phi_a}{\partial t} + \Gamma_{0} \frac{\delta F}{\delta \phi_a}-\orange{\frac{g}{2}}\{\phi_a,n_{bc}\}\frac{\delta F}{\delta n_{bc}} \right) \nonumber \\
    &-\orange{\frac{1}{2}} \tilde{n}_{ab} \left( \frac{\partial n_{ab}}{\partial t} - \gamma \vec{\nabla}^2 \frac{\delta F}{\delta n_{ab}}-g\{n_{ab},\phi_c\}\frac{\delta F}{\delta \phi_c}-\orange{\frac{g}{2}}\{n_{ab},n_{cd}\}\frac{\delta F}{\delta n_{cd}} \right) \nonumber \\
    &+i T \tilde{\phi}_a \Gamma_{0} \tilde{\phi}_a - \orange{\frac{1}{2}} iT \tilde{n}_{ab} \gamma \vec{\nabla}^2 \tilde{n}_{ab}  \bigg] \label{MSR04}
\end{align}
(note the new variant of our shorthand notation $\int_x  = \int_t\int_{\vec x} =\int dt\, d^dx $ here) such that the expectation value of a multi-time observable $O(\phi,n)$ can be calculated from the corresponding path integral,
\begin{align}
    \langle O(\phi,n) \rangle &= \int \mathcal{D}\phi\,\mathcal{D}\tilde{\phi}\,\mathcal{D}n\,\mathcal{D}\tilde{n}\,\mathcal{J}[\phi,n]\, \exp\left\{ iS \right\} \,O(\phi,n) \,.   \label{path_integral}
\end{align}
In the MSR technique, one also needs to introduce a (generally) field-dependent Jacobian determinant $\mathcal{J}[\phi,n]$ to the path integral, which is the determinant of the Jacobian matrix of the map 
\begin{equation}
    E(\phi,n) \equiv \begin{pmatrix}
        E^{\phi}(\phi,n) & 0 \\
        0 & E^{n}(\phi,n)
    \end{pmatrix}
\end{equation}
from the fields to the fluctuationless parts of the equations of motion, with components
\begin{subequations}
\begin{align}
    E_{a}^{\phi}(\phi,n) &\equiv \frac{\partial \phi_a}{\partial t} + \Gamma_{0} \frac{\delta F}{\delta \phi_a}-\orange{\frac{g}{2}}\{\phi_a,n_{bc}\}\frac{\delta F}{\delta n_{bc}}  \, , \label{opForEomPhi} \\
    E_{ab}^{n}(\phi,n) &\equiv \frac{\partial n_{ab}}{\partial t} - \gamma \vec{\nabla}^2 \frac{\delta F}{\delta n_{ab}}-g\{n_{ab},\phi_c\}\frac{\delta F}{\delta \phi_c}-\orange{\frac{g}{2}}\{n_{ab},n_{cd}\}\frac{\delta F}{\delta n_{cd}} \, , \label{opForEomN}
\end{align} \label{eq:opsForEoms}%
\end{subequations}
i.e.\ the operators to which the response fields $\tilde{\phi}$ and $\tilde{n}$ couple linearly in the MSR action \eqref{MSR04}. With these the stochastic equations of motion \eqref{eq:eoms} can then compactly be  written as
\begin{equation}
    E_{a}^{\phi}(\phi,n) = \theta_a \,, \hspace{0.5cm} E_{ab}^{n}(\phi,n) = \vec{\nabla} \!\cdot\! \vec{\zeta}_{ab} \,.
\end{equation}
Before we continue with the construction of the generating functional in the next subsection, in the following, for the convenience of the reader, we briefly review the structure of the MSR action \eqref{MSR04} and the conceptual steps in its derivation.

Generally, a thermal expectation value $\langle O(\phi,n)\rangle$ of an operator $O(\phi,n)$ is given by an average of the observable over an ensemble of solutions $\phi_{\theta,\zeta},n_{\theta,\zeta}$ to the equations of motion \eqref{eq:eoms},
\begin{align}
    \langle O(\phi,n) \rangle &= \int \mathcal{D}\theta\,\mathcal{D}\zeta\,P[\theta,\vec{\zeta}]\,O(\phi_{\theta,\zeta},n_{\theta,\zeta}) \label{eq:thermExpVal}
\end{align}
where the specific noise instances $\theta(t,\vec{x})$, $\vec{\zeta}(t,\vec{x})$ are randomly drawn from a Gaussian distribution with probability (density)
\begin{equation}
    P[\theta,\vec{\zeta}] = \exp\left\{ -\int_x \frac{\theta_a(x) \theta_a(x)}{4\Gamma_{0} T} \right\}\exp\left\{ -\int_x \frac{\zeta_{ab}^i(x) \zeta_{ab}^i(x)}{\orange{8}\gamma T} \right\} \,,
\end{equation}
which implements the white noise statistics \eqref{eq:whiteNoise}, with variances set by the fluctuation-dissipation relation for thermal equilibrium at temperature $T$.
One can now insert a trivial unity into \eqref{eq:thermExpVal},  to introduce (path) integrals over $\phi_a(t,\vec{x})$ and $n_{ab}(t,\vec{x})$,
\begin{equation}
    1 = \int \mathcal{D}\phi\,\mathcal{D}n\,\delta(\phi-\phi_{\theta,\zeta})\,\delta(n-n_{\theta,\zeta}) \,. \label{eq:oneInPathIntegral}
\end{equation}
The $\delta$-functionals enforce the fields $\phi$, $n$ in the path integral to be solutions to the stochastic equations of motion \eqref{eq:eoms}.
With the functional generalization of the identity 
$\delta(x-x_{0}) = |f'(x)|\, \delta(f(x)) $, for a function $f(x)$ with a unique root at $x=x_{0}$, we can rewrite the $\delta$-distributions in \eqref{eq:oneInPathIntegral} to arrive at the somewhat more convenient form
\begin{equation}
    \delta(\phi-\phi_{\theta,\zeta}) \, \delta(n-n_{\theta,\zeta}) = \mathcal{J}[\phi,n] \, \delta(E^{\phi} - \theta) \, \delta(E^{n} - \vec{\nabla} \cdot \vec{\zeta})  \label{eq:deltaFunctionalSubstitution}
\end{equation}
This expresses the fact that only the solutions to the equations of motion contribute to the path integral, i.e.~those that satisfy $E_{a}^{\phi}(\phi,n) - \theta_a = 0$ and $E_{ab}^{n}(\phi,n) - \vec{\nabla} \cdot \vec{\zeta}_{ab} = 0$.
This however necessitates the introduction of the  functional Jacobian $\mathcal{J}[\phi,n] $
(which can be set to unity with a suitable time discretization, as discussed below). Next, one uses Fourier representations of the $\delta$-functionals in \eqref{eq:deltaFunctionalSubstitution}, introducing the auxiliary `response' fields $\tilde{\phi}_a$ and $\tilde{n}_{ab}$,
\begin{equation}
    \delta(E^{\phi}_a - \theta_a) \, \delta(E^{n}_{ab} - \vec{\nabla} \cdot \vec{\zeta}_{ab}) = \int\mathcal{D}\tilde{\phi}\,\mathcal{D}\tilde{n}\,\exp\left\{-i\int_x \tilde{\phi}_a (E_a^{\phi} - \theta_a) - \frac{i}{\orange{2}}\int_x \tilde{n}_{ab}(E_{ab}^{n}-\vec{\nabla}\cdot\vec{\zeta}_{ab}) \right\} \,.
\end{equation}
In total, one then arrives at
\begin{align}
    \langle O(\phi,n) \rangle &= \int \mathcal{D}\theta\,\mathcal{D}\zeta\,\mathcal{D}\phi\,\mathcal{D}n\,\mathcal{D}\tilde{\phi}\,\mathcal{D}\tilde{n}\,\mathcal{J}[\phi,n] \,P[\theta,\vec{\zeta}]\,\times \\ \nonumber
    &\hspace{0.5cm} \exp\left\{-i\int_x \tilde{\phi}_a (E_a^{\phi} - \theta_a) - \frac{i}{\orange{2}}\int_x \tilde{n}_{ab}(E_{ab}^{n}-\vec{\nabla}\cdot\vec{\zeta}_{ab}) \right\} O(\phi,n)
\end{align}
for the thermal expectation value.
The (path) integral over the stochastic noises is now Gaussian and can be performed analytically, yielding
\begin{align}
    \langle O(\phi,n) \rangle &= \int \mathcal{D}\phi\,\mathcal{D}n\,\mathcal{D}\tilde{\phi}\,\mathcal{D}\tilde{n}\,\mathcal{J}[\phi,n]\,\times \\ \nonumber
    &\hspace{0.5cm} \exp\left\{i\int_x \left[ -\tilde{\phi}_a E_a^{\phi} - \orange{\frac{1}{2}} \tilde{n}_{ab} E_{ab}^{n}  +i T \tilde{\phi}_a \Gamma_{0} \tilde{\phi}_a - \orange{\frac{1}{2}} i T \tilde{n}_{ab} \gamma \vec{\nabla}^2 \tilde{n}_{ab} \right] \right\} O(\phi,n)
\end{align}
which is precisely the anticipated expression \eqref{path_integral} with the shorthand notation \eqref{eq:opsForEoms} for the fluctuationless parts in the equations of motion.
The MSR path integral \eqref{path_integral} is thus equivalent to \eqref{eq:thermExpVal}, but more convenient for a field-theoretic treatment.

We close this subsection with a few comments on the Jacobian.
On a fundamental level, the Jacobian ensures the normalization condition $\langle 1 \rangle = 1$ of the MSR path integral \eqref{path_integral}, which is expressed in the definition of a real-time partition function
\begin{equation}
    Z \equiv \int \mathcal{D}\phi\,\mathcal{D}\tilde{\phi}\,\mathcal{D}n\,\mathcal{D}\tilde{n}\,\mathcal{J}[\phi,n]\, \exp\left\{ iS \right\}   = 1 \,,  \label{eq:realTimePartFnc}
\end{equation}
defined as the MSR path integral \eqref{path_integral} when no observables are measured.
Notably, this normalization condition is not just a  choice of an overall (constant) normalization factor, but implies that for field-independent observables in \eqref{path_integral} the path integral becomes `topological'.
We discuss this particular property of the MSR technique  in Appendices \ref{sct:contFormOfIto} and \ref{sct:jacobianAndGhosts}, as it is related to a hidden Becchi-Rouet-Stora-Tyutin (BRST) symmetry.

In the present case, we can achieve $\mathcal{J}[\phi,n] = 1$ by choosing a retarded (Ito) discretization of the stochastic equations of motion.
There are other possible discretizations, however, in which $\mathcal{J}[\phi,n] \neq 1$, e.g.\ in the Stratonovich discretization.
The fact that the choice of discretization is not unique becomes particularly important in systems with \emph{multiplicative} noise, where the stochastic force terms in the equations of motion \eqref{eq:eoms} are field dependent.
Under such circumstances  different discretizations can indeed lead to different continuum limits, which has historically also been known as the Ito–Stratonovich dilemma.
In fact, already in the present case, without multiplicative noise here, one has to be slightly careful in obtaining the correct continuum limit of the Ito discretization \cite{Canet:2011wf}. 
We defer a discussion of this particular subtlety to Appendix \ref{sct:contFormOfIto}, and continue with the construction of a generating functional in the next subsection.

\subsection{Generating functional for systems with reversible mode couplings}
\label{sct:genFunc}

A generating functional for unequal-time correlation functions can be constructed by introducing external source terms in the general definition of an expectation value~\eqref{path_integral}.
First of all, \emph{unphysical} sources $\tilde{H}_a$ and $\tilde{\mathcal{H}}_{ab}$ which are conjugate to the classical fields $\phi_a$ and $n_{ab}$ are simply introduced as linear coupling terms on the level of the MSR action~\eqref{MSR04},
\begin{equation}
    S \to S + \int_x \left( \tilde{H}_a \phi_a + \orange{\frac{1}{2}} \tilde{\mathcal{H}}_{ab} n_{ab} \right) \,. \label{eq:unphyssources}
\end{equation}
The partition function \eqref{eq:realTimePartFnc} then becomes a functional of $\tilde{H}$ and $\tilde{\mathcal{H}}$.
Expectation values of classical fields can then be obtained as usual by functional differentiation of the partition function in presence of these sources, e.g.
\begin{equation}
    \langle \phi_a(x) \rangle = -i\frac{\delta Z[\tilde{H},\tilde{\mathcal{H}}]}{\delta \tilde{H}_a(x)} \bigg\rvert_{\tilde{H}=\tilde{\mathcal{H}}=0} \,, \hspace{0.5cm} \langle n_{ab}(x) \rangle = -i\frac{\delta Z[\tilde{H},\tilde{\mathcal{H}}]}{\delta \tilde{\mathcal{H}}_{ab}(x)} \bigg\rvert_{\tilde{H}=\tilde{\mathcal{H}}=0} \,,
\end{equation}
where $Z$ denotes the generating functional from here on, i.e.\ the partition function in the presence of external sources.
One has to be more careful when one introduces \emph{physical} source terms $H$ and $\mathcal{H}$  conjugate to the response fields. They should be included in a way that the dynamic response functions (i.e.\ the retarded propagators) of the system can be calculated via 
\begin{subequations}
\begin{align}
    G_{\phi,ab}^R(x,x')_{\text{unconn.}} &=\frac{\delta \langle\phi_a(x)\rangle_{H,\mathcal{H}}}{\delta H_b(x')} \bigg\rvert_{H,\mathcal{H}=0} \,, \\
    G_{n,abcd}^R(x,x')_{\text{unconn.}} &=\frac{\delta \langle n_{ab}(x)\rangle_{H,\mathcal{H}}}{\delta \mathcal{H}_{cd}(x')} \bigg\rvert_{H,\mathcal{H}=0} \,. 
\end{align} \label{eq:responseFunctions}%
\end{subequations}
where the subscript `unconn.' indicates that these denote \emph{unconnected} correlation functions, i.e.\ the disconnected components are not subtracted here.
As it turns out, the correct way of introducing such physical source terms is to add them one level earlier: Instead of adding the corresponding source terms directly to the MSR action, as done for the unphysical ones in \eqref{eq:unphyssources}, we add the physical sources to the free energy as in~\eqref{eq:freeEnergyReplacementWithSources} for the static case, i.e.~writing
\begin{equation}
    F\to F -\int_{\vec x}\big(H_a\phi_a+\orange{\frac{1}{2}}\mathcal{H}_{ab}n_{ab}\big) \, ,
\end{equation}
which can now be time dependent, however, to be able to compute unequal-time correlations functions. This formal replacement is then used to obtain correspondingly modified deterministic force terms inside the MSR action in a second step. 
We will elaborate on the reasoning for this choice in more detail below.
Crucially, when introduced in this way, the physical sources do not couple  to the standard response fields $\tilde{\phi}_a$ and $\tilde{n}_{ab}$ directly, but to the following composite operators
\begin{subequations}
\begin{eqnarray}
    \tilde{\Phi}_a &\equiv& \Gamma_0 \tilde{\phi}_a - \orange{\frac{g}{2}} \tilde{n}_{bc} \{n_{bc}, \phi_a\} \,, \label{PhiTildeTrafo} \\
    \tilde{N}_{ab} &\equiv&  -\gamma \vec{\nabla}^2 \tilde{n}_{ab} - \orange{\frac{g}{2}} \tilde{n}_{cd} \{n_{cd},n_{ab}\} - g \tilde{\phi}_c\{\phi_c,n_{ab}\} \, .  \label{NTildeTrafo}
\end{eqnarray} \label{eq:responseFieldTrafo}%
\end{subequations}
This is because they then couple to all terms in the equations of motion \eqref{eq:eoms} that contain ($d$-dimensional) functional derivatives of the free energy. In addition to the dissipative terms, this includes contributions from the reversible mode couplings, i.e.~Poisson-bracket terms in the equations of motion. 
In fact, it has been argued that this is the natural way in which physical source terms \emph{should} appear in effective field theories of dissipative hydrodynamics, see e.g.~Ref.~\cite{Harder:2015nxa}.
One can  straightforwardly verify that by including (time-independent)  source terms in this way, the stationary state of the system is still the Boltzmann distribution (but with the free energy \eqref{eq:freeEnergyReplacementWithSources} which then includes the static sources, of course).
Correspondingly, the fluctuation-dissipation relation (FDR) in its usual form based on Kubo-Martin-Schwinger conditions \cite{doi:10.1143/JPSJ.12.570,PhysRev.115.1342} is maintained, even in the presence of static external sources.
This is not entirely trivial in a system with reversible mode couplings.
In fact, if one were to naively couple the sources directly to the elementary response fields $\tilde{\phi}_a$ and $\tilde{n}_{ab}$ in the MSR action \eqref{MSR04}, such a system would no longer approach the corresponding 
equilibrium Boltzmann distribution. This is shown in Appendix~\ref{traditionalapproach}.

Therefore, introducing  the sources as described, we arrive at the generating functional 
\begin{align}
    Z[H,\mathcal{H},\tilde{H},\tilde{\mathcal{H}}] &= \int \mathcal{D}\phi\,\mathcal{D}\tilde{\phi}\,\mathcal{D}n\,\mathcal{D}\tilde{n}\, \exp\bigg\{ iS[\phi,n,\tilde{\phi},\tilde{n}] + i\int_x \big( \tilde{H}_a \phi_a + \orange{\frac{1}{2}} \tilde{\mathcal{H}}_{ab} n_{ab} \big) \; + \nonumber\\ &\hspace{2.4cm} i\int_x \big( \Gamma_{0} \tilde{\phi}_a + 
    \orange{\frac{g}{2}}\{\phi_a,n_{bc}\}\tilde{n}_{bc} \big) H_a +\\
    &\hspace{1cm} \frac{i}{\orange{2}}\int_x \big( -\gamma\vec{\nabla}^2 \tilde{n}_{ab} + g\{n_{ab},\phi_c\} \tilde{\phi}_c + \orange{\frac{g}{2}}\{n_{ab},n_{cd}\}\tilde{n}_{cd} \big) \mathcal{H}_{ab} \bigg\} \,, \nonumber
\end{align}
where we explicitly see that the sources $H_{a}$ and $\mathcal{H}_{ab}$ couple to the composite fields \eqref{eq:responseFieldTrafo}. 
This formulation of the generating functional will form the basis of our treatment of Model~G within the functional renormalization group in Section~\ref{sct:frg} below.

Note that the composite response fields \eqref{eq:responseFieldTrafo} have the structure of a field-dependent linear transformation of the standard response fields,
\begingroup 
\setlength\arraycolsep{10pt}
\begin{equation}
    \begin{pmatrix}
        \tilde{\Phi}_a \\ \tilde{N}_{bc} 
    \end{pmatrix} =
    \begin{pmatrix}
        \delta_{ad} \Gamma_0  & \frac{g}{\orange{2}}\{\phi_a,n_{ef}\} \\
        g\{n_{bc},\phi_d\} & -\tfrac{1}{2}\left(\delta_{be}\delta_{cf}-\delta_{bf}\delta_{ce}\right) \gamma\vec{\nabla}^2 + \frac{g}{\orange{2}}\{n_{bc},n_{ef}\} 
    \end{pmatrix}
    \begin{pmatrix}
        \tilde{\phi}_d \\ \tilde{n}_{ef}
    \end{pmatrix} \,. \label{eq:responseFieldTrafoInMatrixNotation}
\end{equation}
\endgroup
We emphasize that the matrix in \eqref{eq:responseFieldTrafoInMatrixNotation} depends on the `classical' fields $\phi$ and $n$, but \emph{not} on the response fields $\tilde{\phi}$ and $\tilde{n}$. In particular, the reversible mode couplings with the Poisson brackets \eqref{current2} and \eqref{current3}  only introduce classical fields $\{\phi,n\} \sim \phi$, $\{n,n\}\sim n$, but no response fields. 
Because this field-dependent matrix is used repeatedly throughout this work, we introduce the symbol $J(\psi)$ for it together with a compact `superfield' notation $\psi \equiv (\phi,n)$, $\tilde{\Psi} \equiv (\tilde{\Phi},\tilde{N})$ and $\tilde{\psi} \equiv (\tilde{\phi},\tilde{n})$.
We can then compactly write $\tilde{\Psi} = J(\psi) \tilde{\psi}$, and express the field-dependent matrix $J(\psi)$ as
\begingroup 
\setlength\arraycolsep{10pt} 
\begin{align}
    J(x,x';\psi ) &= \frac{\delta \tilde{\Psi}(x)}{\delta \tilde{\psi}(x')} \\ \nonumber
    &= \begin{pmatrix}
        \delta_{ad} \Gamma_0  & \frac{g}{\orange{2}}\{\phi_a,n_{ef}\}(x) \\
        g\{n_{bc},\phi_d\}(x) & -\tfrac{1}{2}\left(\delta_{be}\delta_{cf}-\delta_{bf}\delta_{ce}\right) \gamma\vec{\nabla}_{\vec{x}}^2 + \frac{g}{\orange{2}}\{n_{bc},n_{ef}\}(x)
    \end{pmatrix} \, \delta(x-x') \, .
\end{align}
\endgroup
Instead of formulating the path integral in terms of the standard response fields $\tilde{\phi}$ and $\tilde{n}$, we can also perform the change of variables \eqref{eq:responseFieldTrafo} directly on the level of the path integral and formulate the action in terms of the composite response fields $\tilde{\Phi}$ and $\tilde{N}$. I.e.,
using the definition of the composite fields in  \eqref{eq:responseFieldTrafo}
as a field transformation in the path integral \eqref{path_integral} then yields
\begin{align}
    Z[H,\mathcal{H},\tilde{H},\tilde{\mathcal{H}}] &= \int \mathcal{D}\phi\,\mathcal{D}\tilde{\Phi}\,\mathcal{D}n\,\mathcal{D}\tilde{N}\, \mathcal{J}'[\phi,n] \, \exp\bigg\{ iS'[\phi,n,\tilde{\Phi},\tilde{N}] \; + \label{eq:genFncAfterFieldTrafo} \\ \nonumber &\hspace{3.0cm}  i\int_x \big( \tilde{H}_a \phi_a + \orange{\frac{1}{2}} \tilde{\mathcal{H}}_{ab} n_{ab} + \tilde{\Phi}_a H_a + \orange{\frac{1}{2}} \tilde{N}_{ab} \mathcal{H}_{ab} \big)  \bigg\} 
\end{align}
with a formally different bare action $S'$ that is obtained by substituting the response fields with the new composite fields,
\begin{equation}
    S'[\phi,n,\tilde{\Phi},\tilde{N}] = S[\phi,n,\tilde{\phi},\tilde{n}] \, , \label{eq:transformedMSRAction}
\end{equation}
and with a field-dependent Jacobian $\mathcal{J}'[\phi,n]$ that arises due to the transformation of the measure.
The transformed action $S'$ is spatially non-local in the new fields since the inverse of the field transformation \eqref{eq:responseFieldTrafo} is needed in its definition \eqref{eq:transformedMSRAction}, which involves inverting the spatial Laplacian $\vec{\nabla}^2$.
Moreover, because the field transformation is non-linear, the Jacobian determinant in \eqref{eq:genFncAfterFieldTrafo}, 
\begin{equation}
    \mathcal{J}'[\phi,n] = \left| \det \begin{pmatrix} \frac{\delta \tilde{\phi}}{\delta \tilde{\Phi}} & \frac{\delta \tilde{\phi}}{\delta \tilde{N}} \\ \frac{\delta \tilde{n}}{\delta \tilde{\Phi}} & \frac{\delta \tilde{n}}{\delta \tilde{N}} \end{pmatrix} \right| = \left| \det J^{-1}[\phi, n] \right| \,  \label{eq:jacobianOfNonlinearFieldTrafo}
\end{equation}
depends on the fields $\phi$ and $n$, and thus needs special attention.
Generally speaking, the sole purpose of such a Jacobian determinant in the MSR path integral is to ensure that the normalization condition $\langle 1 \rangle = 1$ is maintained.
On a diagrammatic level, this requires the cancellation of all acausal diagrams\footnote{These are all diagrams which would otherwise vanish in the Keldysh/MSR formalism by standard causality arguments, for example those diagrams where two points $x$ and $x'$ are connected both by a retarded and an advanced propagator, $G^R(x,x') G^A(x,x') = 0$, or where a retarded/advanced propagator closes into a loop, $G^{R/A}(x,x) = 0$.} by corresponding `ghost' diagrams, which arise from anti-commuting ghost degrees of freedom $\tilde{c}$ and $c$ introduced to represent the Jacobian determinant as an integral over Grassmann variables,
\begin{equation}
    \mathcal{J}'[\phi,n] = \int \mathcal{D}\tilde{c}\,\mathcal{D}c\,\exp\left\{ -\int_{xx'} \tilde{c}(x) \frac{\delta \tilde{\psi}(x)}{\delta \tilde{\Psi}(x')} c(x') \right\} \,, \label{eq:jacobianViaGhosts}
\end{equation}
here again expressed in our superfield notation.
Knowing that this cancellation must occure, in practice, one may therefore simply dismiss the Jacobian together with all acausal diagrams \cite{Gao:2018bxz}. 
For more technical details on the Jacobian, see Appendix~\ref{sct:jacobianAndGhosts}.

The unconnected physical response functions (i.e.\ the unconnected retarded propagators) are  obtained as second functional derivatives of the generating functional
\begin{align}
    G_{\phi,ab}^R(x,x')_{\text{unconn.}} &= -i\frac{\delta^2 Z[H,\mathcal{H},\tilde{H},\tilde{\mathcal{H}}]}{\delta \tilde{H}_a(x) \delta H_b(x')} \bigg\rvert_{H,\mathcal{H}=0} \,, \\
    G_{n,abcd}^R(x,x')_{\text{unconn.}} &= -i\frac{\delta^2 Z[H,\mathcal{H},\tilde{H},\tilde{\mathcal{H}}]}{\delta \tilde{\mathcal{H}}_{ab}(x) \delta \mathcal{H}_{cd}(x')} \bigg\rvert_{H,\mathcal{H}=0} \,.
\end{align}
They describe the dynamic response at spacetime point $x=(t,\vec{x})$ of the system with respect to an external perturbation applied at $x'=(t',\vec{x}')$, cf.~\eqref{eq:responseFunctions}.
\emph{Connected} correlation functions are obtained from the corresponding Schwinger functional
\begin{equation}
    W[H,\mathcal{H},\tilde{H},\tilde{\mathcal{H}}] = -i\log Z[H,\mathcal{H},\tilde{H},\tilde{\mathcal{H}}] \, .
\end{equation}
For example, the connected propagators (here in the presence of arbitrary background sources $H(x)$, $\mathcal{H}(x)$) are given by
\begin{subequations}
\begin{align}
    iF_{\phi,ab}(x,x') &= \frac{\delta^2 W[H,\mathcal{H},\tilde{H},\tilde{\mathcal{H}}]}{\delta \tilde{H}_a(x) \delta \tilde{H}_b(x')} = i\langle \phi_a(x) \phi_b(x') \rangle - i\langle \phi_a(x)\rangle\langle\phi_b(x') \rangle \,, \\
    G_{\phi,ab}^R(x,x') &= \frac{\delta^2 W[H,\mathcal{H},\tilde{H},\tilde{\mathcal{H}}]}{\delta \tilde{H}_a(x) \delta H_b(x')} = i\langle \phi_a(x) \tilde{\Phi}_b(x') \rangle - i\langle \phi_a(x)\rangle\langle\tilde{\Phi}_b(x') \rangle \,, \\
    G_{\phi,ab}^A(x,x') &= \frac{\delta^2 W[H,\mathcal{H},\tilde{H},\tilde{\mathcal{H}}]}{\delta H_a(x) \delta \tilde{H}_b(x')} = i\langle \tilde{\Phi}_a(x) \phi_b(x') \rangle - i\langle \tilde{\Phi}_a(x)\rangle\langle\phi_b(x') \rangle \,, \\
    i\widetilde{F}_{\phi,ab}(x,x') &= \frac{\delta^2 W[H,\mathcal{H},\tilde{H},\tilde{\mathcal{H}}]}{\delta H_a(x) \delta H_b(x')} = i\langle \tilde{\Phi}_a(x) \tilde{\Phi}_b(x') \rangle - i\langle \tilde{\Phi}_a(x)\rangle\langle\tilde{\Phi}_b(x') \rangle \,,
\end{align} \label{eq:phiPropFromW}%
\end{subequations}
and,
\begin{subequations}
\begin{align}
    iF_{n,abcd}(x,x') &= \frac{\delta^2 W[H,\mathcal{H},\tilde{H},\tilde{\mathcal{H}}]}{\delta \tilde{\mathcal{H}}_{ab}(x) \delta \tilde{\mathcal{H}}_{cd}(x')} = i\langle n_{ab}(x) n_{cd}(x') \rangle - i\langle n_{ab}(x)\rangle\langle n_{cd}(x') \rangle \,, \\
    G_{n,abcd}^R(x,x') &= \frac{\delta^2 W[H,\mathcal{H},\tilde{H},\tilde{\mathcal{H}}]}{\delta \tilde{\mathcal{H}}_{ab}(x) \delta \mathcal{H}_{cd}(x')} = i\langle n_{ab}(x) \tilde{N}_{cd}(x') \rangle - i\langle n_{ab}(x)\rangle\langle\tilde{N}_{cd}(x') \rangle \,, \\
    G_{n,abcd}^A(x,x') &= \frac{\delta^2 W[H,\mathcal{H},\tilde{H},\tilde{\mathcal{H}}]}{\delta \mathcal{H}_{ab}(x) \delta \tilde{\mathcal{H}}_{cd}(x')} = i\langle \tilde{N}_{ab}(x) n_{cd}(x') \rangle - i\langle \tilde{N}_{ab}(x)\rangle\langle n_{cd}(x') \rangle \,, \\
    i\widetilde{F}_{n,abcd}(x,x') &= \frac{\delta^2 W[H,\mathcal{H},\tilde{H},\tilde{\mathcal{H}}]}{\delta \mathcal{H}_{ab}(x) \delta \mathcal{H}_{cd}(x')} = i\langle \tilde{N}_{ab}(x) \tilde{N}_{cd}(x') \rangle - i\langle \tilde{N}_{ab}(x)\rangle\langle\tilde{N}_{cd}(x') \rangle \,,
\end{align} \label{eq:nPropFromW}%
\end{subequations}
The effective action, the generating functional of the one-particle irreducible (1PI) correlation functions, is obtained from a Legendre transformation of the Schwinger functional,
\begin{equation}
    \Gamma[\bar{\phi},\bar{n},\bar{\tilde{\Phi}},\bar{\tilde{N}}] = W[H,\mathcal{H},\tilde{H},\tilde{\mathcal{H}}] - \int_x \Big( \tilde{H}_a \bar{\phi}_a + \orange{\frac{1}{2}} \tilde{\mathcal{H}}_{ab} \bar{n}_{ab} + H_a \bar{\tilde{\Phi}}_a + \orange{\frac{1}{2}} \mathcal{H}_{ab} \bar{\tilde{N}}_{ab} \Big) \, \label{eq:effActionDef}
\end{equation}
with
\begin{align}
    \bar{\phi}_a(x) &= \frac{\delta W[H,\mathcal{H},\tilde{H},\tilde{\mathcal{H}}]}{\delta \tilde{H}_a(x)} \,, \hspace{0.5cm} \bar{\tilde{\Phi}}_a(x) = \frac{\delta W[H,\mathcal{H},\tilde{H},\tilde{\mathcal{H}}]}{\delta H_a(x)} \,, \\
    \bar{n}_{ab}(x) &= \frac{\delta W[H,\mathcal{H},\tilde{H},\tilde{\mathcal{H}}]}{\delta \tilde{\mathcal{H}}_{ab}(x)} \,, \hspace{0.5cm} \bar{\tilde{N}}_{ab}(x) = \frac{\delta W[H,\mathcal{H},\tilde{H},\tilde{\mathcal{H}}]}{\delta \mathcal{H}_{ab}(x)} \, .
\end{align}
We have thereby indicated the field expectation values by bars.
Since the classical sources actually couple to the composite operators \eqref{eq:responseFieldTrafo},  the effective action \eqref{eq:effActionDef} is not a functional of the expectation values $\langle \tilde{\phi}_a(x)\rangle$ and $\langle \tilde{n}_{ab}(x)\rangle$ of the standard response fields, but instead a function of the expectation values of the composite operators  \eqref{eq:responseFieldTrafo}, i.e.~of
\begin{subequations}
\begin{eqnarray}
    \bar{\tilde{\Phi}}_a(x) &=& \Gamma_0 \langle \tilde{\phi}_a(x) \rangle - \orange{\frac{g}{2}} \langle \tilde{n}_{bc}(x) \{n_{bc}, \phi_a\}(x) \rangle \, ,\label{PhiTildeExpVal} \\
    \bar{\tilde{N}}_{ab}(x) &=&  -\gamma \vec{\nabla}^2 \langle \tilde{n}_{ab}(x) \rangle - \orange{\frac{g}{2}} \langle \tilde{n}_{cd}(x) \{n_{cd},n_{ab}\}(x) \rangle - g \langle \tilde{\phi}_c(x) \{\phi_c,n_{ab}\}(x)  \rangle \,. \label{NTildeExpVal}
\end{eqnarray} \label{eq:responseFieldsExpVal}%
\end{subequations}
In the following we will drop the bars from the field expectation values again since there will be generally no confusion with the fields appearing in the path integral \eqref{eq:genFncAfterFieldTrafo}.
The various propagators \eqref{eq:phiPropFromW} and \eqref{eq:nPropFromW} are then dependent on the background field expectation values instead of the external sources and can be expressed by 
\begin{align}
    G[\psi,\tilde{\Psi}] = -(\Gamma^{(2)}[\psi,\tilde{\Psi}])^{-1} \label{eq:propAsInvGam2}
\end{align}
where $\Gamma^{(2)}[\psi,\tilde{\Psi}]$ denotes the Hessian matrix of $\Gamma$ in the superfield notation.
In general, we indicate functional derivatives by superscripts, e.g.
\begin{align}
    \Gamma^{\tilde{\Psi}_i \psi_j}[\psi,\tilde{\Psi}](x,y) \equiv \frac{\delta^2 \Gamma[\psi,\tilde{\Psi}]}{\delta \tilde{\Psi}_i(x)\delta \psi_j(y)}
\end{align}
where $i,j$ here are indices in superfield space.
When written in components, the compact relation \eqref{eq:propAsInvGam2} explicitly expands to \cite{Berges:2012ty}
\begin{subequations}
\begin{align}
    G^R &= -\left\{ \Gamma^{\tilde{\Psi}\psi} - \Gamma^{\tilde{\Psi}\tilde{\Psi}}  \circ (\Gamma^{\psi\tilde{\Psi}})^{-1} \circ \Gamma^{\psi\psi} \right\}^{-1} \,, \\
    G^A &= -\left\{ \Gamma^{\psi\tilde{\Psi}} - \Gamma^{\psi\psi}  \circ (\Gamma^{\tilde{\Psi}\psi})^{-1} \circ \Gamma^{\tilde{\Psi}\tilde{\Psi}} \right\}^{-1} \,, \\
    iF &= -\left\{ \Gamma^{\psi\psi} - \Gamma^{\psi\tilde{\Psi}}  \circ (\Gamma^{\tilde{\Psi}\tilde{\Psi}})^{-1} \circ \Gamma^{\tilde{\Psi}\psi} \right\}^{-1} \,, \\
    i\widetilde{F} &= -\left\{ \Gamma^{\tilde{\Psi}\tilde{\Psi}} - \Gamma^{\tilde{\Psi}\psi}  \circ (\Gamma^{\psi\psi})^{-1} \circ \Gamma^{\psi\tilde{\Psi}} \right\}^{-1} \,, 
\end{align} \label{eq:propsAsInvsOfGam2}%
\end{subequations}
for the retarded, advanced, statistical, and anomalous propagators in superfield space.
We have suppressed the two superfield indices and the two spacetime arguments on all two-point functions and propagators in \eqref{eq:propsAsInvsOfGam2}, so that all objects are to be interpreted as matrices (and their inverses) in spacetime and in superfield space.
Moreover, the symbol $\circ$ denotes matrix multiplication in superfield space as well as integration over adjacent spacetime coordinates, e.g.
\begin{equation}
    (A \circ B)_{ij}(x,y) \equiv \int_{z} A_{il}(x,z) B_{lj}(z,y) 
\end{equation}
with superfield indices $i,j,l$, spacetime points $x,y,z$, and implicit summation over $l$.

To conclude this subsection, we furthermore note that our rule for functional 
derivatives with respect to the elements of antisymmetric  $N\!\times\! N$ matrices such as $n_{ab}=-n_{ba}$, with only $N(N-1)/2$ independent components, is defined by the antisymmetric fundamental functional derivative as follows,
\begin{equation}
    \orange{\frac{\delta n_{ab}(x)}{\delta n_{cd}(y)} = (\delta_{ac}\delta_{bd}-\delta_{ad}\delta_{bc}) \, \delta(x-y)} \, . \label{eq:funcDerivDefForN}
\end{equation}

\subsection{Symmetries of the MSR action}
\label{sct:SymmetriesOfMSRAction}

Symmetries generally provide powerful tools for practical calculations. For instance, they give rise to Ward identities of the effective action and hence to non-trivial relations between the $n$-point correlation functions. If the symmetry is continuous Noether's theorem implies an associated conserved current, which means that continuous symmetries can be used as a guideline for the degrees of freedom that are part of an effective hydrodynamic description.
Moreover, as a peculiarity of the real-time formalism, the question whether a system is in thermal equilibrium is also expressed as the presence (or absence) of a discrete symmetry, as discussed in \cite{Sieberer:2015hba}.
In the context of the FRG, one can show that a given symmetry is exactly conserved by the FRG flow~\cite{Gies:2006wv} (assuming that the regulator does not explicitly break the symmetry).
This means that possible truncation schemes are restricted to those which conserve all symmetries that are already present at tree level.
With this motivation in mind, our goal in this subsection is to discuss the various symmetries of the MSR action \eqref{MSR04}.
We will use some of these symmetries in later sections to derive some important general results such as the FRG-scale independence of the reversible mode coupling $g$, or the independence of the `static' FRG flow for the free energy of any real-time quantity, for example. They also allow one to restrict the possible operators that can occur within a given truncation of the FRG flow. 

\paragraph{Thermal equilibrium.}
Thermal equilibrium is generally expressed as a symmetry of the MSR action \cite{Sieberer:2015hba}. The symmetry transformation can be most directly expressed as a transformation of the composite response fields \eqref{eq:responseFieldTrafo}, and is given by
\begin{align}
    \mathcal{T}_\beta \begin{pmatrix}
    \psi_a(\omega,\vec{p})\\
    \tilde{\Psi}_a(\omega,\vec{p})
    \end{pmatrix} &=
    \begin{pmatrix}
    1& \phantom{-}0\phantom{-} \\
    -\beta\omega& \phantom{-}1 \phantom{-}
    \end{pmatrix}
    \begin{pmatrix}
    \epsilon_a \psi_a(-\omega,\vec{p})\\
    \epsilon_a \tilde{\Psi}_a(-\omega,\vec{p})
    \end{pmatrix} \label{eq:thermEqSymm}
\end{align}
where $\epsilon_a = \pm 1$ are the time-reversal parities of the fields
(this symmetry of thermal equilibrium can of course also be expressed as a transformation of the standard response fields $\tilde{\phi}$ and $\tilde{n}$, in which case it is non-linear, however \cite{Janssen1979}).
The occurrence of the time-reversal parity in this transformation is due to the fact that the symmetry expresses detailed balance when the system is in thermal equilibrium, which includes a time-reversal transformation \cite{Sieberer:2015hba}. Since the $\phi$-field is the order parameter of the system and the order parameters are unchanged under the time reversal transformation, one has $\epsilon_\phi =+1$. The $n$-fields on the other hand represent the zero components of conserved currents, and their parity under time reversal thus is $\epsilon_n=-1$. 
We explicitly demonstrate in Appendix~\ref{appendix:fdr} that our bare action is indeed invariant under the symmetry transformation~\eqref{eq:thermEqSymm}.

\paragraph{Temporal gauge and displacement symmetry.}
There is a symmetry that emerges due to the underlying Poisson-bracket structure of the reversible mode couplings. By construction, the free energy \eqref{freeEnergy} and the equations of motion \eqref{eq:eoms} are of course invariant under global $O(N)$ transformations.
In fact, this global $O(N)$ symmetry can even be extended to cover \emph{time-dependent} (but spatially constant) $O(N)$ transformations.
This generalization is possible due to the Poisson-bracket structure of the conservative terms in the equations of motion \eqref{eq:eoms}.
To see this, we first introduce some general notation. Recall that $O(N)$ transformations can generally be written as $O=\exp\{\frac{1}{2}\alpha_{ab} T_{ab}\} \in O(N)$, where the $T_{ab}$'s are the $N(N-1)/2$ generators of $O(N)$.
The latter are elements of the Lie algebra $o(N)$.
One possible representation that complies with the Poisson-bracket relations \eqref{current2} and \eqref{current3} is given by
$(T_{ab})_{cd} = \delta_{ad}\delta_{bc}-\delta_{ac}\delta_{bd}$.
The generators satisfy the commutation relations
\begin{equation}
    [T_{ab},T_{cd}] = \delta_{ac} T_{bd} + \delta_{bd} T_{ac} - \delta_{ad} T_{bc} - \delta_{bc} T_{ad}
    \label{eq:ONCommutationRels}
\end{equation}
which defines the structure constants of the group.
They are normalized according to
\begin{equation}
    \tr(T_{ab} T_{cd}) = 2(\delta_{ad}\delta_{bc}-\delta_{ac}\delta_{bd}) \,.
\end{equation}
The Poisson-bracket relations reflect the transformation behavior of fields under infinitesimal $O(N)$ transformations, so we can express 
\begin{align}
    \{\phi_a, n_{cd}\}=(T_{cd})_{ab}\phi_b \,, \hspace{0.5cm}
    \{n_{ab}, n_{cd}\} =(T_{cd})_{ab,ef}n_{ef} \,, \label{eq:relBetweenPBsAndONGenerators}
\end{align}
in which $(T_{cd})_{ab,ef}$ denote the generators of the adjoint representation of $O(N)$.
They are explicitly given by the structure constants of $O(N)$, which can be read off from the commutation relations \eqref{eq:ONCommutationRels}. 
The order parameter $\phi$, its associated response field $\tilde{\phi}$, and the corresponding noise term $\theta$ are all elements of the fundamental (defining) representation of $O(N)$ and thus transform as $\phi \to O\phi$, $\tilde{\phi} \to O\tilde{\phi}$, and $\theta \to O\theta$ under an action of $O$.
The matrix of charge densities $n \equiv -\frac{1}{2} n_{ab} T_{ab}$, its associated response field $\tilde{n} \equiv -\frac{1}{2} \tilde{n}_{ab} T_{ab}$, and the corresponding noise $\zeta \equiv -\frac{1}{2} \zeta_{ab} T_{ab}$ are elements of the adjoint representation.
They thus transform as $n \to O n O^T$, $\tilde{n} \to O \tilde{n} O^T$, $\zeta \to O\zeta O^T$.
Similarly, the corresponding sources $H,\tilde{H}$ and $\mathcal{H},\tilde{\mathcal{H}}$, for the order parameter and the conserved charges, are elements of the fundamental and the adjoint representation, respectively.

Now we let the transformation depend on time, $O \to O(t)$, i.e.\ we consider $\alpha_{ab} \to \alpha_{ab}(t)$.
One can straightforwardly verify that the equations of motion \eqref{eq:eoms} with the physical source terms included are invariant under the following time-dependent $O(N)$ transformations,
\begin{align}
    \phi(t,\vec{x}) &\to O(t) \phi \nonumber \\
    \theta(t,\vec{x}) &\to O(t) \theta(t,\vec{x}) \nonumber \\
    n(t,\vec{x}) &\to O(t) n(t,\vec{x}) O^T(t) \nonumber \\
    \zeta(t,\vec{x}) &\to O(t) \zeta(t,\vec{x}) O^T(t) \nonumber \\
    \mathcal{H}(t,\vec{x}) &\to O(t) \mathcal{H}(t,\vec{x}) O^T(t) + \frac{1}{g} O(t)\partial_t O^T(t) \label{eq:displacementSymm}
\end{align}
which, importantly, requires an inhomogeneous transformation of the external (physical) source $\mathcal{H}$ of the charge densities by $\frac{1}{g} O(t)\partial_t O^T(t)$. This symmetry is therefore like a purely temporal (i.e. spatially constant) gauge symmetry in a non-Abelian gauge theory, e.g., with an incomplete Coulomb gauge fixing. The source $\mathcal{H}_{ab}$ plays the role of the zero-component $\mathcal{H}_{ab} = A^{0}_{ab}$ of an external non-Ablian gauge field $A^{\mu}_{ab}$, and the gauge symmetry can be interpreted as the residual invariance under purely temporal gauge transformations that remains after choosing Coulomb gauge $\vec{\nabla} \!\cdot\! \vec{A}_{ab} = 0$. This is the non-Abelian generalization of the observation that  sources $A^{\mu}$ for conserved $U(1)$ currents $j^{\mu}$, due to the current conservation $\partial_{\mu} j^{\mu} = 0$, admit an Abelian gauge symmetry \cite{Crossley:2015evo,Harder:2015nxa}. 

In the special case of rigid rotations about a fixed axis, with $\alpha_{ab}(t) = \omega_{ab}\, t$ and constant angular velocity components $\omega_{ab}$ (i.e.~in an Abelian subgroup of the gauge group), the purely temporal gauge transformations reduce to constant shifts or \emph{displacements} of the external source $\mathcal{H}_{ab} $ by  
\begin{align}
    \frac{1}{g} O(t)\partial_t O^T(t) = -\frac{1}{2} \omega_{ab} T_{ab} \, .
\end{align}
We will thus refer to this symmetry as the `displacement symmetry' in the following.
In QED, e.g.~when completing the Coulomb gauge by requiring the zero component of the gauge field to be time independent, or in covariant gauges, an analogous displacement symmetry is left as part of the residual global gauge invariance after gauge fixing \cite{Nakanishi:1990qm,Alkofer:2000wg,Lenz:2000zt}.\footnote{In QED this global displacement symmetry is in fact always spontaneously broken, and the photon has been interpreted as the associated Goldstone boson to explain its masslessness \cite{Nakanishi:1990qm,Lenz:1994tc}.} 

Intuitively speaking, this symmetry states that the sole effect of a shift in the external source $\mathcal{H}$ is to induce a corresponding time-dependent rotation of all fields.
In the language of the antiferromagnet, where the sole effect of a constant uniform magnetic field is to let all spins precess about the magnetic field with an angular frequency proportional to its strength, this phenomenon is known as Larmor precession. 

Such a transformation behavior is usually called an `extended' symmetry in the literature, as the corresponding change $\delta S$ in the MSR action does not generally vanish, but is linear in the (composite) response fields and hence corresponds to a mere shift of the external sources \cite{Canet:2014cta,Tarpin:2017uzn,Tarpin:2018yvs,Floerchinger:2021uyo,Canet2022}.
An extended symmetry poses just as strong constraints on the 1PI effective action as a normal symmetry does.

To formulate the time-dependent $O(N)$ transformations on the level of the MSR action \eqref{MSR04}, we need to transform the (composite) response fields instead of the noises,
\begin{align}
    \tilde{\phi}(t,\vec{x}) &\to O(t) \tilde{\phi}(t,\vec{x})\, , && \tilde{\Phi}(t,\vec{x}) \to O(t) \tilde{\Phi}(t,\vec{x}) \, , \nonumber \\
    \tilde{n}(t,\vec{x}) &\to O(t) \tilde{n}(t,\vec{x}) O^T(t) \, , && \tilde{N}(t,\vec{x}) \to O(t) \tilde{N}(t,\vec{x}) O^T(t) \, .\nonumber
\end{align}
Knowing that our extended symmetry acts on the external sources $\mathcal{H}_{ab}$ as a residual purely temporal $O(N)$ gauge symmetry, we may introduce corresponding covariant time derivatives for $\phi$ and $n$ via
\begin{subequations}
\begin{align}
    D_t \phi_a &\equiv \partial_t \phi_a -\frac{g}{\orange{2}} (T_{cd})_{ab}\mathcal{H}_{cd}\phi_b \,, \label{eq:covarTimeDerivPhi} \\
    D_t n_{ab} &\equiv \partial_t n_{ab} -\frac{g}{\orange{2}} (T_{cd})_{ab,ef}\mathcal{H}_{cd}n_{ef} \,. \label{eq:covarTimeDerivN}
\end{align} \label{eq:covarTimeDerivs}%
\end{subequations}
Including the coupling to the sources $\mathcal{H}$ into the definition of an extended MSR action, \[ S_{\mathcal{H}} \equiv S + \orange{\frac{1}{2}} \int_x \mathcal{H}_{ab} \tilde{N}_{ab} \,, \] 
we can then express $S_{\mathcal{H}}$ using these covariant derivatives \eqref{eq:covarTimeDerivs} compactly as
\begin{align}
    S_{\mathcal{H}} &= \int_x \bigg[ -\tilde{\phi}_a D_t \phi_a - \orange{\frac{1}{2}} \tilde{n}_{ab}D_t n_{ab}-\tilde{\Phi}_a\frac{\delta F}{\delta \phi_a}-\orange{\frac{1}{2}}\tilde{N}_{ab}\frac{\delta F}{\delta n_{ab}} \; +  \\ \nonumber
    &\hspace{9.0cm} \tilde{\phi}_a iT\Gamma_0 \tilde{\phi}_a- \orange{\frac{1}{2}}\tilde{n}_{ab}iT\gamma \vec{\nabla}^2 \tilde{n}_{ab} \bigg] \,. 
\end{align}
Here, the standard response fields $\tilde{\phi}_a$ and $\tilde{n}_{ab}$ are understood as functions of the composite response fields $\tilde{\Phi}_a$ and $\tilde{N}_{ab}$ (implicitly defined by the transformation \eqref{eq:responseFieldTrafo}).
Since the covariant derivatives transform covariantly under the purely temporal  gauge transformations, the MSR action $S_{\mathcal{H}}$ (including the external source $\mathcal{H}$) is clearly seen to be invariant  under these gauge transformations, and thus in particular also under the uniformly  time-dependent $O(N)$ transformations with $\alpha_{ab}(t) = \omega_{ab}\, t$ of the displacement symmetry.  

The transformation behavior of the  MSR action $S$ under such a residual gauge transformation can be compactly expressed as\footnote{Note that we can represent a contraction $A_{ab}B_{ab}$ as a trace, $A_{ab} B_{ab} = - \tr(AB)$ for two elements $A = \frac{1}{2}A_{ab} T_{ab}$, $B=\frac{1}{2}B_{ab} T_{ab}$ in the adjoint representation.}
\begin{align}
    S[O\phi,O\tilde{\phi},OnO^T,O\tilde{n}O^T] -\frac{1}{2} \int_x \tr  \left( O\mathcal{H}O^T + \frac{1}{g}O \partial_t O^T \right) O \tilde{N} O^T &= \\ \nonumber &\hspace{-1.0cm} S[\phi,\tilde{\phi},n,\tilde{n}] -\frac{1}{2}\int_x \tr\left( \mathcal{H} \tilde{N} \right)
\end{align}
with $O=O(t)$ here.
For rigid rotations with infinitesimal angular frequencies (corresponding to infinitesimal shifts of the external source $\mathcal{H}$) we obtain the Ward identity
\begin{align}
	\int_{x}\,\bigg[ &\frac{1}{g} \tilde{N}_{ab}(x) + \frac{\delta \Gamma}{\delta \phi_{[a}(x)} \phi_{b]}(x) \, x^0 + \frac{\delta \Gamma}{\delta \tilde{\Phi}_{[a}(x)}  \tilde{\Phi}_{b]}(x) \, x^0  + \nonumber \\
	&\orange{\frac{1}{2}} \frac{\delta \Gamma}{\delta n_{cd}(x)} \left( \delta_{bc} n_{ad}(x) - \delta_{ac} n_{bd}(x) - \delta_{bd} n_{ac}(x) + \delta_{ad} n_{bc}(x) \right)  x^0 + \nonumber  \\
	&\orange{\frac{1}{2}} \frac{\delta \Gamma}{\delta \tilde{N}_{cd}(x)} \left( \delta_{bc} \tilde{N}_{ad}(x) - \delta_{ac} \tilde{N}_{bd}(x) - \delta_{bd} \tilde{N}_{ac} (x)+ \delta_{ad} \tilde{N}_{bc}(x) \right) x^0
	\bigg] = 0 \label{eq:wardIdWithHDeriv}
\end{align}
for the effective action, which in the context of the FRG below will tightly constrain possible truncation schemes consistent with this displacement symmetry. Here we have already replaced the bare action $S$ by the effective action $\Gamma$ in the Ward identity, because, in absence of anomalies, the effective action admits the same symmetries as the bare action, e.g.~see \cite{Gies:2006wv}.

\paragraph{Charge conservation.}

For vanishing physical external sources $H=\mathcal{H}=0$ the charges are conserved.
This corresponds to a symmetry of the MSR action with respect to constant shifts in the standard response fields of the charges,
\begin{align}
\begin{split}
    \tilde{n}_{ab}(x) &\to \tilde{n}_{ab}(x) +\delta \tilde{n}_{ab}
\end{split}
\end{align}
with a constant but otherwise arbitrary $\delta \tilde{n}_{ab}$.
Upon such a constant displacement the Lagrangian $\mathcal{L}$ in the MSR action \eqref{MSR04} changes by a total time derivative
\begin{equation}
    \delta \mathcal{L} = -\orange{\frac{1}{2}}\delta \tilde{n}_{ab} \partial_t n_{ab} \, , \label{eq:shiftSymmOfTildeN}
\end{equation}
and hence the action is invariant.
Using a formulation of Noether's theorem on the closed-time path~\cite{Sieberer_2016}, we then have
$N(N-1)/2$ independent continuity equations (on average),
\begin{equation}
    \langle \partial_{\mu} j^{\mu}_{ab} \rangle = 0 \label{eq:contEqForCharges}
\end{equation}
corresponding to the conservation of the $N(N-1)/2$ charges
\begin{equation}
    N_{ab} \equiv \int_{\vec{x}} n_{ab}(t,\vec{x}) \,.
\end{equation}
As a consequence of Noether's theorem, the conserved Lorentz vector current $j_{ab}^{\mu}=(j^0_{ab},\vec{j}_{ab})$ can be derived using standard rules from the MSR action and the symmetry transformation \eqref{eq:shiftSymmOfTildeN} as e.g.~illustrated in Sec.~2.4.3 of Ref.~\cite{Sieberer_2016}. It is then given by
\begin{align}
    j^0_{ab} &= n_{ab} \, , \\
    \vec{j}_{ab} &= -\frac{\gamma}{\chi} \vec{\nabla} n_{ab} + g (\phi_a \vec{\nabla} \phi_b - \phi_b \vec{\nabla} \phi_a) + i\gamma T \vec{\nabla} \tilde{n}_{ab} \, , \label{eq:veccur}
\end{align}
Since $\langle \tilde{n}_{ab} \rangle = 0$ due to causality, in fact,
the last term in \eqref{eq:veccur} does not contribute to the average in  \eqref{eq:contEqForCharges}.

For non-vanishing external source fields $H$ for the order parameter (which correspond to non-vanishing external magnetic fields in the language of the Heisenberg antiferromagnet, or to non-vanishing current quark masses in QCD with two light flavors), Eq.~\eqref{eq:contEqForCharges} is replaced by  the familiar \emph{partial} conservation of the currents,
\begin{equation}
    \langle \partial_{\mu} j^{\mu}_{ab} \rangle = g\langle \phi_a H_b - \phi_b H_a \rangle \label{eq:partialContEqForCharges}
\end{equation}
which is analogous to the PCAC relation in QCD.

\paragraph{BRST symmetry.}
The fundamental normalization condition $\langle 1 \rangle = 1$ of the MSR path integral implies
\begin{equation}
    Z[J] \equiv 
    Z[J,\tilde J =0] = 1 \label{eq:normCond}
\end{equation}
for arbitrary classical sources here collected in the (classical) superfield source $J=(H,\mathcal{H})$.
I.e.~for vanishing external response sources $\tilde J= (\tilde{H}, \tilde{\mathcal{H}})$  the generating functional is equal to unity, regardless of the values for the classical source terms, which means that  the path integral becomes topological \cite{Birmingham:1991ty} in this limit. Such a condition is generally a consequence of the conservation of probability, i.e.~the probability distribution of the field configurations $\psi$ remains normalized 
under unitary time evolution of the corresponding probability amplitudes.
Of course, this does \emph{not} require the underlying evolution equations to be of conservative (Hamiltonian) form.
This general property of the path integral on the CTP is enforced by a hidden BRST symmetry \cite{Crossley:2015evo}:
One can show that the total action that appears in the generating functional \eqref{eq:genFncAfterFieldTrafo}, when the Jacobian is expressed via anticommuting ghosts fields as in \eqref{eq:jacobianViaGhosts}, is invariant under the following BRST transformation
\begin{equation}
    \delta \psi_i = \epsilon c_i \,,\, \, \, \, \,\delta c_i=0 \,,\,\,\,\, \, \delta \tilde{c}_i=-i\epsilon \tilde{\Psi}_i \,,\,\,\,\, \,\delta\tilde{\Psi}_i =0 \,, \,\,\,\, \, \delta J_i = 0 \,, \label{eq:BRSTTrafo}
\end{equation}
where $\epsilon $ is an anti-commuting number in the Grassmann algebra of the ghosts, and $i$ an index in superfield space. The normalization condition \eqref{eq:normCond} is then a consequence of this BRST symmetry, see Appendix~\ref{sct:jacobianAndGhosts}.  It is generated by the corresponding BRST charge
\begin{equation}
    Q = c_i \frac{\delta}{\delta \psi_i}-i \tilde{\Psi}_i \frac{\delta}{\delta \tilde{c}_i}\,, \,\,\,\,\, Q^2=0 \,.
\end{equation}
In fact, the MSR action is BRST exact: In superfield notation, the total MSR action $S_J$ including the classical sources $J_i$ and the ghost action $S_{gh}$ from \eqref{eq:jacobianViaGhosts}
\begin{equation}
    S_J = S' + S_{gh} + J_i \tilde{\Psi}_i \label{eq:SJ}
\end{equation}
can be expressed as a BRST variation,\footnote{Note, however, that the form of the BRST invariant action is not unique, and there are in general multiple BRST invariant actions that lead to the same MSR action \eqref{MSR04}, cf.~Sec.~1.5 of \cite{Crossley:2015evo}.}
\begin{align}
    S_J &= Q\left\{ -i\tilde{c}_i \left( {J^{-1}_{ij}}^T(\psi) {\partial_t} \psi_j + \frac{\delta F}{\delta \psi_i} - J_i -iT {J^{-1}_{ij}}^T(\psi) \tilde{\Psi}_j \right) \right\} \,. \label{eq:BRSTMSR} 
\end{align}
We explain how \eqref{eq:SJ} is obtained from \eqref{eq:BRSTMSR} in Appendix \ref{sct:jacobianAndGhosts}. In the form \eqref{eq:BRSTMSR}, the BRST invariance of the MSR action immediately follows from the nilpotency of the BRST charge $Q$. With vanishing response sources $\tilde{J}=(\tilde{H},\tilde{\mathcal{H}})$, the MSR path integral computes the Witten index of a topological field theory and is hence independent of the classical sources $J=(H,\mathcal{H})$. 

\section{FRG for dynamical systems with reversible mode couplings}\label{sec:FRG_rev_model_coup}
\label{sct:frg}

Instead of solving the path integral of a given theory in some approximation scheme defined for fluctuations of all scales at once, the general strategy of the functional renormalization group (FRG) is to implement the Wilsonian idea of successively integrating out fluctuations momentum shell by momentum shell.
In the pioneering formulation by Wetterich \cite{Wetterich:1992yh} this is achieved with adding an infrared (IR) cutoff to the theory, called the regulator $R_k$, which suppresses fluctuations of modes with momenta  $p$ smaller than the current `FRG scale' $k$. At every given fixed scale $k$, fluctuations with $p\gtrsim k$, on the other hand, are assumed to have been integrated out. Then, instead of solving the path integral \eqref{eq:genFncAfterFieldTrafo} directly, one can equivalently follow the FRG flow down to $k=0$ in some truncation, where all relevant  fluctuations are then fully taken into account. At the initial reference scale $k = \Lambda$ in the ultraviolet (UV) these fluctuations are assumed to be sufficiently suppressed so that the bare MSR action \eqref{MSR04} provides a suitable starting point for the FRG flow.
In this way, the problem of solving a functional integral is converted into the problem of solving a functional differential equation.
This works particularly well for the characteristic low-frequency long-wavelength fluctuations in the vicinity of a second-order phase transition.
Because the regulator suppresses all modes with momenta $p\lesssim k$, the critical contributions from these generally large fluctuations to any observable then gradually build up as the FRG scale $k$ is
successively lowered by solving the corresponding FRG flow equations.

A rather non-trivial task in devising a regulator suitable for systems with reversible mode couplings is to maintain all the symmetries of the MSR action we have discussed in Sec.~\ref{sct:SymmetriesOfMSRAction} above. In particular, simply adding the regulator as a quadratic form in the fields directly to the MSR (or Keldysh) action, which works for systems without reversible mode couplings \cite{Roth:2023wbp}, here one immediately observes that some of the symmetries are then necessarily violated. For example, one can no-longer guarantee that the equilibrium distribution will be a Boltzmann distribution during the flow, and hence the usual formulation of the FDR no longer holds. Instead, as in the case of the physical sources in our construction of the generating functional in Sec.~\ref{sct:genFunc}, we therefore add the regulator one step earlier to the LGW free energy,\footnote{We assume the regulator to be diagonal in field space to preserve the global $O(N)$ symmetry.}
\begin{equation}
    F \rightarrow F + \frac{1}{2} \int_{\vec{x}\vec{y}} \left( \phi_a(\vec{x})R^{\phi}_k(\vec{x},\vec{y})\phi_a(\vec{y}) + \orange{\frac{1}{2}} n_{ab}(\vec{x})R^n_k(\vec{x},\vec{y})n_{ab}(\vec{y}) \right) \, , \label{eq:regToFreeEnergy}
\end{equation}
and obtain the MSR action with regulator terms from that.
This guarantees that none of the symmetries from Sec.~\ref{sct:SymmetriesOfMSRAction} are violated by the regulator terms.

Because of the Poisson brackets in the equations of motion \eqref{eq:eoms}, this gives rise to regulator terms in the MSR action that are no-longer quadratic in the original fields but couple to the \emph{composite} response fields $\tilde{\Phi}_a$ and $\tilde{N}_{ab}$, 
\begin{equation}
    S \to S + \Delta S_k \,, \hspace{0.5cm}
    \Delta S_k = -\int_{xy} \left( \tilde{\Phi}_a(x)R^{\phi}_k(x,y)\phi_a(y) + \orange{\frac{1}{2}} \tilde{N}_{ab}(x)R^n_k(x,y)n_{ab}(y) \right) \,, \label{eq:DelSk}
\end{equation}
where
\begin{align}
    R_{k}^{\phi}(x,y) \equiv R_{k}^{\phi}(\vec{x},\vec{y}) \,\delta(x^0-y^0) \,, \text{  and  }\;   R_{k}^{n}(x,y) \equiv R_{k}^{n}(\vec{x},\vec{y}) \,\delta(x^0-y^0) \,. \label{eq:regWithSpacetimeArgs}
\end{align}
The crucial generalization of the standard procedure here is that the regulators $R_{k}^{\phi}$ and $R_{k}^{n}$ couple to the composite response fields $\tilde{\Phi}$ and $\tilde{N}$ in \eqref{eq:DelSk}. On the other hand this then necessarily implies that the regulator term $\Delta S_k$ added to the MSR action, when expressed in terms of the standard response fields $\tilde{\phi}_a$ and $\tilde{n}_{ab}$, involves products of \emph{three} fields.

As a side remark, also note that in \eqref{eq:regWithSpacetimeArgs} we have introduced equal-time regulators whose Fourier transforms are hence frequency independent. In this way, the causal structure of the MSR action  \cite{kamenev_2011} is maintained trivially by the regulators which is sufficient for our purposes here. If frequency dependent regulators are needed, on the other hand, a general construction scheme for causal regulators analogous to the one given in Sec.~2 of Ref.~\cite{Roth:2023wbp} can be used here as well. In this construction 
the regulators are introduced as self-energies from (suitably subtracted) Kramers-Kronig relations of fictitious (unphysical) heat baths whose FRG-scale-dependent spectral distributions provide the frequency-dependent regularization in a causal way by construction.

The presence of the regulator term $\Delta S_k$ in addition to the bare MSR action lets the generating functional $Z \to Z_k$ and the Schwinger functional $W \to W_k$ depend on the FRG scale $k$. 
These can be viewed as their `coarse-grained' counterparts where all fluctuations with momenta $p \lesssim k$ have been suppressed.
In the FRG, the central object is the effective \emph{average} action $\Gamma_k$, which can be similarly interpreted as a coarse-grained effective action. $\Gamma_k$ is as usual defined by a modified Legendre transformation (as in \eqref{eq:effActionDef} but with an additional subtraction of the regulator term) of the Schwinger functional $W_k$. 

In line with the developments for systems with reversible mode couplings above, this modified Legendre transformation must be done here with respect to the physical sources inside the LGW free energy, leading to the corresponding (expectation values of the) \emph{composite} response fields in the effective average action,
\begin{align}
    \Gamma_k[{\phi},{n},{\tilde{\Phi}},{\tilde{N}}] &= W_k[H,\mathcal{H},\tilde{H},\tilde{\mathcal{H}}] - \Delta S_k[{\phi},{n},{\tilde{\Phi}},{\tilde{N}}] \label{eq:effAvgActionDef} \\ \nonumber
    &\hspace{1.0cm} - \int_x \Big( \tilde{H}_a {\phi}_a + \orange{\frac{1}{2}} \tilde{\mathcal{H}}_{ab} {n}_{ab} + H_a {\tilde{\Phi}}_a + \orange{\frac{1}{2}} \mathcal{H}_{ab} {\tilde{N}}_{ab} \Big) \, .
\end{align}
By expressing the regulator as a matrix in the two-dimensional $(\psi,\tilde{\Psi})$ MSR field space,
\begin{equation}
    \hat{R}_k = \begin{pmatrix}
        0&R_k \\ R_k&0
    \end{pmatrix}\, , \label{eq:regInSuperfieldSpace}
\end{equation}
the flow equation for the effective average action formulated in our compact superfield notation is given by
\begin{equation}
    \partial_k \Gamma_k[\psi,\tilde{\Psi}] = \frac{i}{2} \tr\left\{ \partial_k \hat{R}_k \circ G_k[\psi,\tilde{\Psi}] \right\} \label{eq:treeLevelFlowEq}
\end{equation}
which is derived in Appendix \ref{derivationfloweq}.
The full FRG-scale and field-dependent propagators \eqref{eq:phiPropFromW} and \eqref{eq:nPropFromW} appearing in the flow equation \eqref{eq:treeLevelFlowEq} are as usual related to the regulated two-point function via~\cite{Berges:2012ty}
\begin{equation}
    G_k[\psi,\tilde{\Psi}] = -\left( \Gamma_k^{(2)}[\psi,\tilde{\Psi}] - \hat{R}_k \right)^{-1} \,. \label{eq:fullPropFRGAsInvOfGam2}
\end{equation}
At the origin $\phi=n=0$ and $\tilde{\Phi}=\tilde{N}=0$ in field space the various propagators of the order parameter are diagonal and hence one can introduce scalar propagators to write them as
\begin{subequations}
\begin{align}
    G^R_{\phi_a \phi_b,k}(x,y) &= G_{\phi,k}^R(x,y) \delta_{ab}\,, \\
    G^A_{\phi_a \phi_b,k}(x,y) &= G_{\phi,k}^A(x,y) \delta_{ab}\,, \\
    iF_{\phi_a \phi_b,k}(x,y) &= iF_{\phi,k}(x,y) \delta_{ab}\,,
\end{align} \label{eq:phiPropDiag}%
\end{subequations}
and similarly for the propagators of the charge densities,
\begin{subequations}
\begin{align}
    G_{n_{ab} n_{cd},k}^R(x,y) &= G_{n,k}^R(x,y) \, \orange{(\delta_{ac} \delta_{bd} - \delta_{ad}\delta_{bc})}\,, \\
    G_{n_{ab} n_{cd},k}^A(x,y) &= G_{n,k}^A(x,y) \, \orange{(\delta_{ac} \delta_{bd} - \delta_{ad}\delta_{bc})}\,, \\
    iF_{n_{ab} n_{cd},k}(x,y)  &= iF_{n,k}(x,y) \, \orange{(\delta_{ac} \delta_{bd} - \delta_{ad}\delta_{bc})}\,,
\end{align} \label{eq:nPropDiag}%
\end{subequations}
where we have used the antisymmetric identity in the adjoint representation of $O(N)$ corresponding to the antisymmetric fundamental functional derivative \eqref{eq:funcDerivDefForN}.
The corresponding scalar functions in front $\delta_{ab}$ in \eqref{eq:phiPropDiag} and in front of $\orange{(\delta_{ac} \delta_{bd} - \delta_{ad}\delta_{bc})}$ in \eqref{eq:nPropDiag} can be derived from \eqref{eq:fullPropFRGAsInvOfGam2} and are explicitly given by
\begin{subequations}
\begin{align}
    G_{\phi,k}^R(x,y) \delta_{ab} &=-\left(\frac{\delta^2 \Gamma_k}{\delta \tilde{\Phi}_a(x)\delta\phi_b(y)} \bigg\rvert_{0} - R^{\phi}_k(\vec x,\vec y) \delta_{ab} \right)^{-1}, \\
    G_{\phi,k}^A(x,y) \delta_{ab} &=-\left(\frac{\delta^2 \Gamma_k}{\delta \phi_a(x)\delta\tilde{\Phi}_b(y)} \bigg\rvert_{0} - R^{\phi}_k(\vec x,\vec y) \delta_{ab} \right)^{-1} ,\\
    iF_{\phi,k}(x,y) \delta_{ab} &= \int_{zw} G_{\phi,k}^R(x,z)  \frac{\delta^2 \Gamma_k}{\delta \tilde{\Phi}_a (z)\delta\tilde{\Phi}_b(w)} \bigg\rvert_{0} G_{\phi,k}^A(w,y)
    \, ,
\end{align} \label{eq:phiProp}%
\end{subequations}
in case of the order parameter, and by
\begin{subequations}
\begin{align}
    G_{n,k}^R(x,y) \, \orange{(\delta_{ac} \delta_{bd} - \delta_{ad}\delta_{bc})} &=-\left(\frac{\delta^2 \Gamma_k}{\delta \tilde{N}_{ab}(x)\delta n_{cd}(y)} \bigg\rvert_{0} - R^{n}_k(\vec x,\vec y) \, \orange{(\delta_{ac} \delta_{bd} - \delta_{ad}\delta_{bc})} \right)^{-1} ,\\
    G_{n,k}^A(x,y) \, \orange{(\delta_{ac} \delta_{bd} - \delta_{ad}\delta_{bc})} &=-\left(\frac{\delta^2 \Gamma_k}{\delta n_{ab}(x)\delta \tilde{N}_{cd}(y)} \bigg\rvert_{0} - R^{n}_k(\vec x,\vec y) \, \orange{(\delta_{ac} \delta_{bd} - \delta_{ad}\delta_{bc})} \right)^{-1} ,\\
    iF_{n,k}(x,y) \, \orange{(\delta_{ac} \delta_{bd} - \delta_{ad}\delta_{bc})} &= \int_{zw} G_{n,k}^R(x,z)  \frac{\delta^2 \Gamma_k}{\delta \tilde{N}_{ab} (z)\delta \tilde{N}_{cd}(w)} \bigg\rvert_{0} G_{n,k}^A(w,y)\, ,
\end{align} \label{eq:nProp}%
\end{subequations}
in case of the charge densities.
Note that the tensor structures $\delta_{ab}$ and $\delta_{ac} \delta_{bd} - \delta_{ad}\delta_{bc}$ match on both sides at the origin in field space, and hence the corresponding scalar propagators as $G_{\phi,k}^R(x,y)$ and  $G_{n,k}^R(x,y)$ can be read off by a comparison of coefficients.
For non-vanishing field expectation values (or finite external fields $H \neq 0$) the full expressions for the propagators in general involve various other $O(N)$ tensor structures besides the identities here, and hence become evidently rather cumbersome.
For this reason we restrict ourselves to the origin in field space $\phi=n=0$ for the scope of this work.

In thermal equilibrium, the symmetry of detailed balance \eqref{eq:thermEqSymm} implies a fluctuation-dissipation relation (FDR) for the propagators in Fourier space (for more details on the derivation of the FDR in the context of dynamical systems with reversible mode couplings see Ref.~\cite{Janssen1979}),
\begin{equation}
    iF_{ij,k}(\omega,\vec{p}) = \frac{T}{\omega} \left(G_{ij,k}^R(\omega,\vec{p}) - G_{ij,k}^A(\omega,\vec{p})\right) \, ,\label{eq:FDRForPropagators}
\end{equation}
and for the two-point functions
\begin{equation}
    \Gamma_{ij,k}^{\tilde{\Psi}\tilde{\Psi}}(\omega,\vec{p})  = \frac{T}{\omega} \left(\frac{1+\epsilon_i\epsilon_j}{2}\right) \left(\Gamma_{ij,k}^{\tilde{\Psi}\psi}(\omega,\vec{p}) - \Gamma_{ij,k}^{\psi\tilde{\Psi}}(\omega,\vec{p})\right) \, , \label{eq:FDRFor2PtFncs}
\end{equation}
where we have used the superfield notation $\psi=(\phi,n)$ and $\tilde{\Psi}=(\tilde{\Phi},\tilde{N})$ again, to encompass the FDR for both types of fields concisely in one equation.
Importantly, \eqref{eq:FDRForPropagators} would not hold in this particularly simple form if the non-composite standard response fields $\tilde{\phi}$ and $\tilde{n}$ were used to define the propagators.

Besides two-point functions, one can also derive generalized fluctuation-dissipation relations between the higher order $n$-point functions in the spirit of Ref.~\cite{Wang:1998wg}, as illustrated in Appendix~\ref{appendix:fdr}. 
Given some of these generalized FDRs, we can prove another central result for our formulation of the FRG for dynamical systems with reversible mode couplings:

As we show in Appendix~\ref{sct:CorrespondenceRealTimeAndEuclFlows}, the flow of the scale-dependent free energy $F_k$ (extracted generally in a suitable way from the effective average MSR action $\Gamma_k$, see Eq.~\eqref{eq:defOfFk} below), satisfies its own closed flow equation and hence decouples from the remaining system of flow equations. In particular, the flow of $F_k$ decouples from the flow of purely dynamic quantities like kinetic coefficients or the reversible mode couplings.
This proves, as an important result, that the static universal critical behavior is not altered by changes in the dynamics, for instance whether the dynamics is relaxational, diffusive, or whether it includes reversible mode couplings or not, as long as the symmetry of detailed balance \eqref{eq:thermEqSymm} holds.
Moreover, this decoupling of $F_k$ from the rest can be used as a powerful tool, since it means that one can derive flow equations for static quantities without having to invoke the entire complexity of the real-time formalism.
We will discuss the exact statement of this decoupling of the flow of the free energy in the following Sec.~\ref{sct:staticFlow} below.

If the regulator couples to composite fields such as \eqref{eq:responseFieldTrafo} one has to make sure that the effective average action still converges to the bare action in the limit $k \to \Lambda$, i.e.~that the flow properly integrates out all fluctuations.
For standard flows where the regulator term is a quadratic form in the elementary fields, this is usually shown using a saddle-point approximation. However, if the regulator couples to composite fields the situation is not as clear anymore.
For a general discussion on this issue see e.g.~Sec.~III C 4 of Ref.~\cite{Pawlowski:2005xe}.
For the way the FRG flow is set up in the present work, we show in Appendix \ref{convergenceofeffectiveaction} that in the UV limit the effective average action converges to
\begin{align}
    \Gamma_k[\psi,\tilde{\Psi}] \xrightarrow{k \to \Lambda} S[\psi,\tilde{\psi}] - i \log | \det J^{-1}(\psi) | + \text{const.} \label{effAvgActionUVConvergence}
\end{align}
which, again, involves the Jacobian matrix \eqref{eq:jacobianOfNonlinearFieldTrafo}.
Roughly speaking, the additional Jacobian determinant in \eqref{effAvgActionUVConvergence} compensates for the fact that the bare action depends on the elementary response fields $\tilde{\phi}$ and $\tilde{n}$, whereas the effective average action actually depends on the expectation values of our \emph{composite} response fields $\tilde{\Phi}$ and $\tilde{N}$.  
As such, this is just another reflection of the Jacobian matrix \eqref{eq:jacobianOfNonlinearFieldTrafo} in our non-linear field transformation, since exponentiating \eqref{effAvgActionUVConvergence} yields
\begin{align}
    e^{i\Gamma_k[\psi,\tilde{\Psi}]} \xrightarrow{k \to \Lambda}  \left|\det J^{-1}(\psi) \right|\,e^{iS[\psi,\tilde{\psi}]} =  \mathcal{J}'[\psi]\,e^{iS[\psi,\tilde{\psi}]}
\end{align}
(up to an irrelevant overall normalization factor).
This thus coincides with the integrand in \eqref{eq:genFncAfterFieldTrafo}.

\subsection{Diagrammatics}
Next, we want to derive the flow equations. In order to develop a straightforward way to derive the flow equations, it is much more convenient to carry out derivatives using Feynman rules. Here we want to give the diagrammatical representation of propagators, regulators, and vertices.
The propagators are represented by
\begin{align*}
	\begin{split}
		G^{R}_{\phi,k}(x,y) =
		\begin{tikzpicture}[baseline=-0.5ex,samples=120]
		      \draw[pblue] (1,0) -- (0,0) node[anchor=east,black] {$x$};
		 	\draw[pred] (2,0) node[anchor=west,black] {$y$} -- (1,0);
		\end{tikzpicture},
		\\
		G^{A}_{\phi,k}(x,y) =
		\begin{tikzpicture}[baseline=-0.5ex]
			\draw[pred] (1,0) -- (0,0) node[anchor=east,black] {$x$};
		 	\draw[pblue] (2,0) node[anchor=west,black] {$y$} -- (1,0);
		\end{tikzpicture},
		\\
		iF_{\phi,k}(x,y) =
		\begin{tikzpicture}[baseline=-0.5ex]
			\draw[pblue] (1,0) -- (0,0) node[anchor=east,black] {$x$};
		 	\draw[pblue] (2,0) node[anchor=west,black] {$y$} -- (1,0);
		\end{tikzpicture},
	\end{split}
	\begin{split}
		G^{R}_{n,k}(x,y) =
		\begin{tikzpicture}[baseline=-0.5ex,samples=120]
                \draw[pblue,smooth,domain=0:1,variable=\x] node[anchor=east,black] {$x$}  plot (\x,{0.1*sin(6*0.5*2*3.14159*\x r)});
                \draw[pred,smooth,domain=0:1,variable=\x] plot (1+\x,{0.1*sin(6*0.5*2*3.14159*(1+\x) r)}) node[anchor=west,black] {$y$} ;
		\end{tikzpicture},
		\\
		G^{A}_{n,k}(x,y) =
		\begin{tikzpicture}[baseline=-0.5ex]
                \draw[pred,smooth,domain=0:1,variable=\x] node[anchor=east,black] {$x$}  plot (\x,{0.1*sin(6*0.5*2*3.14159*\x r)});
                \draw[pblue,smooth,domain=0:1,variable=\x] plot (1+\x,{0.1*sin(6*0.5*2*3.14159*(1+\x) r)}) node[anchor=west,black] {$y$} ;
		\end{tikzpicture},
		\\
		iF_{n,k}(x,y) =
		\begin{tikzpicture}[baseline=-0.5ex]
                \draw[pblue,smooth,domain=0:1,variable=\x] node[anchor=east,black] {$x$}  plot (\x,{0.1*sin(6*0.5*2*3.14159*\x r)});
                \draw[pblue,smooth,domain=0:1,variable=\x] plot (1+\x,{0.1*sin(6*0.5*2*3.14159*(1+\x) r)}) node[anchor=west,black] {$y$} ;
		\end{tikzpicture}.
	\end{split}
\end{align*}
The regulators are diagrammatically represented by
\begin{align*}
\begin{split}
    \partial_k R_{\phi,k} (x,y)&= \begin{tikzpicture}[baseline=-0.5ex,samples=120]
			\draw[pblue] (1,0) -- (0,0) node[anchor=east,black] {$x$};
		 	\draw[pred] (2,0) node[anchor=west,black] {$y$} -- (1,0);
    \filldraw[white] (1,0) circle (0.1414213562);
	\draw (1,0) circle (0.1414213562);
		 	\draw[black] (1-0.1,-0.1) -- (1+0.1,+0.1);
		 	\draw[black] (1-0.1,+0.1) -- (1+0.1,-0.1);
		\end{tikzpicture}, \\
  &= \begin{tikzpicture}[baseline=-0.5ex]
			\draw[pred] (1,0) -- (0,0) node[anchor=east,black] {$x$};
		 	\draw[pblue] (2,0) node[anchor=west,black] {$y$} -- (1,0);
    \filldraw[white] (1,0) circle (0.1414213562);
	\draw (1,0) circle (0.1414213562);
		 	\draw[black] (1-0.1,-0.1) -- (1+0.1,+0.1);
		 	\draw[black] (1-0.1,+0.1) -- (1+0.1,-0.1);
		\end{tikzpicture},
\end{split}
\begin{split}
    \partial_k R_{n,k} (x,y)&=\begin{tikzpicture}[baseline=-0.5ex]
                \draw[pblue,smooth,domain=0:1,variable=\x] node[anchor=east,black] {$x$}  plot (\x,{0.1*sin(6*0.5*2*3.14159*\x r)});
                \draw[pred,smooth,domain=0:1,variable=\x] plot (1+\x,{0.1*sin(6*0.5*2*3.14159*(1+\x) r)}) node[anchor=west,black] {$y$} ;
    \filldraw[white] (1,0) circle (0.1414213562);
	\draw (1,0) circle (0.1414213562);
		 	\draw[black] (1-0.1,-0.1) -- (1+0.1,+0.1);
		 	\draw[black] (1-0.1,+0.1) -- (1+0.1,-0.1);
		\end{tikzpicture}, \\
  &= \begin{tikzpicture}[baseline=-0.5ex]
                \draw[pred,smooth,domain=0:1,variable=\x] node[anchor=east,black] {$x$}  plot (\x,{0.1*sin(6*0.5*2*3.14159*\x r)});
                \draw[pblue,smooth,domain=0:1,variable=\x] plot (1+\x,{0.1*sin(6*0.5*2*3.14159*(1+\x) r)}) node[anchor=west,black] {$y$} ;
    \filldraw[white] (1,0) circle (0.1414213562);
	\draw (1,0) circle (0.1414213562);
		 	\draw[black] (1-0.1,-0.1) -- (1+0.1,+0.1);
		 	\draw[black] (1-0.1,+0.1) -- (1+0.1,-0.1);
		\end{tikzpicture}.
\end{split}
\end{align*}
Since regulators almost always appear in sums over all their possible combinations, we introduce a corresponding shorthand notation for a regulator inserted between two propagators (with the color of the external legs fixed) for the $\phi$'s,
\begin{subequations}
\begin{align}
    B^F_{\phi,k}&=-G^R_{\phi,k} \circ \partial_k R_{\phi,k} \circ  iF_{\phi,k}-iF_{\phi,k} \circ  \partial_k R_{\phi,k} \circ G^A_{\phi,k} \,, \\
    B^R_{\phi,k}&=-G^R_{\phi,k}\circ \partial_k R_{\phi,k}\circ  G^R_{\phi,k} \,, \label{eq:BPropsRet} \\
    B^A_{\phi,k}&=-G^A_{\phi,k}\circ \partial_k R_{\phi,k}\circ G^A_{\phi,k} \,,\label{eq:BPropsAdv}
\end{align} \label{eq:BProps}%
\end{subequations}
and the $n$'s,
\begin{subequations}
\begin{align}
    B^F_{n,k}&=-G^R_{n,k}\circ \partial_k R_{n,k}\circ iF_{n,k}-iF_{n,k}\circ \partial_k R_{n,k}\circ G^A_{n,k} \,, \\
    B^R_{n,k}&=-G^R_{n,k}\circ \partial_k R_{n,k}\circ G^R_{n,k} \,, \\
    B^A_{n,k}&=-G^A_{n,k}\circ \partial_k R_{n,k}\circ G^A_{n,k} \,.
\end{align}
\end{subequations}
Diagrammatically, we represent the $B$'s via boxes,
\begin{align*}
	\begin{split}
		B^{R}_{\phi,k}(x,y) =
		\begin{tikzpicture}[baseline=-0.5ex,samples=120]
		      \draw[pblue] (1,0) -- (0,0) node[anchor=east,black] {$x$};
		 	\draw[pred] (2,0) node[anchor=west,black] {$y$} -- (1,0);
                \fill[black] (1-0.12,0-0.12) rectangle ++(0.24,0.24);
		\end{tikzpicture},
		\\
		B^{A}_{\phi,k}(x,y) =
		\begin{tikzpicture}[baseline=-0.5ex]
			\draw[pred] (1,0) -- (0,0) node[anchor=east,black] {$x$};
		 	\draw[pblue] (2,0) node[anchor=west,black] {$y$} -- (1,0);
                \fill[black] (1-0.12,0-0.12) rectangle ++(0.24,0.24);
		\end{tikzpicture},
		\\
		B^{F}_{\phi,k}(x,y) =
		\begin{tikzpicture}[baseline=-0.5ex]
			\draw[pblue] (1,0) -- (0,0) node[anchor=east,black] {$x$};
		 	\draw[pblue] (2,0) node[anchor=west,black] {$y$} -- (1,0);
                \fill[black] (1-0.12,0-0.12) rectangle ++(0.24,0.24);
		\end{tikzpicture},
	\end{split}
	\begin{split}
		B^{R}_{n,k}(x,y) =
		\begin{tikzpicture}[baseline=-0.5ex,samples=120]
                \draw[pblue,smooth,domain=0:1,variable=\x] node[anchor=east,black] {$x$}  plot (\x,{0.1*sin(6*0.5*2*3.14159*\x r)});
                \draw[pred,smooth,domain=0:1,variable=\x] plot (1+\x,{0.1*sin(6*0.5*2*3.14159*(1+\x) r)}) node[anchor=west,black] {$y$} ;
                \fill[black] (1-0.12,0-0.12) rectangle ++(0.24,0.24);
		\end{tikzpicture},
		\\
		B^{A}_{n,k}(x,y) =
		\begin{tikzpicture}[baseline=-0.5ex]
                \draw[pred,smooth,domain=0:1,variable=\x] node[anchor=east,black] {$x$}  plot (\x,{0.1*sin(6*0.5*2*3.14159*\x r)});
                \draw[pblue,smooth,domain=0:1,variable=\x] plot (1+\x,{0.1*sin(6*0.5*2*3.14159*(1+\x) r)}) node[anchor=west,black] {$y$} ;
                \fill[black] (1-0.12,0-0.12) rectangle ++(0.24,0.24);
		\end{tikzpicture},
		\\
		B^{F}_{n,k}(x,y) =
		\begin{tikzpicture}[baseline=-0.5ex]
                \draw[pblue,smooth,domain=0:1,variable=\x] node[anchor=east,black] {$x$}  plot (\x,{0.1*sin(6*0.5*2*3.14159*\x r)});
                \draw[pblue,smooth,domain=0:1,variable=\x] plot (1+\x,{0.1*sin(6*0.5*2*3.14159*(1+\x) r)}) node[anchor=west,black] {$y$} ;
                \fill[black] (1-0.12,0-0.12) rectangle ++(0.24,0.24);
		\end{tikzpicture}.
	\end{split}
\end{align*}
We employ the same shorthand notation as \cite{Huelsmann:2020xcy}, i.e.\ that 
green lines denote a sum over all possible combinations of red and blue.
With these, the flow equation of the scale dependent effective action can be compactly expressed as
\begin{equation}
    \partial_k \Gamma_k = \frac{i}{2} \bigg\{ \begin{gathered}
            \includegraphics[height=2.5cm]{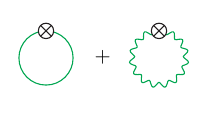}
        \end{gathered}  \bigg\} \,,
\end{equation}
where in comparison to Refs.~\cite{Huelsmann:2020xcy,Roth:2021nrd,Roth:2023wbp} the global minus sign is here absorbed into our definition of the regulator.

\subsection{Static flow}
\label{sct:staticFlow}
Generally, one can assign a free energy $F_k=F_k[\phi,n]$ to the effective average MSR action $\Gamma_k=\Gamma_k[\phi,n,\tilde{\Phi},\tilde{N}]$ by matching the functional derivative of $F_k$ with respect to the classical fields $\phi,n$ to a functional derivative of $\Gamma_k$ with respect to the composite response fields $\tilde{\Phi}, \tilde{N}$, 
\begin{equation}
    \frac{\delta F_k}{\delta \phi_a} \equiv -\frac{\delta \Gamma_k}{\delta \tilde{\Phi}_a} \bigg\rvert_{\substack{\tilde{\Phi} \,=\, \tilde{N} \,=\, 0 \,, \hspace{0.3cm} \\ \partial_t \phi \,=\, \partial_t n \,=\, 0}} \,, \hspace{0.5cm} \frac{\delta F_k}{\delta n_{ab}} \equiv -\frac{\delta \Gamma_k}{\delta \tilde{N}_{ab}} \bigg\rvert_{\substack{\tilde{\Phi} \,=\, \tilde{N} \,=\, 0 \,, \hspace{0.3cm} \\ \partial_t \phi \,=\, \partial_t n \,=\, 0}} \,. \label{eq:defOfFk}
\end{equation}
Using this matching procedure one can define an FRG-scale-dependent LGW functional $F_k[\phi,n]$ (up to an irrelevant additive constant), as motivated in more detail below.

We emphasize a central feature of our formulation here, namely that the such defined free energy $F_k$ generally satisfies a closed flow equation on its own,
\begin{equation}
    \partial_k F_k = \frac{T}{2} \tr\left\{ \partial_k R_k \circ \left( F_k^{(2)} + R_k \right)^{-1} \right\} \,, \label{eq:staticFlowEq}
\end{equation}
with the second functional derivative $F_k^{(2)}$, which is in superfield space $\psi=(\phi,n)$ given by \[ F_{ij,k}^{(2)}(\vec{x},\vec{y}) \equiv \frac{\delta^2 F}{\delta \psi_i(\vec{x}) \delta \psi_j(\vec{y})} \,, \]
where $i$ and $j$ are again superfield indices, and with the regulator $R_k = \mathrm{diag}(R_k^\phi,R_k^n)$ in superfield space.
Remarkably, the right-hand side of \eqref{eq:staticFlowEq} only depends on the free energy $F_k$ again, but not on any genuine real-time quantity like the kinetic coefficients or the reversible mode coupling constant.
We therefore say that the `static' part of the flow given by the free energy~\eqref{eq:defOfFk}, decouples from the `dynamic' part of the flow, given by any remaining parts of the effective average MSR action which are not contained in~\eqref{eq:defOfFk} (as e.g.~the kinetic coefficients $\Gamma_0$, $\gamma \vec{\nabla}^2$, or the reversible mode coupling~$g$).
Notably, Eq.~\eqref{eq:staticFlowEq} coincides with the standard flow equation in $d$-dimensional Euclidean spacetime, so one can employ the well-developed methods from the standard formulation of the FRG in Euclidean space to find suitable truncation schemes for \eqref{eq:staticFlowEq}, for instance by expanding $F_k$ in the number of derivatives~\cite{Berges:2000ew}.
There the lowest order would correspond to a local potential approximation (LPA) of the free energy.

To elaborate more on \eqref{eq:defOfFk} and \eqref{eq:staticFlowEq}, we
first discuss the reasoning for why the flow equation \eqref{eq:staticFlowEq} of the free energy $F_k$ should be closed (i.e.~independent of the dynamics).
Since we add the regulators $R_k^{\phi}$ and $R_k^{n}$ as well as the classical sources $H$ and $\mathcal{H}$ directly to the free energy as in \eqref{eq:regToFreeEnergy} and \eqref{eq:freeEnergyReplacementWithSources}, we do not violate the symmetry of thermal equilibrium \eqref{eq:thermEqSymm} throughout the flow, and hence the equilibrium distribution is given by a (regulated) Boltzmann distribution at all FRG scales.\footnote{This can be shown via a Fokker-Planck equation to which the MSR path integral is equivalent, see e.g.~\cite{taeuber_2014} in the context of Model~G.}
As a non-trivial feature of our formulation, this implies that we have a physical theory (but with a regulated free energy \eqref{eq:regToFreeEnergy}) at \emph{all} FRG scales~$k$.
Hence, equal-time correlation functions can be obtained from the usual equilibrium generating functional \eqref{eq:equilibriumGenFunc} as averages over the Boltzmann distribution \eqref{eq:boltzmannDist}, but with the regulated free energy \eqref{eq:regToFreeEnergy}.
More specifically, in the equilibrium formulation the FRG-scale-dependent free energy $F_k$ is given by a (modified) Legendre transform of the equilibrium generating functional $Z_k^{eq}$, i.e.
\begin{align}
    F_k \equiv \sup_{H,\mathcal{H}} \left\{ \int_{\vec{x}} \left( H_a \phi_a + \orange{\frac{1}{2}} \mathcal{H}_{ab} n_{ab} \right) - \Delta F_k - T\log Z_k^{eq} \right\} \label{eq:eqModLegendreTransform}
\end{align}
where $\Delta F_k$ denotes the regulator term from \eqref{eq:regToFreeEnergy}.
Following the standard derivation from~\cite{Wetterich:1992yh},
one can straightforwardly deduce that $F_k$ satisfies the standard Euclidean flow equation \eqref{eq:staticFlowEq}.

Next, we motivate the definition \eqref{eq:defOfFk} by relating $F_k$, as defined in \eqref{eq:eqModLegendreTransform}, to the effective average MSR action $\Gamma_k$  as defined in \eqref{eq:effAvgActionDef}:
For a purely classical and time-independent source configuration ($\tilde{H}=\tilde{\mathcal{H}}=0$ and $\partial_t H=\partial_t \mathcal{H}=0$)
equal-time correlation functions can be computed in two equivalent ways: From functional derivatives of $F_k$ in the equilibrium formalism, i.e.~as an average over the Boltzmann distribution \eqref{eq:boltzmannDist}, or from functional derivatives of $\Gamma_k$ in the real-time MSR formalism. In the equilibrium formalism, on the one hand, time-independent sources $H,\mathcal{H}$ are related to the first functional derivative of $F_k$ via 
\begin{equation}
    H_a = \frac{\delta F_k}{\delta \phi_a} + R_k^\phi \phi_a \,, \hspace{0.5cm}  \mathcal{H}_{ab} = \frac{\delta F_k}{\delta n_{ab}} + R_k^n n_{ab} \,, \label{eq:sourceCalc1}
\end{equation}
as can be straightforwardly derived from the modified Legendre transform \eqref{eq:eqModLegendreTransform}.
On the other hand, such a classical source configuration $H,\mathcal{H}$ is given
as a special case of the real-time MSR formalism by setting $\tilde{H}=\mathcal{\tilde{H}}=0$ and $\partial_t H=\partial_t \mathcal{H}=0$ in the effective equations of motion (where the latter are given by the first functional derivative of the modified Legendre transform \eqref{eq:effAvgActionDef} with respect to $\tilde{H}_a$, and $\tilde{\mathcal{H}}_{ab}$, respectively)
\begin{equation}
    H_a = -\frac{\delta \Gamma_k}{\delta \tilde{\Phi}_a} \bigg\rvert_{\substack{\tilde{\Phi} \,=\, \tilde{N} \,=\, 0 \,, \hspace{0.3cm} \\ \partial_t \phi \,=\, \partial_t n \,=\, 0}} + R_k^\phi \phi_a \,, \hspace{0.5cm} \mathcal{H}_{ab} = -\frac{\delta \Gamma_k}{\delta \tilde{N}_{ab}} \bigg\rvert_{\substack{\tilde{\Phi} \,=\, \tilde{N} \,=\, 0 \,, \hspace{0.3cm} \\ \partial_t \phi \,=\, \partial_t n \,=\, 0}} + R_k^n n_{ab} \,.  \label{eq:sourceCalc2}
\end{equation}
Here we have used that the vanishing of response-field expectation values $\tilde{\Phi}=\tilde{N}=0$ is a solution to the effective equations of motion for vanishing response sources $\tilde{H}=\tilde{\mathcal{H}}=0$ which follows from the normalization condition $Z=1$ of the MSR path integral, see e.g.~\cite{kamenev_2011}, and that if the external classical source $\partial_t H = \partial_t \mathcal{H} = 0$ is time independent, so is the classical field expectation value $\partial_t \phi=\partial_t n=0$ since our MSR action $S$ does not carry any other explicit time dependence.

Since both the equilibrium \eqref{eq:sourceCalc1} and the real-time MSR approach \eqref{eq:sourceCalc2} have to yield the same static classical source $H_a$ and $\mathcal{H}_{ab}$,
we find the anticipated relation \eqref{eq:defOfFk} between the free energy $F_k$ and the effective average MSR action $\Gamma_k$ by identifying \eqref{eq:sourceCalc1} and \eqref{eq:sourceCalc2}.
This relation
allows us to generally extract the free energy $F_k$ from a given $\Gamma_k$ at all FRG scales. Then, by virtue of \eqref{eq:staticFlowEq}, we know that its flow equation is closed.

In addition to this general argument, we provide an explicit derivation of \eqref{eq:staticFlowEq} from \eqref{eq:defOfFk} and the known flow equation \eqref{eq:treeLevelFlowEq} of $\Gamma_k$ in Appendix~\ref{sct:CorrespondenceRealTimeAndEuclFlows}.

\section{Truncation and flow equations}\label{sec:truncation}

As a starting point for practical applications within the FRG, one has to truncate the effective average action in a way that respects all relevant symmetries (here listed in Sec.~\ref{sct:SymmetriesOfMSRAction}).
In the context of critical phenomena, an expansion of the effective average action in terms of derivatives (frequently referred to as `derivative expansion') has proven to converge rather quickly towards quantitatively accurate results for critical exponents~\cite{Litim:2001dt,DePolsi:2020pjk}, universal amplitude ratios~\cite{DePolsi:2021cmi}, and recently, the location of Yang-Lee edge singularities~\cite{Connelly:2020gwa,Rennecke:2022ohx,Johnson:2022cqv}.
This suggests that a similar expansion in terms of derivatives is a sensible choice for a systematic truncation scheme also for Model~G here.
In the context of critical \emph{dynamics}, a derivative expansion has been already employed in previous works in the purely relaxational Models~A \cite{Canet:2011wf} and~C \cite{Mesterhazy:2013naa}, for example.
However, in the presence of reversible mode couplings, it is rather difficult to set up a derivative expansion systematically since the temporal-gauge and its accompanying residual displacement symmetry from Sec.~\ref{sct:SymmetriesOfMSRAction} prohibits a naive introduction of terms with ordinary time derivatives
(rather, the covariant time derivative~\eqref{eq:covarTimeDerivs} that is manifestly invariant under the displacement symmetry must thus be used instead of the ordinary one in a systematic derivative expansion). 
For the calculation of spectral functions, the combined expansion scheme for the effective average action of Refs.~\cite{Roth:2021nrd,Roth:2023wbp}, first in terms of 1PI vertex functions and a second in the number of loops taken into account on the level of the flow equations (if possible self-consistently), has proven to be well suited to study their  dynamic critical behavior in Models A, B and C~\cite{Roth:2023wbp}. 
For the scope of this work, we take the more pragmatic approach of \emph{postulating} a suitable ansatz for the effective average action based on previous experience.

In particular, we construct our truncation by considering a generalized form of the bare MSR action~\eqref{eq:eoms} where all couplings are promoted to be running (i.e.\ to depend on the FRG scale~$k$),
\begin{align}
    \Gamma_k = \int_x \bigg[ &-\tilde{\phi}_{a,k} \left( Z_{\phi,k}^{\omega} \frac{\partial \phi_a}{\partial t} + \gamma_{\phi,k}(\vec{\nabla}) \frac{\delta F_k}{\delta \phi_a}-\frac{g_k^{\phi n}}{\orange{2}}\{\phi_a,n_{bc}\}\frac{\delta F_k}{\delta n_{bc}} \right) \nonumber \\
    &-\orange{\frac{1}{2}}\tilde{n}_{ab,k} \left( Z_{n,k}^{\omega} \frac{\partial n_{ab}}{\partial t} + \gamma_{n,k}(\vec{\nabla}) \frac{\delta F_k}{\delta n_{ab}}-g_k^{n\phi}\{n_{ab},\phi_c\}\frac{\delta F_k}{\delta \phi_c}-\frac{g_k^{nn}}{\orange{2}} \{n_{ab},n_{cd}\}\frac{\delta F_k}{\delta n_{cd}} \right) \nonumber \\
    &+Z_{\phi,k}^\omega i T \tilde{\phi}_{a,k} \gamma_{\phi,k}(\vec{\nabla}) \tilde{\phi}_{a,k} + \orange{\frac{1}{2}} Z_{n,k}^\omega i T \tilde{n}_{ab,k} \gamma_{n,k}(\vec{\nabla}) \tilde{n}_{ab,k}  \bigg] \,. \label{eq:effAvgActionInStandardResponseFields}
\end{align}
We have included an arbitrary scale-dependent free-energy functional $F_k[\phi,n]$, temporal wave function renormalization factors $Z_{\phi,k}^{\omega}$ and $Z_{n,k}^{\omega}$, generalized kinetic coefficients $\gamma_{\phi,k}(\vec{\nabla})$ and $\gamma_{n,k}(\vec{\nabla})$ that involve arbitrary spatial gradients, and three different instances $g_k^{\phi n}$, $g_k^{n \phi}$ and $g_k^{nn}$ of the reversible mode couplings corresponding to the three different Poisson brackets.
To guarantee the conservation of the charges, the spatial Fourier transform of the kinetic coefficient $\gamma_{n,k}(\vec{p})$ must vanish in the limit $\vec{p} \to 0$, and thus its expansion in gradients must start at order $\vec{\nabla}^2$.

Moreover, note that the temporal wave function renormalization factors $Z_{\phi,k}^\omega$ and $Z_{n,k}^\omega$ are not independent of the other couplings, but can be eliminated via a redefinition of the response fields $\tilde{\phi}_k \to \tilde{\phi}_k/Z_{\phi,k}^\omega$, $\tilde{n}_k \to \tilde{n}_k/Z_{n,k}^\omega$ together with a corresponding rescaling of the kinetic coefficients $\gamma_{\phi,k}(\vec{p})/Z_{\phi,k}^\omega \to \gamma_{\phi,k}(\vec{p})$, $\gamma_{n,k}(\vec{p})/Z_{n,k}^\omega \to \gamma_{n,k}(\vec{p})$ and reversible mode couplings, $g_k^{\phi n}/Z_{\phi,k}^\omega \to g_k^{\phi n}$, $g_k^{n \phi}/Z_{n,k}^\omega \to g_k^{n\phi}$ and $g_k^{nn}/Z_{n,k}^\omega \to g_k^{nn}$. We can therefore safely set $Z_{\phi,k}^\omega = Z_{n,k}^\omega = 1$ from now on.

Finally, reversibility (i.e.~requiring that the discrete transformation \eqref{eq:thermEqSymm} of thermal equilibrium remains a symmetry of the effective average action) implies that the matrix of reversible mode couplings is symmetric, i.e.~that $g_k^{\phi n} = g_k^{n \phi}$, here.

The standard response fields $\tilde{\phi}$ and $\tilde{n}$ are related to the composite response fields $\tilde{\Phi}$ and $\tilde{N}$ by an FRG-scale-dependent generalization of \eqref{eq:responseFieldsExpVal},
\begin{subequations}
\begin{eqnarray}
    \tilde{\Phi}_a &\equiv& \gamma_{\phi,k}(\vec{\nabla}) \tilde{\phi}_{a,k} - \frac{g_k^{n\phi}}{\orange{2}} \tilde{n}_{bc,k} \{n_{bc}, \phi_a\} \,, \label{PhiTildeTrafoScaleDep} \\
    \tilde{N}_{ab} &\equiv&  \gamma_{n,k}(\vec{\nabla}) \tilde{n}_{ab,k} - \frac{g_k^{nn}}{\orange{2}} \tilde{n}_{cd,k} \{n_{cd},n_{ab}\} - g_k^{\phi n} \tilde{\phi}_{c,k} \{\phi_c,n_{ab}\}  \,. \label{NTildeTrafoScaleDep}
\end{eqnarray} \label{eq:responseFieldTrafoScaleDep}%
\end{subequations}
In particular, this implies that in order to keep the composite response fields $\tilde{\Phi}$ and $\tilde{N}$ independent of the FRG scale $k$, the standard response fields $\tilde{\phi}_k$ and $\tilde{n}_k$ must necessarily depend on $k$.
In the more compact superfield notation, this definition reads
\begin{equation}
    \tilde{\Psi} = J_k(\psi) \tilde{\psi}_k
\end{equation}
with the cale-dependent Jacobian $J_k(\psi)$ given by
\begingroup 
\setlength\arraycolsep{10pt} 
\begin{equation}
   J_k(\psi) =
    \begin{pmatrix}
        \delta_{ad}\, \gamma_{\phi,k}(\vec{\nabla})  & \frac{g_k^{n\phi}}{\orange{2}}\{\phi_a,n_{ef}\} \\
        g_k^{\phi n} \{n_{bc},\phi_d\} & -\tfrac{1}{2}\left(\delta_{be}\delta_{cf}-\delta_{bf}\delta_{ce}\right) \gamma_{n,k}(\vec{\nabla}) + \frac{g_k^{nn}}{\orange{2}}\{n_{bc},n_{ef}\} 
    \end{pmatrix}\,. 
\end{equation}
\endgroup
In our superfield notation the ansatz for the effective average action reads
\begin{align}
    \Gamma_k &= -\tilde{\psi}_k^T \partial_t \psi + iT \tilde{\psi}_k^T \gamma_k \tilde{\psi}_k - \tilde{\psi}_k^T J_k^T(\psi)  \frac{\delta F_k}{\delta\psi} && \text{(old fields)} \label{effAvgActionCompactOldFields} \\
    &= -\tilde{\Psi}^T {J_k^{-1}}^T(\psi) \partial_t \psi + iT \tilde{\Psi}^T J_k^{-1}(\psi) \tilde{\Psi} - \tilde{\Psi}^T \frac{\delta F_k}{\delta\psi} \label{effAvgActionCompactNewFields} && \text{(new fields)}
\end{align}
where also the Jacobian depends on the FRG scale~$k$, and with the inverse kinetic coefficient $\gamma_k$ in superfield space denoted by
\begin{align}
    \gamma_k(\vec{\nabla}) \equiv \begin{pmatrix}
        \gamma_{\phi,k}(\vec{\nabla}) & \\
        & \gamma_{n,k}(\vec{\nabla})
    \end{pmatrix} \,.
\end{align}
To extract vertices and propagators from \eqref{effAvgActionCompactNewFields}, we introduce small but spacetime-dependent perturbations around  (possibly) FRG-scale-dependent background fields, $\phi(x) = \phi_{0,k} + \delta\phi(x)$, $n(x) = n_{0,k} +  \delta n(x)$,
\begin{align}
    J_k^{-1}(\psi) &=J_k^{-1}(\psi_{0,k} + \delta \psi) \equiv \left( J_k(\psi_{0,k}) - \delta J_k(\delta\psi) \right)^{-1}
\end{align}
where the last equality defines the perturbation $\delta J(\delta\psi)$ of the Jacobian, which vanishes $\delta J(\delta \psi = 0) = 0$ for  unperturbed background fields $\psi(x)=\psi_0$.
Then we can expand the inverse Jacobian in a Neumann series in  terms of powers of the field perturbations,
\begin{align}
    J_k^{-1}(\psi) &= J_k(\psi_{0,k}) + J_k(\psi_{0,k}) \circ \delta J_k(\delta\psi) \circ J_k(\psi_{0,k}) + \cdots \,, \label{JExpansion}
\end{align}
where the circle $\circ$ denotes convolution again, i.e.~summation over indices and integration over adjacent spacetime coordinates. The $n$-point vertices are given by taking the corresponding $n^{\text{th}}$ functional derivative of the effective average action.
In this work we expand around vanishing field expectation values, so we set the background fields to $\phi_0 = n_0 = 0$.
With this choice, the FRG-scale-dependent propagators \eqref{eq:phiProp} and \eqref{eq:nProp} are given by (with $\chi_{\phi,k}(\vec p)$ and $\chi_{n,k}(\vec p)$ the corresponding FRG-scale-dependent static susceptibilities as extracted from the free energy, see Eq.~\eqref{eq:phiAndNStaticSusceps} below)
\begin{subequations}
\begin{align}
    G_{\phi,k}^R(\omega,\vec{p}) &= -\frac{\gamma_{\phi,k}(\vec p)}{i\omega - \gamma_{\phi,k}(\vec p)\,\chi_{\phi,k}^{-1}(\vec p) } \,, \label{eq:phiPropTruncGR} \\
    G_{\phi,k}^A(\omega,\vec{p}) &=-\frac{\gamma_{\phi,k}(\vec p)}{-i\omega - \gamma_{\phi,k}(\vec p)\,\chi_{\phi,k}^{-1}(\vec p) } \,, \label{eq:phiPropTruncGA} \\
    iF_{\phi,k}(\omega,\vec{p}) &=\frac{2i\gamma_{\phi,k}(\vec p) T}{\omega^2 + \left( \gamma_{\phi,k}(\vec p)\,\chi_{\phi,k}^{-1}(\vec p) \right)^2 } \,, \quad \text{(fixed from the FDR)} \label{eq:phiPropTrunciF}
\end{align} \label{eq:phiPropTrunc}%
\end{subequations}
for the order parameter, and by
\begin{subequations}
\begin{align}
    G_{n,k}^R(\omega,\vec{p}) &= -\frac{\gamma_{n,k}(\vec p)}{i\omega - \gamma_{n,k}(\vec p)\, \chi_{n,k}^{-1}(\vec p) } \,, \\
    G_{n,k}^A(\omega,\vec{p}) &=-\frac{\gamma_{n,k}(\vec p)}{-i\omega - \gamma_{n,k}(\vec p)\, \chi_{n,k}^{-1}(\vec p) } \,, \\
    iF_{n,k}(\omega,\vec{p}) &=\frac{2i\gamma_{n,k}(\vec p) T}{\omega^2 + \left( \gamma_{n,k}(\vec p)\,\chi_{n,k}^{-1}(\vec p) \right)^2 } \,, \quad \text{(fixed from the FDR)} \label{eq:nPropTrunciF}
\end{align} \label{eq:nPropTrunc}%
\end{subequations}
for the charge densities.
The static susceptibilities are defined as the equal-time correlation functions
\begin{align}
    \chi_{\phi_a \phi_b,k}(\vec p) &= \int \frac{d^dx}{(2\pi)^d} \,e^{-i\vec{p}\cdot\vec{x}} \langle \phi_a(t,\vec{x}) \phi_b(t,\vec{0}) \rangle \,, \\
    \chi_{n_{ab} n_{cd},k}(\vec p) &= \int \frac{d^dx}{(2\pi)^d} \,e^{-i\vec{p}\cdot\vec{x}} \langle n_{ab}(t,\vec{x}) n_{cd}(t,\vec{0}) \rangle \,,
\end{align}
and are hence related to the statistical functions \eqref{eq:phiPropTrunciF} and \eqref{eq:nPropTrunciF} via
\begin{align}
    \int \frac{d\omega}{2\pi} \,iF_{\phi,k}(\omega,\vec{p}) = T \chi_{\phi,k}(\vec p) \, ,\quad\text{and}\quad \int \frac{d\omega}{2\pi} \,iF_{n,k}(\omega,\vec{p}) = T \chi_{n,k}(\vec p) \, .
\end{align}
For vanishing field expectation values, the static susceptibilities are diagonal in field space, i.e.~we can parameterize them as
\begin{align}
    \chi_{\phi_a \phi_b,k}(\vec p) = \chi_{\phi,k}(\vec p) \,\delta_{ab} \quad\text{and}\quad 
    \chi_{n_{ab} n_{cd},k}(\vec p) = \chi_{n,k}(\vec p)\, \orange{(\delta_{ac} \delta_{bd} - \delta_{ad}\delta_{bc})} \, ,
\end{align} \label{eq:staticSuscepDiagInFieldSpace}%
analogous to the propagators \eqref{eq:phiPropDiag} and \eqref{eq:nPropDiag}.
They are related directly to the second functional derivatives of the FRG-scale-dependent free energy $F_k$ via
\begin{subequations}
\begin{align}
    \chi_{\phi,k}^{-1}(\vec p) \,\delta_{ab} &= \frac{\delta^2 F_k}{\delta \phi_a(-\vec p)\,\delta \phi_b(\vec p)} \bigg\rvert_{\phi=n=0} + R_{\phi,k}(\vec p)\,\delta_{ab} \label{eq:phiStaticSusc} \\
    \chi_{n,k}^{-1}(\vec p)\, \orange{(\delta_{ac} \delta_{bd} - \delta_{ad}\delta_{bc})} &= \frac{\delta^2 F_k}{\delta n_{ab}(-\vec p)\,\delta n_{cd}(\vec p)} \bigg\rvert_{\phi=n=0}  + R_{n,k}(\vec p) \, \orange{(\delta_{ac} \delta_{bd} - \delta_{ad}\delta_{bc})} \label{eq:nStaticSusc}
\end{align} \label{eq:staticSuscs}%
\end{subequations}
The explicit expressions are given in Eq.~\eqref{eq:phiAndNStaticSusceps} below, in the following subsection for the truncation of the free energy that we introduce next.

To conclude this subsection, we mention that
at least in principle, one can straightforwardly generalize the present scheme to (e.g.)~an expansion around the scale-dependent minimum where the background field for the order parameter $\phi_0 = \phi_{0,k}$ becomes scale-dependent.
However, the corresponding Neumann series \eqref{JExpansion} and subsequent functional derivatives of $\Gamma_k$ then become rather tedious, so we restrict ourselves to the choice $\phi_0 = n_0 = 0$ mainly for convenience here.

\subsection{Static couplings}
\label{sct:staticCouplings}

We implement the (arguably simplest possible)  truncation for the free energy by promoting the squared mass and the quartic coupling to depend on the FRG scale, i.e.\ we consider
\begin{equation}
    F_k=\int_{\vec{x}} \left\{ \frac{1}{2}(\partial^i \phi_a)(\partial^i \phi_a)+\frac{m_k^2}{2} \phi_a \phi_a + \frac{\lambda_k}{4!N}(\phi_a\phi_a)^2 +\frac{1}{\orange{4}\chi}n_{ab}n_{ab} \right\} \, .\label{eq:freeEnergyTrunc}
\end{equation}
The flow of the possibly scale-dependent static susceptibility $\chi_k \equiv \chi$ vanishes trivially since $n$ only appears quadratically in the free energy and does not couple to $\phi$ therein.
Recall that the flow of the free energy decouples from the rest of the flow, as we have discussed in Sec.~\ref{sct:staticFlow} above. This allows us to study the flow of $F_k$ in closed form, separately from the dynamics.
In particular, the flow equations for the squared mass and the quartic coupling can be derived using well-established methods from the standard FRG for a $\phi^4$-theory in $d=3$ Euclidean spacetime dimensions.
As a first step, one needs the propagators, which are here given by the static susceptibilities \eqref{eq:phiStaticSusc} and \eqref{eq:nStaticSusc}.
For the truncation \eqref{eq:freeEnergyTrunc}, they read
\begin{align}
    \chi_{\phi,k}(\vec p) &= \frac{1}{m_k^2 + \vec p^2 + R_{k}^{\phi}(\vec p)} \quad\text{and}\quad
    \chi_{n,k}(\vec p) = \frac{1}{\chi^{-1} + R_{k}^{n}(\vec p)} \, . \label{eq:phiAndNStaticSusceps}
\end{align}
for vanishing field expectation values $\phi_0=n_0=0$.
Moreover, one has the quartic interaction vertex
\begin{equation}
    \frac{\delta^4 F_k}{\delta \phi_a(\vec{x}) \delta \phi_b(\vec{y}) \delta \phi_c(\vec{z}) \delta \phi_d(\vec{w})}  = \frac{\lambda_k}{6N} \, \left(\delta_{ab}\delta_{cd}+\delta_{ac}\delta_{bd}+\delta_{ad}\delta_{bc}\right) \, \delta(\vec{x}-\vec{y})\delta(\vec{y}-\vec{z})\delta(\vec{z}-\vec{w})\, .
\end{equation}
Inserting the propagators and vertices into appropriate functional derivatives of the general flow equation \eqref{eq:staticFlowEq} for the FRG-scale-dependent free energy $F_k$ evaluated at vanishing field expectation values $\phi_0=n_0=0$
results in the flow equations (here for an arbitrary momentum-dependent regulator $R_{k}^{\phi}(\vec{q})$)
\begin{align}
    \partial_k m_k^2 &= - \frac{(N+2)\lambda_k T}{6N} \int \!\! \frac{d^d q}{(2\pi)^d}   \frac{\partial_k R_{k}^{\phi}(\vec{q})}{\big( m_k^2 + \vec{q}^2 + R_{k}^{\phi}(\vec{q})\big)^2}\, , \label{flowm2k} \\
    \partial_k \lambda_k  &=  \frac{(N+8)\lambda_k^2 T}{3N} \int \!\! \frac{d^d q}{(2\pi)^d}  \frac{\partial_k R_{k}^{\phi}(\vec{q})}{\big( m_k^2 + \vec{q}^2 + R_{k}^{\phi}(\vec{q})\big)^3}\, . \label{flowlambdak}
\end{align}
By dimensional reduction at finite temperature, the flow equations \eqref{flowm2k} and \eqref{flowlambdak} for $m_k^2$ and $\lambda_k T$ are the same as those for a corresponding $d$-dimensional $O(N)$ symmetric scalar quantum field theory in Euclidean spacetime at zero temperature, which are well known in the literature, see for instance \cite{Berges:2000ew}.
The present truncation corresponds to the lowest order in a systematic expansion of the free energy in spatial gradients (usually called the local potential approximation, LPA) combined with a Taylor expansion of the effective potential up to the minimally necessary second order in the field invariant $\rho=\phi_a \phi_a$.

We emphasize here that the truncation used in this work can be straightforwardly improved by employing a more sophisticated truncation for the free energy $F_k$.
Such sophisticated truncations are well-developed for the $d$-dimensional Euclidean $O(N)$ model.
For the derivative expansion see e.g.\ \cite{Litim:2001dt,DePolsi:2020pjk,DePolsi:2021cmi}.

\subsection{Kinetic coefficients}
\label{sct:flowOfKineticCoeffs}

\begin{figure}
    \centering
    \begin{align*}
    \partial_k \Gamma_k^{\phi\tilde{\Phi}} &= -i\bigg\{
    \begin{gathered}
        \includegraphics{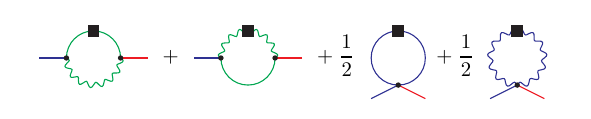}
    \end{gathered}
    \bigg\} \\
    \partial_k \Gamma_k^{n\tilde{N}} &= -i\bigg\{
    \begin{gathered}
        \includegraphics{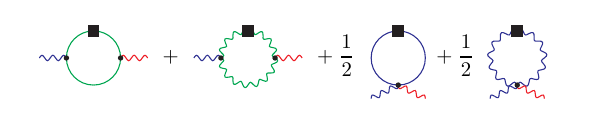}
    \end{gathered}
    \bigg\}
    \end{align*}
    \begin{align*}
    \partial_k \Gamma_k^{\tilde{\Phi}\tilde{\Phi}} &= -i\bigg\{
    \begin{gathered}
        \includegraphics{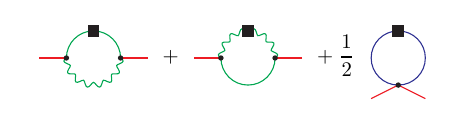}
    \end{gathered}
    \bigg\} \\
    \partial_k \Gamma_k^{\tilde{N}\tilde{N}} &= -i\bigg\{ 
    \begin{gathered}
        \includegraphics{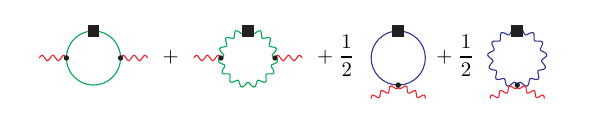}
    \end{gathered}
    \bigg\}
    \end{align*}
    \caption{Flow of the 2-point function at the IR minimum $\phi = n = 0$, $\tilde{\Phi}=\tilde{N}=0$ in the symmetry-restored phase $T>T_c$ without explicit symmetry breaking. Drawn using \textit{JaxoDraw}~\cite{Binosi:2008ig}.}
    \label{fig:flow2PtFncIRMinHighT}
\end{figure}

Using the fluctuation-dissipation relation \eqref{eq:FDRFor2PtFncs} between the two-point functions, one can extract the kinetic coefficients 
$\gamma_{\phi,k}(\vec{p})$ and $\gamma_{n,k}(\vec{p})$ generally from the low-frequency limit, $\omega \to 0$ of the two-point functions $\Gamma_k^{\tilde{\Phi}\tilde{\Phi}}(\omega,\vec{p})$ and $\Gamma_k^{\tilde{N}\tilde{N}}(\omega,\vec{p})$, respectively, taken at non-vanishing spatial momentum $\vec{p} \neq 0$,
\begin{align}
    \gamma_{\phi,k}(\vec{p}) &= \lim_{\omega \to 0} \frac{2iT}{\Gamma_{k}^{\tilde{\Phi}\tilde{\Phi}}(\omega,\vec{p})} \,, \label{flowphi}\\
    \gamma_{n,k}(\vec{p}) &= \lim_{\omega \to 0} \frac{2iT}{\Gamma_{k}^{\tilde{N}\tilde{N}}(\omega,\vec{p})} \,. \label{flown}
\end{align}
Using the product rule, we find that their flow equations are hence given by
\begin{align}
    \partial_{k} \gamma_{\phi,k}(\vec{p}) &= - \frac{\gamma_{\phi,k}^2(\vec{p})}{2iT} \lim_{\omega \to 0} \partial_{k} \Gamma_{k}^{\tilde{\Phi}\tilde{\Phi}}(\omega,\vec{p}) \,, \\
    \partial_{k} \gamma_{n,k}(\vec{p}) &= - \frac{\gamma_{n,k}^2(\vec{p})}{2iT} \lim_{\omega \to 0} \partial_{k} \Gamma_{k}^{\tilde{N}\tilde{N}}(\omega,\vec{p}) \,.
\end{align}
The diagrammatic representation of the flow of two-point functions is given in Fig.~\ref{fig:flow2PtFncIRMinHighT}.
We evaluate these flow equations at vanishing field expectation values $\phi=\tilde{\Phi}=n=\tilde{N}=0$, which is valid in the symmetry-restored phase, i.e.\ for $T \geq T_c$, and for vanishing explicit symmetry breaking $H=0$, corresponding to vanishing current quark masses in QCD, in the chiral limit.

Using the truncation given in \eqref{effAvgActionCompactNewFields}, and the method of expanding the Jacobian in powers of the fields, one can then compute the three and four-point vertices and find that the nonzero vertices are $\tilde{\Phi}\phi n$, $\tilde{N}\phi \phi$, $\tilde{N}nn$, $\tilde{\Phi}\tilde{\Phi}\phi \phi$, $\tilde{N}\tilde{N}\phi\phi$ and $\tilde{N}\tilde{N}nn$. Although the resulting expressions for the vertices are straightforward to derive (at least in principle), they are lengthy and not particularly illuminating, so instead of explicitly writing them here, we list them in Appendix~\ref{sct:vertexFunctions} for completeness.

Because of their diffusive dynamics, the kinetic coefficient $\gamma_{n,k}(\vec{p})$ for the charge densities is momentum dependent already at tree level. Moreover, we expect it to develop a critical power-law form as the system is tuned close to the critical point.
We therefore resolve its full momentum dependence $\gamma_{n,k}(\vec{p}) = \gamma_{n,k}(|\vec{p}|)$ as a function of the momentum's magnitude within our truncation.
However, the order-parameter modes have a finite mass in the symmetric phase $T > T_c$, and hence their kinetic coefficient $\gamma_{\phi,k}(\vec{p} = 0) > 0$ is non-vanishing at zero spatial momentum, so we resort to the approximation that we only keep their kinetic coefficient $\gamma_{\phi,k}(\vec{p}) \approx \gamma_{\phi,k}(\vec{p}=0) \equiv \Gamma_{k}^{\phi}$ evaluated at vanishing spatial momentum.

\noindent
The diagrams contributing to $\partial_k \Gamma_k^{\tilde{\Phi}\tilde{\Phi}}$ are given by:
\begin{align}
    \begin{gathered}
        \includegraphics[height=2cm]{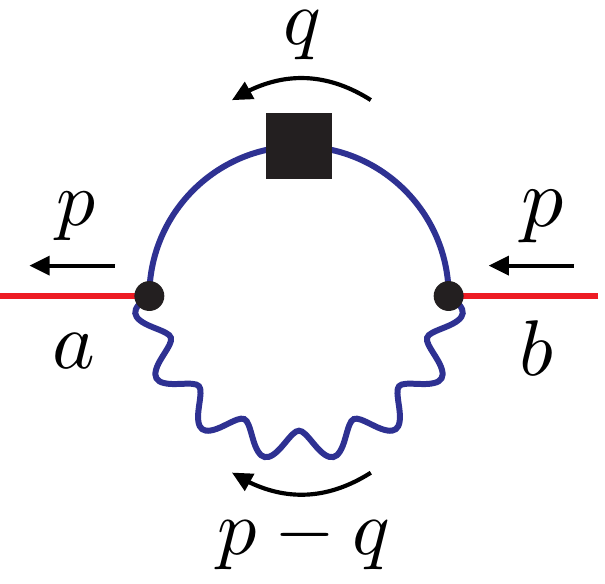}
    \end{gathered} &=  (N-1) \delta_{ab} \int_q \left( \frac{g_k (p^0-q^0)}{\Gamma_{k}^{\phi}\,\gamma_{n,k}(\vec{p}-\vec{q})} \right)^2  B_{\phi,k}^F(q) iF_{n,k}(p-q) \, , \label{eq:feynDiagPhiPhi1}\\
    \begin{gathered}
        \includegraphics[height=2cm]{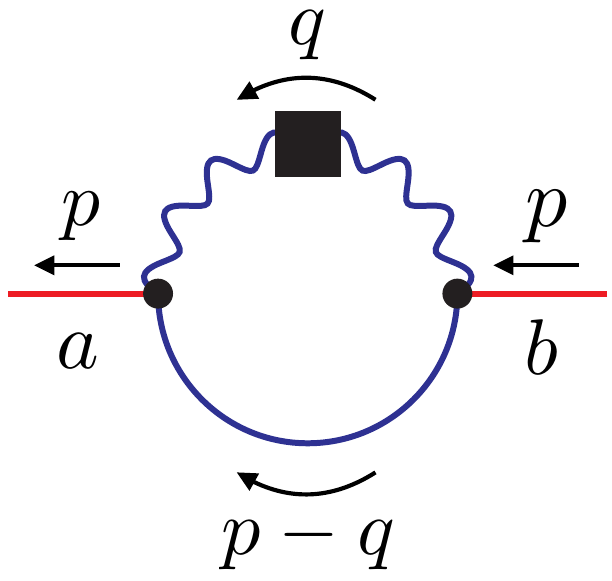}
    \end{gathered} &=  (N-1) \delta_{ab} \int_q \left( \frac{g_k q^0}{\Gamma_{k}^{\phi}\,\gamma_{n,k}(\vec{q})} \right)^2  iF_{\phi,k}(p-q) B_{n,k} ^F(q) \, ,\label{eq:feynDiagPhiPhi2} \\
    \begin{gathered}
        \includegraphics[height=2cm]{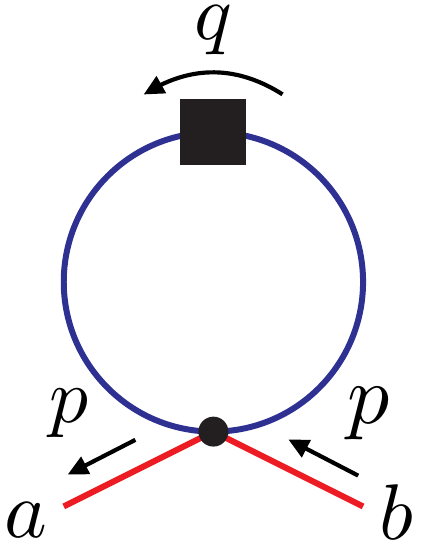}
    \end{gathered} &=  (N-1) \delta_{ab} \left( - \frac{2g_k^2 iT}{(\Gamma_{k}^{\phi})^2} \right) \int_q \left( \frac{1}{\gamma_{n,k}(\vec{q}-\vec{p})} + \frac{1}{\gamma_{n,k}(\vec{q}+\vec{p})} \right) B_{\phi,k}^F(q) \, .\label{eq:feynDiagPhiPhi3}
\end{align}

\noindent
The diagrams contributing to $\partial_k \Gamma_k^{\tilde{N}\tilde{N}}$ are given by:
\begin{align}
    \begin{gathered}
        \includegraphics[height=2cm]{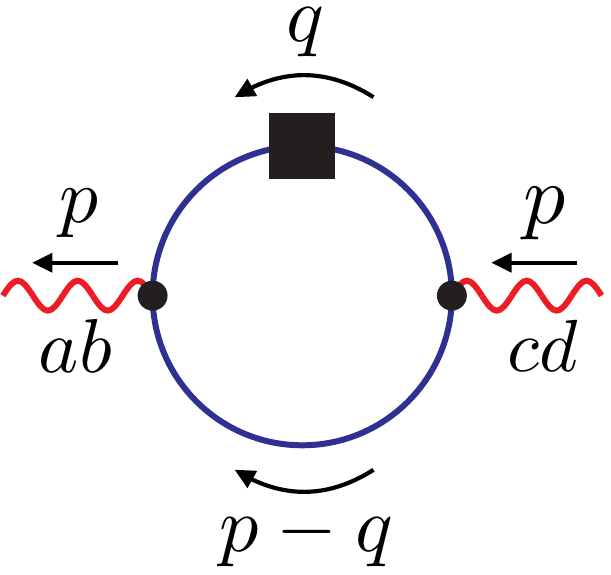}
    \end{gathered} &= 2(\delta_{ac}\delta_{bd}-\delta_{ad}\delta_{bc})  \label{eq:feynDiagNN1} 
    \int_q \left( \frac{g_k (p^0-2q^0)}{\Gamma_{k}^{\phi}\,\gamma_{n,k}(\vec{p})} \right)^2 B_{\phi,k}^F(q) iF_{\phi,k}(p-q) \, ,\\
    \begin{gathered}
        \includegraphics[height=2cm]{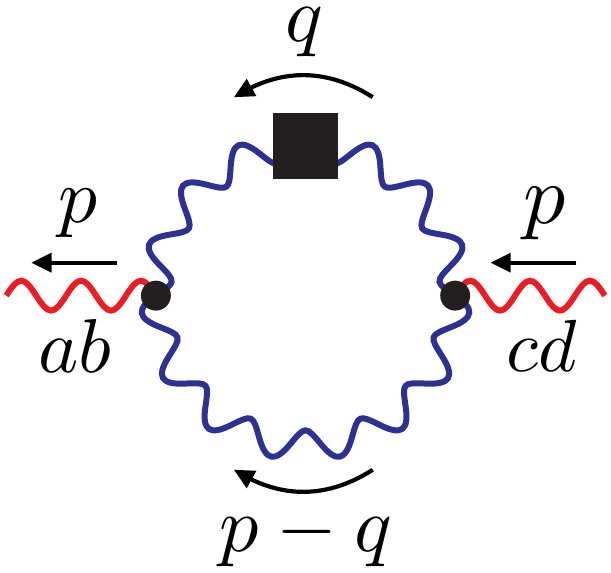}
    \end{gathered}  &= 2(N-2)(\delta_{ac}\delta_{bd}-\delta_{ad}\delta_{bc}) \times \label{eq:feynDiagNN2} \\[-2.0em] \nonumber &\hspace{0.5cm} \int_q \left[\frac{g_k}{\gamma_{n,k}(\vec{p})}\left( \frac{q^0}{ \gamma_{n,k}(\vec{q})} + \frac{q^0-p^0}{\gamma_{n,k}(\vec{q-p})} \right)\right]^2 B^F_{n,k}(q)iF_{n,k}(p-q)  \, , \\
    \begin{gathered}
        \includegraphics[height=2cm]{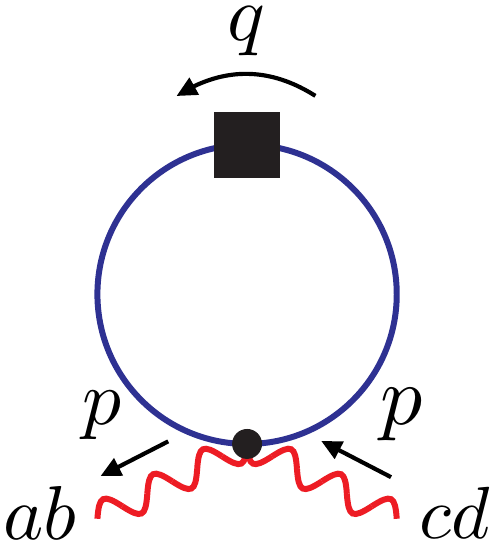}
    \end{gathered}  &= - (\delta_{ac}\delta_{bd}-\delta_{ad}\delta_{bc}) \frac{8g_k^2 iT}{\Gamma_{k}^{\phi} \, \gamma_{n,k}^2(\vec{p})} \int_q B_{\phi,k}^F(q) \, ,  \label{eq:feynDiagNN3} \\
    \begin{gathered}
        \includegraphics[height=2cm]{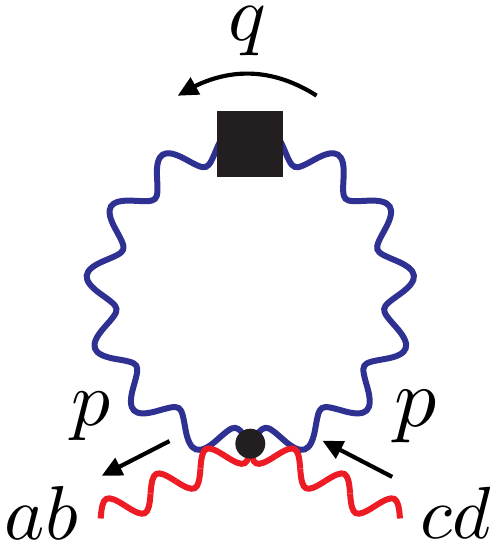}
    \end{gathered} &= -(N-2)(\delta_{ac}\delta_{bd}-\delta_{ad}\delta_{bc})  \frac{4g_k^2 iT}{\gamma_{n,k}^2(\vec{p})} \times  \label{eq:feynDiagNN4} \\[-1.5em] \nonumber &\hspace{2.5cm}
    \int_q \left(\frac{1}{\gamma_{n,k}(\vec{p}+\vec{q})} + \frac{1}{\gamma_{n,k}(\vec{p}-\vec{q})}\right) B_{n,k}^F(q)
\end{align}

\bigskip

\subsection{Non-renormalization of reversible mode couplings}
in this subsection we show that the reversible mode coupling coefficients are protected from renormalization by symmetry, which means $g^{\phi n}_k=g^{n\phi}_k= g^{nn}_k=g$ at any scale. The starting point of our discussion is the temporal (non-Abelian) gauge 
symmetry of the MSR action of Model G, which is preserved by the FRG flow. Based on the discussion in Sec.~\ref{sct:SymmetriesOfMSRAction}, the change of the effective action under time-dependent (but spatially independent) rotations $O(t) \in O(N)$ of the fields  has to obey
\begin{align}
    \Gamma_k[O\phi,O\tilde{\phi},OnO^T,O\tilde{n}O^T] - \Gamma_k[\phi,\tilde{\phi},n,\tilde{n}] &= \frac{1}{2} \int_x \tr  \left[\left( \frac{1}{g}O \partial_t O^T \right) O \tilde{N} O^T \right] \label{eq:wardIdReformulatedAsChangeInEffAction}
\end{align}
Based on the explicit ansatz for the effective average action in Eq.~\eqref{effAvgActionCompactOldFields}, the change on the left-hand side in \eqref{eq:wardIdReformulatedAsChangeInEffAction} can be expressed as
\begin{align}
    &\Gamma_k[O\phi,O\tilde{\phi},OnO^T,O\tilde{n}O^T]  -\Gamma_k[\phi,\tilde{\phi},n,\tilde{n}] = \\ \nonumber &\hspace{3.0cm}  \int_x \left[ -\tilde{\phi}_a(O^T\partial_t O)_{ab}\phi_b -\tilde{n}_{ab}(O^T\partial_t O)_{ab,cd}n_{cd} \right]
\end{align}
Considering an infinitesimal transformation $O\approx 1  + \frac{1}{2}\alpha_{ab}(t)T_{ab}$, with small $\alpha_{ab}(t)$, 
\begin{equation}
    O^T \partial_t O = \frac{1}{2}\dot{\alpha}_{ab}(t)T_{ab}
    + \mathcal{O}(\alpha^2) \, ,\label{eq:formulaForInfONTrafo}
\end{equation}
and the corresponding infinitesimal change of the effective action can then be evaluated as
\begin{align}
    &\Gamma_k[O\phi,O\tilde{\phi},OnO^T,O\tilde{n}O^T] - \Gamma_k[\phi,\tilde{\phi},n,\tilde{n}] = \\ \nonumber &\hspace{1.0cm} \int_x \left[ -\frac{1}{2}\tilde{\phi}_a\dot{\alpha}_{cd}(t)(T_{cd})_{ab}\phi_b -\frac{1}{2}\tilde{n}_{ab}\dot{\alpha}_{ef}(t)(T_{ef})_{ab,cd}n_{cd} \right] + \mathcal{O}(\alpha^2) \,.
\end{align}
Re-expressing the generators $T_{ab}$ in terms of the Poisson-bracket relations of the fields $\phi$ and $n$ in Eq.~\eqref{eq:relBetweenPBsAndONGenerators}, one then obtains the following result for the left-hand side in \eqref{eq:wardIdReformulatedAsChangeInEffAction}
\begin{align}
    & \Gamma_k[O\phi,O\tilde{\phi},OnO^T,O\tilde{n}O^T]  - \Gamma_k[\phi,\tilde{\phi},n,\tilde{n}] = \label{eq:gNonRenormResultForLHS} \\ \nonumber &\hspace{1.0cm} \int_x \left[ -\frac{1}{2}\tilde{\phi}_a\dot{\alpha}_{cd}(t)\{\phi_a, n_{cd}\} -\frac{1}{2}\tilde{n}_{ab}\dot{\alpha}_{ef}(t)\{n_{ab},n_{ef}\} \right] + \mathcal{O}(\alpha^2) \,.
\end{align}
Next we consider the right-hand side in~\eqref{eq:wardIdReformulatedAsChangeInEffAction} which, for an infinitesimal $O(N)$ rotation ~\eqref{eq:formulaForInfONTrafo}, can be expressed as
\begin{equation}
    \frac{1}{2}\int_x \tr \frac{1}{g}O\partial_t O^T O\tilde{N}O^T = \int_x \frac{1}{2g}\dot{\alpha}_{ab}(t)\tilde{N}_{ab} + \mathcal{O}(\alpha^2) \,.\nonumber
\end{equation}
Now, substituting the explicit expression in Eq.~\eqref{NTildeTrafoScaleDep} for the composite field $\tilde{N}_{ab}$ in terms of the standard response field $\tilde{\phi}$ and $\tilde{n}$, this can be written as\footnote{Note that we can drop a total derivative term proportional to $-\gamma_k \nabla^2 \tilde{n}_{ab}$ here, since it vanishes upon performing the spatial integral.}
\begin{align}
    &\frac{1}{2}\int_x \tr \frac{1}{g}O\partial_t O^T O\tilde{N}O^T = \label{eq:gNonRenormResultForRHS}  \\ \nonumber
    &\hspace{1.0cm} \int_x \bigg[ -\frac{g_k^{\phi n}}{2g}\tilde{\phi}_a\dot{\alpha}_{cd}(t)\{\phi_a, n_{cd}\}-\frac{g_k^{n n}}{2g}\tilde{n}_{ab}\dot{\alpha}_{ef}(t)\{n_{ab},n_{ef}\} \bigg] + \mathcal{O}(\alpha^2) \,, 
\end{align}
which is identical to the results in Eq.~\eqref{eq:gNonRenormResultForLHS} except for the appearance of the factors $g_{k}^{\phi n}/g$ and $g_{k}^{nn}/g$. Since in order to comply with the displacement symmetry, the expressions in Eq.~\eqref{eq:gNonRenormResultForLHS} and \eqref{eq:gNonRenormResultForRHS}, which correspond to the left-hand side and right-hand side of Eq.~\eqref{eq:wardIdReformulatedAsChangeInEffAction}, have to agree, one concludes that the reversible mode couplings do not get renormalized, i.e.~one has
$$g^{\phi n}_k=g^{n\phi}_k= g^{nn}_k=g$$
at any FRG scale $k$.  We note that the non-renormalization of the mode coupling coefficients $g$ was also exploited in~\cite{Rajagopal:1992qz} to show that the dynamic critical exponent is given by $z=d/2$. In practice, this means that in our calculations we no longer need to compute the flow equation for the reversible mode coupling coefficient. Instead, one can simply set it to a constant value $g_k=g=\text{const.}$ during the flow.

\section{Numerical implementation}\label{sec:numeric}

In this section, we elaborate on the numerical methods we employ to solve the closed system of flow equations derived in Sec.~\ref{sec:truncation}.
For all our numerical calculations we choose the exponential regulator
\begin{equation}
    R_{\phi,k}(\vec{p}) = R_{n,k}(\vec{p}) = \frac{\vec{p}^2}{e^{\vec{p}^2/k^2}-1}. \label{eq:expReg}
\end{equation}
for both $\phi$'s and $n$'s.
For the static couplings, we solve the flow equations of $m^2_k$ and $\lambda_k$, as in Eq.~\eqref{flowm2k} and Eq.~\eqref{flowlambdak}. For the dynamic couplings, we solve the flow equations given by Eq.~\eqref{flowphi} and Eq.~\eqref{flown}.  To resolve the momentum dependence of the kinetic coefficient $\gamma_{n,k}(\vec{p})$ for the charge densities, we discretize $\gamma_{n,k}(\vec{p})$ with a logarithmic grid spacing on a finite interval $[p_{\text{min}},p_{\text{max}}]$. 
We need a small enough lower bound $p_{\text{min}}$ to properly resolve the behavior of the flow in the IR. We choose the upper bound $p_{\text{max}}$ of the grid large enough such that the flow of the diffusion coefficient $\gamma_{n,k}(\vec p)/\vec{p}^2$ approximately goes to zero at $p\to p_{\text{max}}$. In each step of the FRG flow, we evaluate $m^2_k$, $\lambda_k$, $\gamma_{\phi,k}$ (the latter being momentum independent in our truncation), and the function $\gamma_{n,k}(p)$ at each grid point, and then we interpolate $\log \gamma_{n,k}(p)$ as a function of $\log p$ with a GSL interpolator \cite{gsl}. We use a simple forward Euler method to solve the flow equation with respect to the FRG flow time $t=\log {k/\Lambda}$.

The integrals involved in our loop diagrams are generally of the form
\begin{align}
    \int \frac{dq^0}{2\pi} \int \frac{d^d q}{(2\pi)^d} \,f(q^0,|\vec{q}|,p^0-q^0,|\vec{p}-\vec{q}|) \,. \label{eq:generalLoopInt}
\end{align}
Frequency integrals are solved analytically using the residue theorem. To evaluate the momentum integral, we first exploit spatial isotropy to reduce the $d$-dimensional integration in \eqref{eq:generalLoopInt} to two integrals over the momentum's magnitude $q\equiv |\vec{q}|$ and over the angle $\theta$ between $\vec{p}$ and $\vec{q}$, i.e.
\begin{align}
    \frac{S_{d-2}}{(2\pi)^d} \int_{0}^{\infty} dq\,q^{d-1} \int_{0}^{\pi} \sin^{d-2}\theta \, f\left(q^0,q,p^0-q^0,\sqrt{p^2+q^2-2pq\cos \theta}\right)\, ,
\end{align}
where $S_{n}$ is the surface area of the 
of the $n$-dimensional sphere $S^{n}$, with
\begin{equation}
    S_{n-1} =  \frac{2\pi^{n/2}}{\Gamma(n/2)} \end{equation}
(e.g.\ $S_{1} = 2\pi$) and $\Gamma(z)$ denotes the $\Gamma$-function.
To solve these remaining two integrals, we employ two different methods, namely a Gauss quadrature method from the GSL \cite{gsl} and the Cuhre function in the Cuba library \cite{Hahn_2005}. 

Using the GSL we solve the integral over the momentum magnitude using a Gauss-Legendre quadrature on a logarithmically spaced grid in $q$. By performing a variable substitution $u=\cos\theta$ to obtain the necessary weight function, we evaluate the angular integrals using a Gauss-Jacobi quadrature,
\begin{equation}
    \int_{0}^{\pi} d\theta\,\sin^{d-2}\theta\,f(\cos\theta) = \int_{0}^{\pi} d\theta\,\left(1-\cos^2 \theta\right)^{(d-2)/2}\,f(\cos\theta) = \int_{-1}^{1} du\,(1-u^2)^{(d-3)/2} f(u) \,. \label{eq:angularIntegral}
\end{equation}
Gauss-Jacobi quadratures have a weight function of the form $(1-x)^\alpha (1+x)^\beta$ for $\alpha,\beta>-1$ and thus generalize the Gauss-Chebychev quadratures which use a weight function of $1/\sqrt{1-x^2}$.
Our angular integral \eqref{eq:angularIntegral} is conveniently evaluated using the choice $\alpha=\beta=(d-3)/2$.

The Cuhre function from the Cuba library is suited for two-dimensional integrations which is what we need for our purposes since the momentum integration in \eqref{eq:generalLoopInt} can be written as
\begin{align}
    \int \frac{d q }{(2\pi)^d} \int_{-1}^{1}  d u \, \Omega_{d-2} \, g\,(q,u) \,,
\end{align}
similar to the Gauss-Jacobi quadrature in \eqref{eq:angularIntegral} above.
The integrand has singularities at $\vec{p}=\vec{q}$ and $\vec{p}=-\vec{q}$, but these are not problematic because they cancel each other. In practice, we remove a small sphere around $\vec{p}=\vec{q}$ and $\vec{p}=-\vec{q}$ and take the radius of the sphere to zero (resp.~sufficiently small) afterwards.

We have explicitly verified that both numerical methods discussed in the present section lead to the same results within numerical precision.

\begin{table*}[t] 
    \centering
    \begin{tabular}[t]{ |c|c|c|c|c|c||c| } \hline
    $m^2_\Lambda/|m^2_\Lambda|$ & $\lambda_\Lambda/|m_{\Lambda}|^{3-d}$ & $\chi/|m_{\Lambda}|^{d-1}$ & $g$ & $\Gamma^{\phi}_{\Lambda}/|m_{\Lambda}|^{-1}$ & $\gamma_{\Lambda}/|m_{\Lambda}|^{d-2}$ & $T_c/|m_{\Lambda}|$ \\ \hline
    $-1$ & $1$ & $1$ & $1$ & $1$ & $1$ & $17.371845$ \\ \hline 
    \end{tabular}
    \caption{Initial conditions used at the UV initial scale $k=\Lambda = 5.67|m_{\Lambda}|$. The resulting critical temperature $T_c$ in $d=3$ spatial dimensions is shown in the last column.
    \label{tab:uvParamsFlow}}
\end{table*}

For our numerical results in the next section, we generally employ the UV initial conditions listed in Table~\ref{tab:uvParamsFlow}. Moreover, we will quote all dimensionful quantities implicitly in units of the mass parameter $|m_{\Lambda}|$ at the UV cutoff scale $\Lambda$, which is thereby chosen such that we obtain approximately the same critical temperature $T_c$ in $d=3$ spatial dimensions as in the classical-statistical lattice simulations of Ref.~\cite{Schlichting:2019tbr}.

\section{Dynamic critical behavior of $O(4)$ Model G}\label{sec:result}
Based on the FRG flow equations for the static and dynamic couplings of the $O(N)$ Model~G, we will now proceed to study the static and dynamic critical behavior. By numerically solving the FRG flow equations and analyzing the fixed point for the $N=4$ case,  we will first discuss the static critical behavior, which should match the $O(4)$ criticality, and then discuss the dynamic critical behavior of the $O(4)$ Model G.
\subsection{Static critical behavior of the $O(4)$ model}
We follow common procedure and extract the static critical behavior from
a standard fixed point analysis. By introducing dimensionless couplings $\bar{m}_k^2 \equiv k^{-2} m_k^2$, $\bar{\lambda}_k \equiv k^{d-4} T \lambda_k$, and by rewriting the regulator  $R_{\phi,k}(q) = q^2 r_{\phi}(q/k)$ using a dimensionless function $r_{\phi}(\bar{q})$ (with derivative $r_{\phi}'(\bar{q}) = dr_{\phi}(\bar{q})/d\bar{q}$), one can de-dimensionalize the static flow equations \eqref{flowm2k} and \eqref{flowlambdak} and obtain the corresponding $\beta$-functions for the dimensionless couplings $\bar{m}^2$ and $\bar{\lambda}$,
\begin{align}
    k \partial_k \bar{m}_k^2 &= -2\bar{m}^2_k+ \frac{(N+2) \bar{\lambda}_k}{6N} \frac{S_{d-1}}{(2\pi)^d}  \int_{0}^{\infty} \!\! \frac{d\bar{q}\,\bar{q}^{d+2}\, r_{\phi}'(\bar{q}) }{\left( \bar{m}_k^2 + \bar{q}^2(1 + r_{\phi}(\bar{q}))\right)^2} \equiv \beta_{\bar{m}^2}(\bar{m}_k^2,\bar{\lambda}_k) \label{static_beta_m} \\
    k \partial_k \bar{\lambda}_k  &=  (d-4) \bar{\lambda}_k-\frac{(N+8)\bar{\lambda}_k^2}{3N} \frac{S_{d-1}}{(2\pi)^d}  \int_{0}^{\infty} \!\!   \frac{d\bar{q}\,\bar{q}^{d+2}\, r_{\phi}'(\bar{q})}{\left( \bar{m}_k^2 + \bar{q}^2(1 +  r_{\phi}(\bar{q})) \right)^3}  \equiv \beta_{\bar{\lambda}}(\bar{m}_k^2,\bar{\lambda}_k) \label{static_beta_lambda}
\end{align}
To determine the static critical exponent $\nu$ as usual in standard RG approaches, one first determines the Wilson-Fisher fixed point $(\bar{m}^2_k,\bar{\lambda}_k) = (\bar{m}^2_{*},\bar{\lambda}_{*})$ by setting the $\beta$-functions to zero $\beta_{\bar{m}^2}(\bar{m}_{*}^2,\bar{\lambda}_{*}) = \beta_{\bar{\lambda}}(\bar{m}_{*}^2,\bar{\lambda}_{*}) = 0$, solves the resulting system of equations for the fixed-point couplings $(\bar{m}_{*}^2,\bar{\lambda}_{*})$, and then expands the right-hand side of Eqs.~\eqref{static_beta_m}, \eqref{static_beta_lambda} around the fixed point up to first order in $\delta \bar{m}_k^2 \equiv \bar{m}_k^2 - \bar{m}_{*}^2$ and $\delta \bar{\lambda}_k \equiv \bar{\lambda}_k - \bar{\lambda}_{*}$. Collecting the dimensionless $\beta$-functions in a vector, its Jacobian matrix (usually called the stability matrix)
\begin{equation}
    (M_{ij}) = \begin{pmatrix}
        \partial \beta_{\bar{m}^2}/\partial \bar{m}^2 & \partial \beta_{\bar{m}^2}/\partial \bar{\lambda} \\ 
        \partial \beta_{\bar{\lambda}}/\partial \bar{m}^2 & \partial \beta_{\bar{\lambda}}/\partial \bar{\lambda} 
    \end{pmatrix}_{(\bar{m}^2,\bar{\lambda}) = (\bar{m}^2_{*},\bar{\lambda}_{*})}
\end{equation}
determines how the small perturbations $\delta \bar{m}_k$ and $\delta \bar{\lambda}_k$ behave at the fixed point. Since we are working within the local potential approximation, we have no spatial wave function renormalization factor, and hence our truncation yields a vanishing anomalous scaling dimension $\eta$ of the order parameter, $\eta=0$. However, in our general analysis of the scaling relations below, we will keep $\eta$ arbitrary, and only set it to $\eta=0$ in the end. By diagonalizing the stability matrix, the single negative eigenvalue $\omega_0$ is then related to the critical exponent $\nu$ by $\nu=-1/\omega_0$. 
Specifically, for the $N=4$ component theory in $d=3$ spatial dimensions, and the exponential regulator \eqref{eq:expReg}, this procedure yields a value of $\nu = 0.553945$. This result lies between the mean-field value $\nu=1/2$ and the value $\nu=0.7377(41)$ which was determined very precisely from ab-initio simulations of the $O(4)$ spin model \cite{Engels:2014bra}.
The deviation of our result from the ab-initio result gives us an estimate of the systematic error induced by our admittedly rather simple truncation.
In the context of FRG, it is clear that one has to include more Taylor coefficients of the effective potential or even go to higher orders in the derivative expansion to improve the accuracy of the resulting critical exponents \cite{DePolsi:2020pjk}.
However, we emphasize again at this point that in the present work we are primarily interested in the conceptual question of how the dynamic critical behavior of systems with reversible mode-mode couplings, such as Model~G can \emph{in principle} be studied using the FRG framewwork. Hence, we consider the more qualitative result for $\nu$ from above therefore as sufficient for the scope of the present work. 

\begin{figure}[t]
    \centering
    \begin{subfigure}{0.48\textwidth}
    \includegraphics[width=\linewidth]{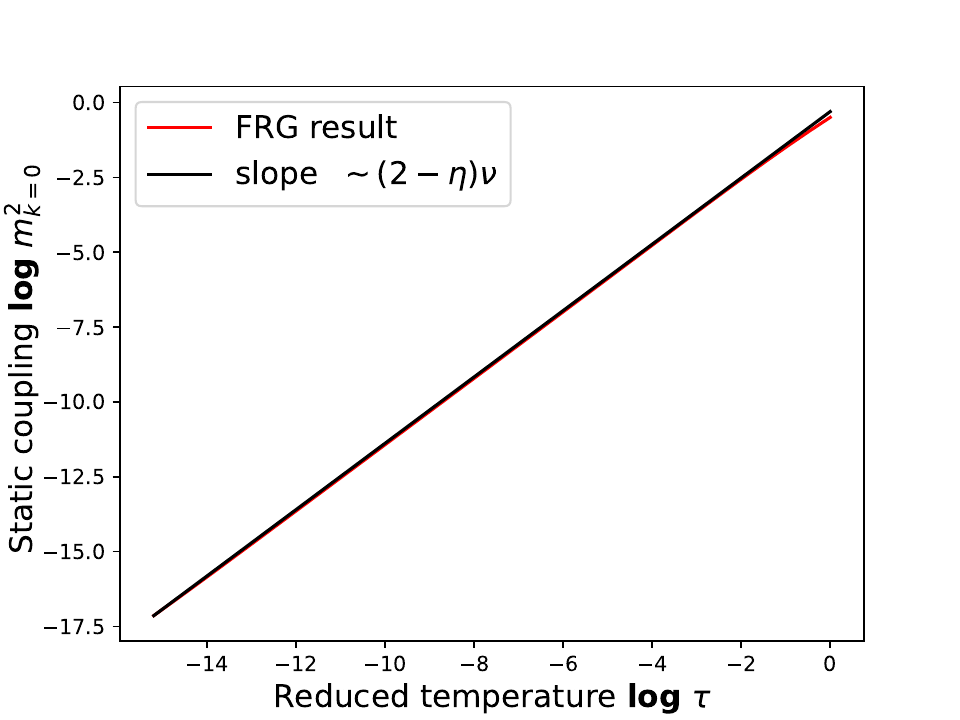}
    \end{subfigure}
    \begin{subfigure}{0.48\textwidth}
    \includegraphics[width=\linewidth]{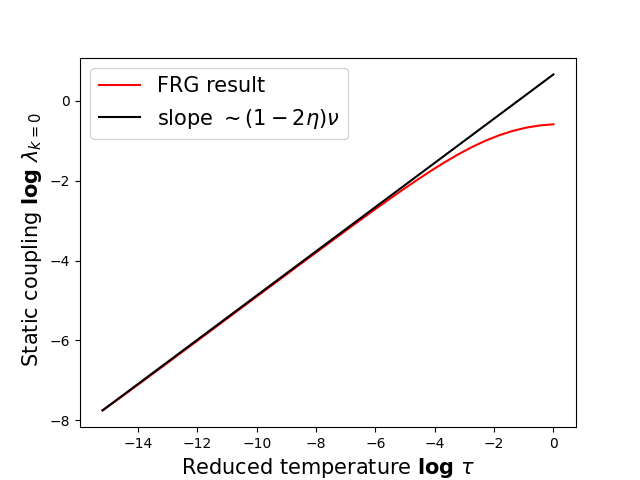}
    \end{subfigure}
    \caption{Static couplings $m_k^2$, $\lambda_k$ in the IR ($k=0$) versus reduced temperature $\tau = (T-T_c)/T_c$. The critical temperature here is given by $T_c= 17.371845$.}
            \label{fig:stat_couplingsvT}
\end{figure}

Beyond this fixed point analysis, it is also possible to approach the static critical behavior from a more physical perspective, where close to the critical temperature of a  second-order phase transitions, the static critical exponent $\nu$ determines the divergence of the equilibrium correlation length $(\xi)$ of order-parameter excitations as $\xi \sim \tau^{-\nu}$, where $\tau=(T-T_c)/T_c$ is the reduced temperature. By investigating the static limit of the $iF$ propagator, one finds that the critical scaling of the infrared mass $m_{k=0}^2$ is in general given by $m^2_{k=0}\sim \xi^{-2+\eta}$, and one can thus deduce that close to the critical temperature, it behaves as $m^2_{k=0}\sim \tau^{(2-\eta)\nu}$. Similar arguments of dimensional analysis can also be applied to the quartic coupling which lead to $\lambda_{k=0}\sim \xi^{-4+d+2\eta}$, such that close the the critical temperature one finds $\lambda_{k=0}\sim \tau^{(4-d-2\eta)\nu}$. Specifically, for our setup in $d=3$ dimensions, we thus expect to see $m^2_{k=0} \sim \tau^{(2-\eta)\nu}$ and $\lambda_{k=0} \sim \tau^{(1-2\eta)\nu}$ in our numerical results below, with the value of $\nu$ extracted above from the eigenvalues of the stability matrix at the Wilson-Fisher fixed point, and with $\eta=0$ in the local potential approximation of the FRG flow equations.
In Fig.~\ref{fig:stat_couplingsvT} the static coupling in the infrared, $m^2_{k=0}$ and $\lambda_{k=0}$, are shown as a function of the reduced temperature $\tau$ on a logarithmic scale. The static critical exponent $\nu$ can be deduced from the slope of the FRG-simulated data. A line with the slope indicating the theoretical value of $\nu = 0.553945$ within our truncation is also shown in the plot for comparison. One can see that close to the critical temperature, the slope of the line obtained from FRG simulation indeed match the slope indicating the theoretical value of $\nu$ for our truncation.

Beyond the reduced temperature dependence of the static coefficients, it is also interesting to investigate the dependence on the FRG scale $k$.
Since sufficiently close to the critical point the relevant infrared cutoff during the FRG flow is always set by the FRG scale $k\neq0$,  the correlation length is effectively determined as $\xi_{k} \sim 1/k$ and static and dynamic quantities exhibit scaling as a function of $k$. Based on the above discussion, one thus expects that the $k$ dependence follows 
\begin{equation}
    m_k^2 \sim k^{2-\eta} \;\;\text{and}\;\; \lambda_k \sim k^{4-d-2\eta} \label{eq:m2AndLamScaling}
\end{equation}
in the vicinity of the critical point. Within our truncation, where the anomalous dimension $\eta=0$, this yields 
\begin{equation}
m^2_k \sim k^2 \;\;\text{and}\;\; \lambda_k \sim k^{4-d} \label{eq:m2AndLamScalingTrunc}
\end{equation}
as can easily be deduced from the fact that in $d$ spatial dimensions, the dimensionless couplings are given by $\bar{m}_k^2 \equiv k^{-2} m_k^2$, $\bar{\lambda}_k = k^{d-4} T \lambda_k$, along with the fact that $\partial_k\bar{m}_k=0$, $\partial_k\bar{\lambda}_k=0$ at the Wilson-Fisher fixed point.

\begin{figure}[t]
    \centering
    \begin{subfigure}{0.48\textwidth}
    \includegraphics[height=5cm]{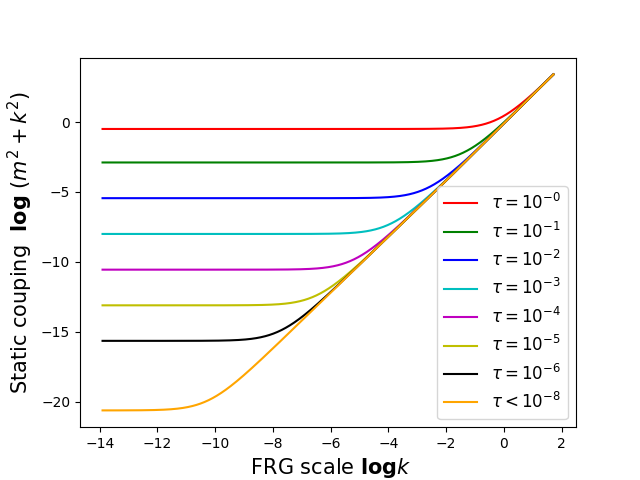}
    \end{subfigure}
    \begin{subfigure}{0.48\textwidth}
    \includegraphics[height=5cm]{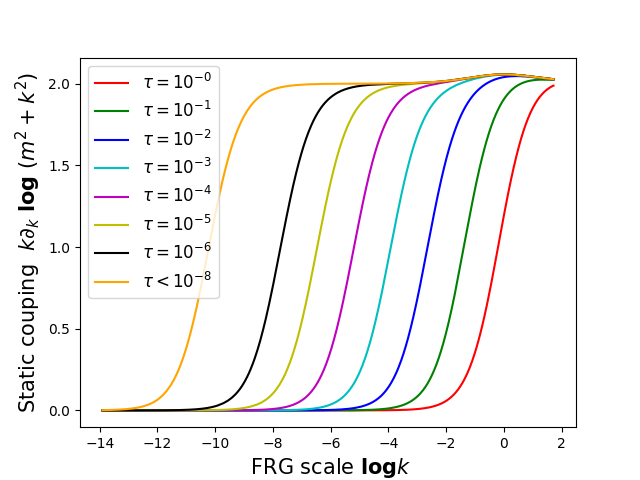}
    \end{subfigure}\\\vspace*{-1mm}
    \begin{subfigure}{0.48\textwidth}
    \includegraphics[height=5cm]{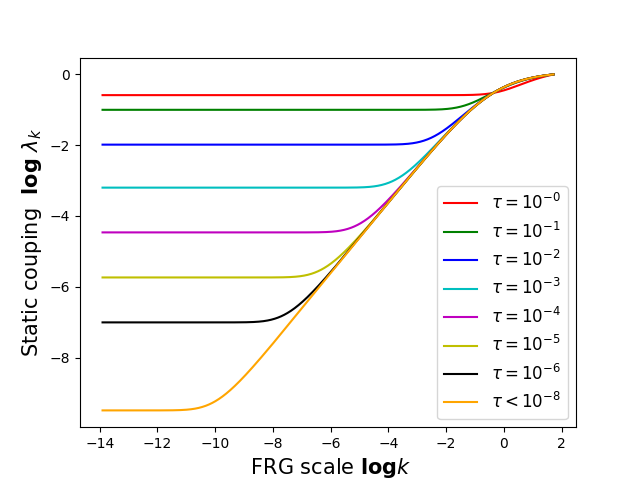}
    \end{subfigure}
    \begin{subfigure}{0.48\textwidth}
    \includegraphics[height=5cm]{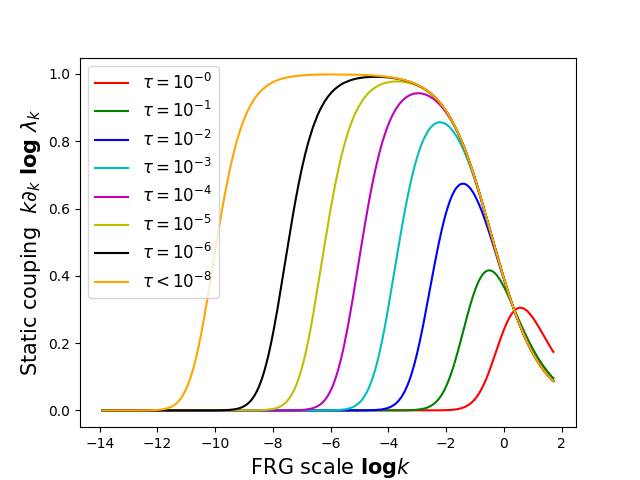}
    \end{subfigure}
    \caption{FRG scale $k$ dependence of the static couplings $\log (m_k^2+k^2)$, $\log \lambda_k$, as the temperature $T$ is approaching the critical temperature $T_c$ from above. The plots on the right show the respective derivatives of the couplings on the left, which coincide with the corresponding critical exponents in the scaling regime.}
            \label{fig:stat_couplingsvk}
\end{figure}

\begin{figure}[t]
    \centering
    \begin{subfigure}{0.48\textwidth}
    \includegraphics[width=\linewidth]{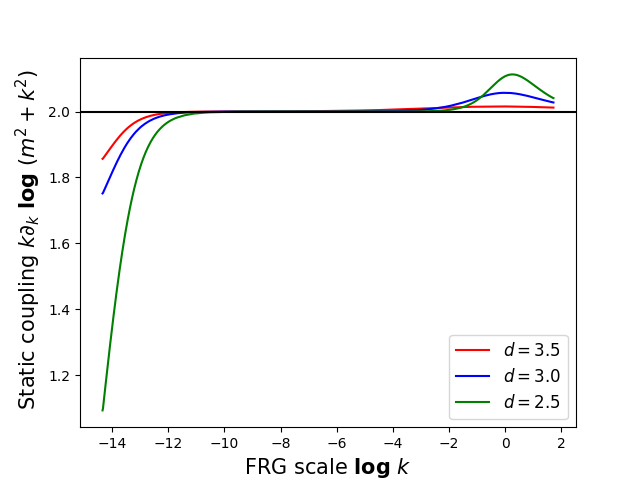}
    \end{subfigure}
    \begin{subfigure}{0.48\textwidth}
    \includegraphics[width=\linewidth]{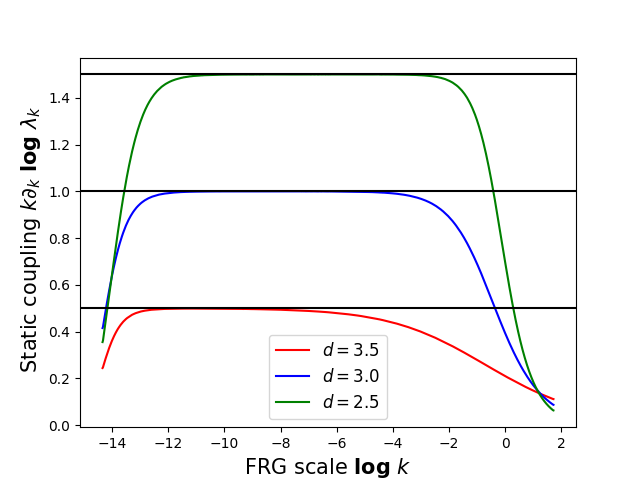}
    \end{subfigure}
    \caption{Static critical exponents in different dimensions. Here the reduced temperature is $\tau < 10^{-12}$. }
            \label{fig:stat_vary_d}
\end{figure}

In Fig.~\ref{fig:stat_couplingsvk}, the scale dependent effective mass $m^2_k+k^2$ and coupling $\lambda_k$ are plotted with respect to the FRG scale $k$ on a logarithmic scale for $d=3$ spatial dimensions. In order to see the critical exponents more easily, we take one further derivative with respect to $\log k$, as depicted in the right panels, where $k\partial_k \log(m^2_k+k^2)$ and $k\partial_k \log \lambda_k$ are shown. In these plots, the scaling region corresponds to a plateau and the corresponding critical exponents are indicated. From Fig.~\ref{fig:stat_couplingsvk}, one can see that in the scaling region the power laws $m^2_k \sim k^2$ and $\lambda_k \sim k$, which correspond to Eq.~\eqref{eq:m2AndLamScalingTrunc} with $d=3$, indeed hold. One can also see that by approaching the critical temperature, the scaling region gets larger. When generalising to different numbers $d$ of spatial dimensions, one still has $m^2_k \sim k^2$ in the scaling region, cf.~Eq.~\eqref{eq:m2AndLamScalingTrunc}, but  $\lambda_k \sim k^{4-d}$ for the static four-point coupling. This is shown in Fig.~\ref{fig:stat_vary_d}, with the plateau in $k\partial_k \log \lambda_k$ with respect to $\log k$ forming around $k\partial_k \log \lambda_k = 0.5, 1.0$ and $ 1.5$ correspondingly in $d=3.5, 3$ and $2.5$.

\subsection{Dynamic critical behavior}
Next we proceed with investigating the dynamic critical behavior of the $O(4)$ Model G, which in addition to the static quantities is governed by the behavior of the kinetic coefficients $\Gamma_{k}^{\phi}$ and $\gamma_{n,k}(\vec{p})$ in the vicinity of the critical point. We first consider the divergence of the kinetic coefficient $\Gamma_{k}^{\phi}$ of the order parameter $\phi$, which determines the relaxation rate of the order parameter  $\omega^{\rm rel}_{k} \sim \Gamma^{\phi}_{k} m^2_{k}$, as can be deduced from the pole of the retarded propagator in Eq.~\eqref{eq:phiPropTruncGR}. Since in the vicinity of the critical point the relaxation rate is expected to diverge as $\omega^{\rm rel}_{k=0} \sim \xi^{-z_{\phi}} \sim \tau^{\nu z_{\phi}}$, one can extract the dynamic critical exponent $z_{\phi}$ of the order parameter field from the reduced temperature ($\tau$) dependence of the kinetic coefficient $\Gamma^{\phi}_{k=0}$ evaluated in the infrared ($k=0$). By using the relation $m_{k=0}^{2} \sim \tau^{(2-\eta)\nu}$, the critical behavior of the kinetic coefficient $\Gamma_{k=0}^{\phi}$
is then given by
\begin{eqnarray}
\label{eq:GKDiv}
\Gamma_{k=0}^{\phi} (\tau)= \Gamma_{+}^{\phi}~\tau^{-\nu(2-\eta-z_{\phi})}.
\end{eqnarray}
where $\Gamma_{+}^{\phi}$ is a non-universal amplitude. Similarly, one can also infer the critical behavior of the kinetic coefficient $\gamma_{n,k}(\vec{p})$ of the charge density, by considering the long-wavelength limit of the charge diffusion coefficient
\begin{equation}
    D^{n}_{k}(\vec{p},\tau) = \gamma_{n,k}(\vec{p},\tau)/\vec{p}^2 \, , \label{eq:defOfDiffCoeff}
\end{equation}
where in the context of QCD in the chiral limit, this would be the diffusion coefficient of the isovector and isoaxial charge densities $V_0^i$ and $A_0^i$. The diffusion coefficient $D^{n}_{k}(\vec{p})$ determines the relaxation rate of charge density perturbations as $\omega^{\rm rel}_{k}(\vec{p}) \sim D^{n}_{k}(\vec{p}) \vec{p}^2$. Evaluated at the characteristic scale of the inverse correlation length $\vec{p} \sim 1/\xi \sim \tau^{\nu}$, the relaxation rate is again expected to behave as $\omega^{\rm rel}_{k}(\vec{p} \sim 1/\xi) \sim \xi^{-z_{n}} \sim \tau^{\nu z_{n}}$, which then gives rise to the following critical divergence,
\begin{eqnarray}
\label{eq:DnDiv}
\left. D^{n}_{k=0}(\vec{p}) \right|_{|\vec{p}|=1/\xi} = D_{+}^{n} ~\tau^{-\nu(2-z_{n})}, \end{eqnarray}
where in our truncation the correlation length is simply determined by $\xi=1/m_{k=0}$.

\begin{figure}[t]
    \centering
    \begin{subfigure}{0.48\textwidth}
    \includegraphics[width=\linewidth]{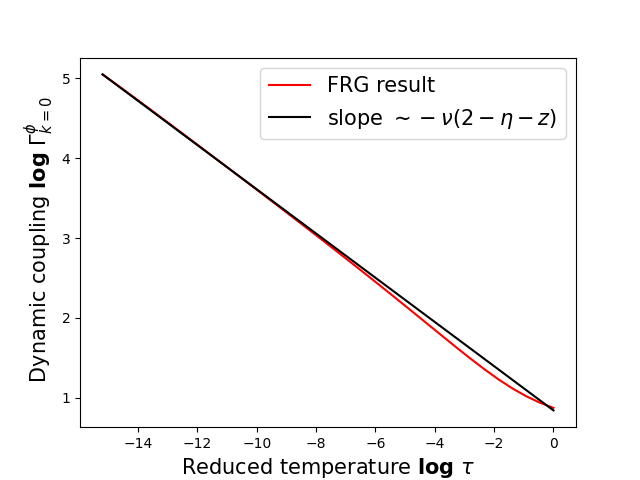}
    \end{subfigure}
    \begin{subfigure}{0.48\textwidth}
    \includegraphics[width=\linewidth]{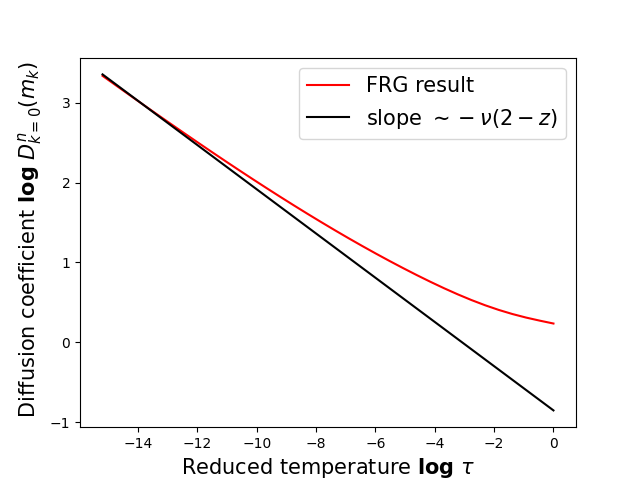}
    \end{subfigure}
    \caption{Double-logarithmic plots of the dynamic couplings $\Gamma^\phi_k$ and $D^n_k(m_k)$ in the IR ($k=0$) over the reduced temperature $\tau$.}
            \label{fig:dyn_couplingsvT}
\end{figure}

Our numerical FRG results for $\log \Gamma^\phi_{k=0}$ and $\log (D^{n}_{k=0}(|\vec{p}|=1/\xi))$ are presented in Fig.~\ref{fig:dyn_couplingsvT} as functions of (the logarithm of) the reduced temperature, $\log \tau$. Data obtained from our numerical FRG calculations is compared to the theoretical expectations in Eqs.~(\ref{eq:GKDiv},\ref{eq:DnDiv}), as indicated by the black lines with slopes $-\nu (2-\eta-z)$ (for $\Gamma_{k=0}^\phi$) and $-\nu(2-z)$ (for $D^{n}_{k=0}(|\vec{p}|=1/\xi)$), where we employ the static critical exponents $\nu=0.55394$ and $\eta=0$ determined within our FRG truncation, and the dynamic critical exponent $z=d/2=1.5$, which is the theoretically expected  value of the dynamic critical exponent   from strong scaling in three spatial dimensions. From the results in Fig.~\ref{fig:dyn_couplingsvT}, one can clearly observe that the slopes from our numerical FRG results approach their theoretically expected values as criticality is approached in the temperature.
As an alternative to the reduced temperature dependence of the kinetic coefficients, we can also extract the dynamic critical exponents from their FRG scale $k$ dependence sufficiently close to criticality. Similar to our discussion above, the scale dependent relaxation rate of the order parameter $\omega_{k}^{\rm rel}$ is determined by the kinetic coefficient $\Gamma_k^{\phi}$ and the mass $m_{k}^{2}$ as $\omega_k \sim \Gamma_k^{\phi}m_k^2$. Since close to the critical point one expects $m_k^2 \sim k^{2-\eta}$ and $\omega_k^{\rm rel} \sim \xi_{k}^{z_{\phi}} \sim k^{z_\phi}$, one can deduce that the scaling of the kinetic coefficient of the order parameter with respect to the FRG scale is given by $\Gamma_k \sim k^{z_\phi-2+\eta}$. Similarly, one finds that the charge diffusion coefficient $D^n_k(k)$ exhibits the following critical dependence $D^n_k(k)\sim k^{z_n-2}$ on the FRG scale $k$, where $z_n$ is the dynamic critical exponent for the charge diffusion constant, and one can thus extract $z_{\phi}$ and $z_{n}$ from the FRG scale dependence of the kinetic coefficients $\Gamma_k$ and $D^n_k(k)$ close to criticality.

\begin{figure}[t]
    \centering
    \begin{subfigure}{0.48\textwidth}
    \includegraphics[height=5cm]{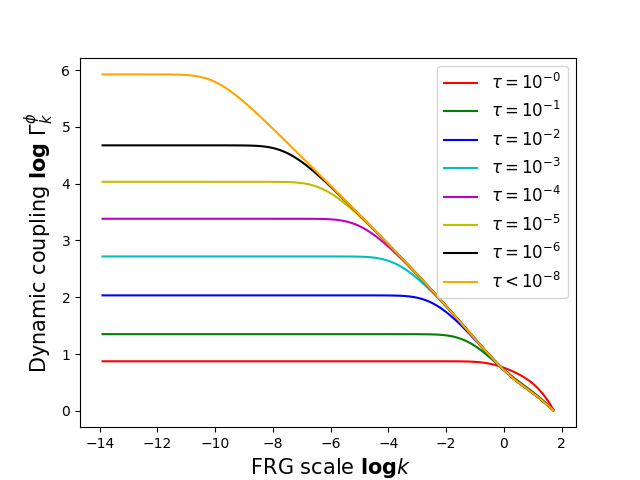}
    \end{subfigure}
    \begin{subfigure}{0.48\textwidth}
    \includegraphics[height=5cm]{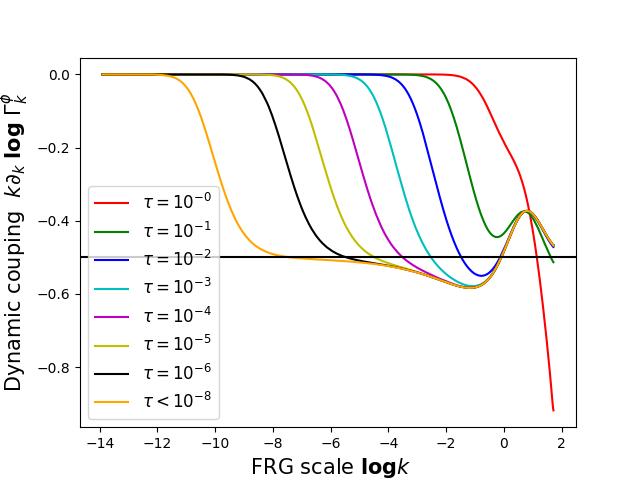}
    \end{subfigure}\\\vspace*{-1mm}
    \begin{subfigure}{0.48\textwidth}
    \includegraphics[height=5cm]{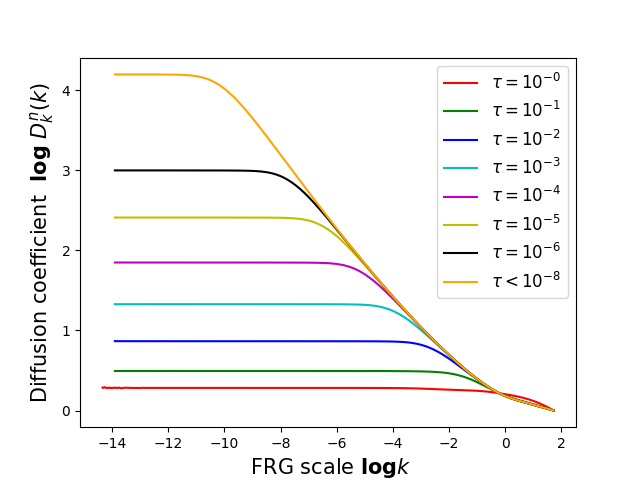}
    \end{subfigure}
    \begin{subfigure}{0.48\textwidth}
    \includegraphics[height=5cm]{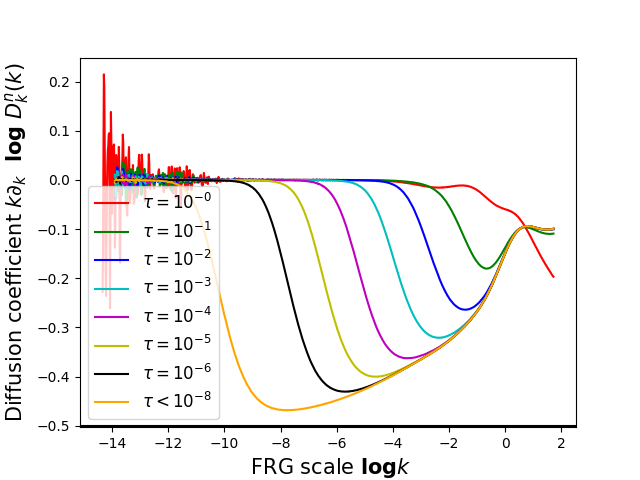}
    \end{subfigure}
    \caption{FRG scale dependence of the dynamic couplings $\log \Gamma^\phi_k$ and $\log D_{k}^{n}(k)$, as the temperature $T$ is approaching the critical temperature $T_c$ from above. The plots on the right show the respective derivatives of the couplings on the left, which coincide with the corresponding critical exponents in the scaling regime.}
            \label{fig:dyn_couplingsvk}
\end{figure}

We present this analysis in Fig.~\ref{fig:dyn_couplingsvk}, where the left panels show  the dependence of the kinetic coefficients $\log \Gamma^\phi_k$ of the order parameter and the conserved charges $\log D^n_k(k)$ on the FRG scale $\log k$. Differently colored curves correspond to results obtained at different values of the reduced temperature in the symmetric phase of the $O(4)$ model, and one can clearly observe the build up of the critical divergence when lowering the reduced temperature. 

Similar to our discussion of the static critical exponents, the dynamic critical exponents $z_{\phi}$ and $z_{n}$ can be inferred from the right panel of 
Fig.~\ref{fig:dyn_couplingsvk}, where we present the logarithmic derivatives $k\partial_k \log \Gamma^\phi_k$ and $k\partial_k \log D^n_k(k)$ of the kinetic coefficients.
In the scaling region, which becomes broader and broader when approaching the critical temperature, the logarithmic derivative $k\partial_k \log \Gamma^\phi_k$ approaches a constant value which yields the exponent $z_\phi-2+\eta$. With $\eta=0$, the value of $-0.5$ indicates that the dynamic critical exponent for the order parameter modes is around $z_\phi = 1.5$. Similarly, from the plot of $k\partial_k \log D^n_k(k)$ with respect to $\log k$, it can be seen that the dynamic critical exponent for the conserved charge density modes is also given by $z=1.5$, since in the scaling region it also approaches $k\partial_k \log D^n_k(k) = -0.5$, as the temperature approaches $T_c$. 

\begin{figure}[t]
    \centering
    \begin{subfigure}{0.48\textwidth}
    \includegraphics[width=\linewidth, height=5cm]{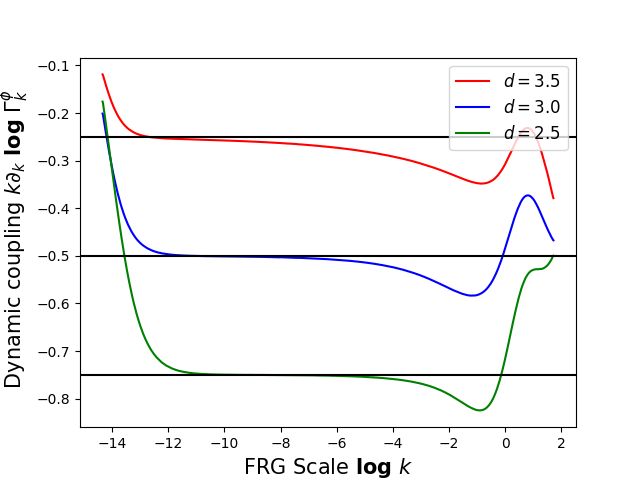}
    \end{subfigure}
    \begin{subfigure}{0.48\textwidth}
    \includegraphics[width=\linewidth, height=5cm]{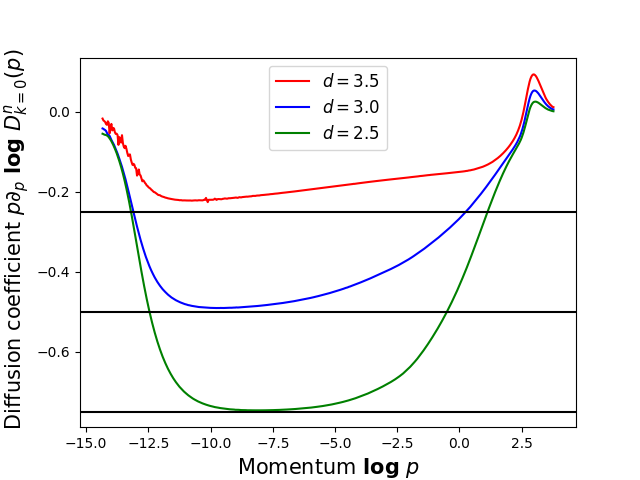}
    \end{subfigure}
    \caption{Dynamic critical exponents in different dimensions, see text. Here the reduced temperature is $\tau < 10^{-12}$. }
            \label{fig:dyn_vary_d}
\end{figure}

So far we have only considered the dynamic critical behavior in $d=3$ spatial dimensions, where our results are consistent with the theoretical value of the dynamic critical exponent $z_\phi = z_n = 3/2$, obtained from the strong-scaling prediction. To verify the expected dependence on the dimensionality of the system, we have also checked the flow of the kinetic coefficient $\Gamma^\phi_{k}$
of the order parameter field in $d=2.5$ and $d=3.5$ spatial dimensions, with the result shown in the left panel of Fig.~\ref{fig:dyn_vary_d}. One can see from the plot of $k\partial_k \log \Gamma^\phi_k$ that for $d=2.5$ the scaling region yields a plateau at the corresponding $y$-axis value of $z_\phi - 2 +\eta = -0.75$. When the spatial dimension is given by $d=3.5$, the $y$-axis value corresponding to the plateau is given by $-0.25$. Similarly, from  the right panel of Fig.~\ref{fig:dyn_vary_d}, which shows the momentum $p$ dependence of the kinetic coefficient $D^n_{k=0}(p)$, by considering $p\partial_p \log D^n_{k=0}(p)$ as a function of $\log p$ in various spatial dimensions, one can see that the kinetic coefficients both approach a power behavior with values of  $-0.75$, $-0.5$ and $-0.25$ for the exponent $d/2-2$ in $d=2.5$, $d=3.0$ and $d=3.5$, respectively. 

\begin{figure}[t]
    \centering
    \includegraphics[width=0.5\textwidth]{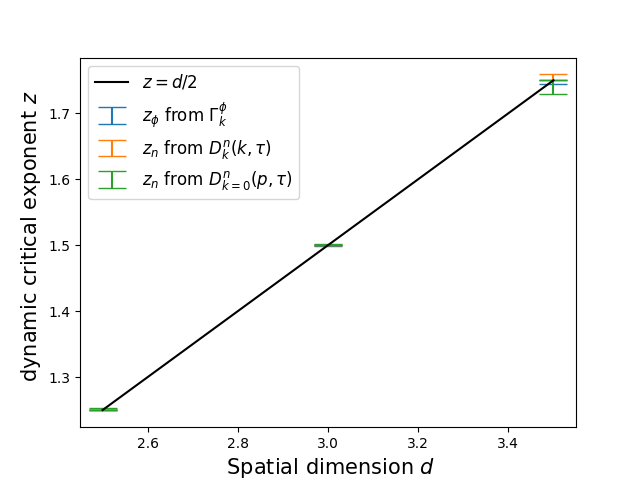}
    \caption{Extraction of the dynamic critical exponent $z$ for various spatial dimensions $d$. The blue data point is extracted from the inflection point of $k\partial_k \log \Gamma^\phi_k$ versus $\log k$, the orange data point is extracted from the minimum point of $k\partial_k \log D^n_k(k)$ versus $\log k$, the green data point is extracted from the minimum point of $p\partial_p \log D^n_{k=0}(p)$ versus $\log p$.}
            \label{fig:z_error}
\end{figure}

Beyond comparing with the theoretical prediction for the dynamic critical exponents $z_{\phi}$ and $z_{n}$, we can also infer their values directly from our FRG calculations by determining the inflection point and respectively the minima of the logarithmic derivatives $k\partial_k \log \Gamma^\phi_k$, $k\partial_k \log D^n_k(k)$ and $p\partial_p \log D^n_{k=0}(p)$ in Figs.~\ref{fig:dyn_couplingsvk} and \ref{fig:gammankpvp}, for a range of reduced temperatures between $\tau = 10^{-8}$ and $\tau = 10^{-12}$, and subsequently extrapolating the results to the critical point at $\tau =0$. Details of this procedure are provided in Appendix~\ref{extraction_app}.  The results for the dynamic critical exponents $z_{\phi}$ and $z_{n}$ are compactly summarized in Fig.~\ref{fig:z_error}. This clearly indicates that our real-time FRG formalism does indeed produce the correct dynamic critical exponent $z_\phi =z_n =d/2$ expected for Model~G. 

\subsection{Universal scaling of charge diffusion coefficients}
In addition to the determination of the dynamic critical exponents $z_\phi $ and $z_n$, the real-time FRG framework also allows us to determine universal dynamic scaling functions, that describe the real-time dynamics in the vicinity of the critical point. In fact, the dynamic scaling hypothesis implies that sufficiently close to the critical point the resulting kinetic coefficient, e.g., of charge densities in the IR (at $k=0$) is a homogeneous function of spatial momentum $\vec{p}$ and reduced temperature $\tau$, i.e.
\begin{equation}
    \gamma_{n}(\vec{p},\tau) = s^{-z_n} \gamma_{n}(s\vec{p},s^{1/\nu}\tau) \, .\label{eq:scalingOfChargeKineticCoeff}
\end{equation}
Based on an appropriate choice of the scaling variable $s$, this implies that the combined temperature and momentum dependence of $\gamma_{n}$ can be described by a power law times a universal scaling function, which we can extract from our FRG calculations as described in the following. Using~\eqref{eq:scalingOfChargeKineticCoeff}, we can infer the analogous scaling relation for the diffusion coefficient defined in Eq.~(\ref{eq:defOfDiffCoeff}), which evaluated at $k=0$ assumes the form
\begin{equation}
    D_{n}(\vec{p},\tau) = s^{2-z_n} D_{n}(s\vec{p},s^{1/\nu}\tau) \, . \label{eq:scalingOfDiffCoeff}
\end{equation}
Setting the scale parameter $s$ so that  $s^{1/\nu} \tau = 1$ in \eqref{eq:scalingOfDiffCoeff}, the momentum and temperature dependence of the diffusion coefficient in Eq.~\eqref{eq:defOfDiffCoeff} can equivalently be written as
\begin{equation}
    D_{n}(\vec{p},\tau) = \tau^{-\nu(2-z_n)} D_{n}(\tau^{-\nu} \vec{p}, 1) \label{eq:univScalFnc1}\, .
\end{equation}
Expressing the dimensionless quantity $\tau^{-\nu}$ in terms of the correlation length
\begin{equation} \xi(\tau) = f^{+}_{\xi} \tau^{-\nu} + \text{less singular}\;,
\end{equation} 
one then finds that the reduced temperature and momentum dependence of the diffusion constant can be expressed in terms of a universal scaling function $\mathcal{L}(x)$ as
\begin{equation}
    D_{n}(\vec{p},\tau) = D_{n}^{+} \left(\frac{\xi(\tau)}{f^{+}_{\xi}}\right)^{2-z_n} \mathcal{L}\left(\xi(\tau)p\right) \, , \label{eq:univScalFnc3}
\end{equation}
up to a non-universal amplitude $D_{n}^{+}$, which upon adopting the normalization condition $\mathcal{L}(1)=1$, is determined from the critical divergence of the diffusion coefficient as
\begin{equation}
    D_{n}^{+}= \lim_{\tau \to 0^{+}} \tau^{\nu(2-z_n)} D_{n}(p=1/\xi(\tau),\tau)\;.
\end{equation}

When considering $p \ll 1/\xi(\tau)$, the universal scaling function $\mathcal{L}(x \ll 1 )$ approaches a constant  such that one recovers the critical divergence of the diffusion coefficient,
\begin{equation}
D_n(\vec{0},\tau) \sim \tau^{-\nu(2-z_n)} \, ,
\end{equation}
at zero spatial momentum. Conversely, for $p \gg 1/\xi(\tau)$ the universal scaling function $\mathcal{L}(x\gg 1)$ approaches a power law behavior $\mathcal{L}(x) \sim x^{-(2-z_n)}$, such that at the critical temperature one recovers the critical momentum dependence 
\begin{equation}
D_n(\vec{p},\tau=0) \sim p^{z_n-2}\;,
\end{equation}
which then also implies the expected scaling of the relaxation rate of the charge density $\omega_{\rm rel}^{n} \sim D_n(\vec{p})\vec{p}^2 \sim p^{z_n}$ at the critical point.

\begin{figure}[t]
    \centering
    \begin{subfigure}{0.48\textwidth}
    \includegraphics[width=\linewidth]{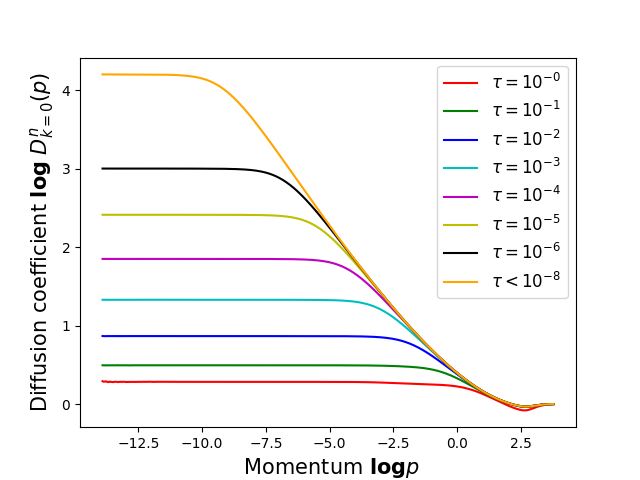}
    \end{subfigure}
    \begin{subfigure}{0.48\textwidth}
    \includegraphics[width=\linewidth]{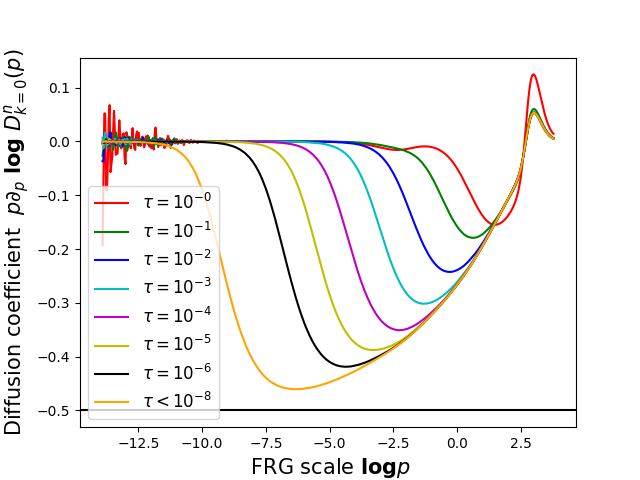}
    \end{subfigure}
    \caption{Diffusion coefficient $\log D^{n}_{k=0}(p)$ for the charge densities at IR ($k=0$) versus the logarithm of the external momentum scale $\log p$, as the temperature $T$ is approaching the critical temperature $T_c$ from above. The plot on the right shows the derivative of the plot on the left, i.e.\ the logarithmic derivative $p \partial_p \log D^{n}_{k=0}(p)$ of the diffusion coefficient with respect to spatial momentum, which coincides with the corresponding critical exponent in the scaling regime.}
            \label{fig:gammankpvp}
\end{figure}

Now that we have established the theoretical basis, we proceed to take a look at the numerical results from our FRG calculation. We first note, that by combining the data of the temperature and momentum-dependent diffusion coefficient $D_{n}(p,\tau)$, as depicted in Fig.~\ref{fig:gammankpvp}, one also extract the dynamic critical exponent $z_{n}$ for the conserved charge modes. By taking the logarithmic derivative with respect to $p$, one clearly observes the emergence of a scaling window as the reduced temperature is lowered, indicating the expected $D_n(p)\sim p^{z_n-2}$ scaling at criticality.

We continue with the extraction of the scaling function $\mathcal{L}(x)$ from our FRG calculation, which by means of Eq.~\eqref{eq:univScalFnc3} can be achieved with plotting 
\begin{equation}
\mathcal{L}\left(\xi(\tau)p\right)  = \frac{\left(f^{+}_{\xi}\right)^{2-z_n}}{ D_{n}^{+}} \xi(\tau)^{-(2-z_n)} D_{n}(\vec{p},\tau) \label{eq:extractScalingFnc}
\end{equation}
against the scaling variable $x=\xi(\tau) p $.
We determine the specific combination $(f^{+}_{\xi})^{2-z_n}/D_{n}^{+}$ of non-universal amplitudes by choosing the normalization $\mathcal{L}(1)=1$, which for our set of parameters results in the value given in Table~\ref{t_amp}.

\begin{figure}[t]
    \centering
    \includegraphics[width=0.5\textwidth]{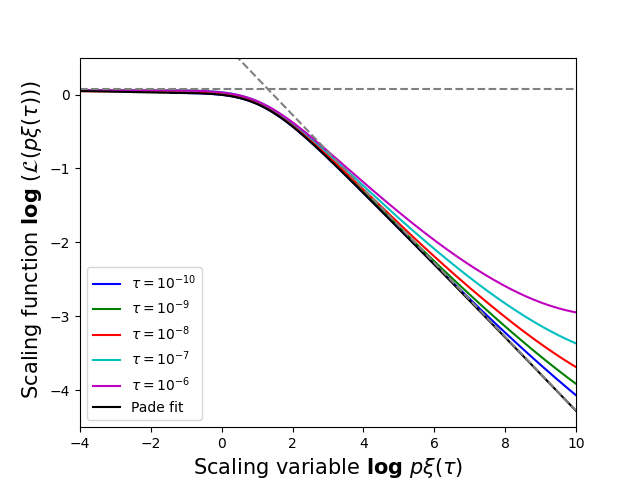}
    \caption{The universal scaling function $\mathcal L(x) $ from the scaled diffusion coefficient at different reduced temperatures. The universal scaling behavior is well fitted by $\mathcal{L}_\mathrm{fit}(x)=(a+bx^{0.5}+cx)/(1+dx^{0.5}+ex+fx^{1.5})$, with $a=1.073, b=-0.065, c=0.444, d=0.143, e=0.099, f=0.215$. The dotted grey lines indicate the asymptotic behavior with $\mathcal{L}_\mathrm{fit}(0)=a$ for $x\to 0$, and $\mathcal{L}_\mathrm{fit}(x) \to (c/f)\,  x^{-0.5}$ for $x\to\infty$. }
            \label{fig:univ_scaling_func}
\end{figure}

In the long-wavelength limit $\xi p \ll 1$, the universal scaling function approaches a constant $\mathcal{L}(x \ll 1) \to \mathcal{L}(0)$. In the short-wavelength limit $\xi p\gg 1$, the universal scaling function approaches $\mathcal{L}(x\gg 1) \sim x^{-(2-z_n)}$, which gives the correct power law $D_n(p,\tau)\sim p^{z_n-2}$ for the momentum dependence of the diffusion coefficient at $\tau=0$. In Fig.~\ref{fig:univ_scaling_func} we plot the quantity defined by the right hand side of \eqref{eq:extractScalingFnc} for different reduced temperatures~$\tau $ using the IR ($k=0$) results from our FRG calculation. We see that the curves indeed beautifully converge towards a unique universal scaling function $\mathcal{L}(x)$ in the limit $\tau \to 0$. The quantitative shape of the resulting scaling function can rather accurately be determined by performing a fit to a Pad{\'e} approximant of the order $[m/m+1]$ in the variable $x^{2-z_n}$,
\begin{equation}
    \mathcal{L}(x) = \frac{\sum_{j=0}^m a_j \left(x^{2-z_n}\right)^j}{1+\sum_{k=1}^{m+1} b_k \left(x^{2-z_n}\right)^k} \,,
\end{equation}
which yields $\mathcal L(0) = a_0$, and ensures the asymptotic behavior for $x \gg 1$ with $\mathcal{L}(x) \to (a_m/b_{m+1})\,   x^{-(2-z_n)}$ for large values of~$x$. For our results in Fig.~\ref{fig:univ_scaling_func} we find that already for $m=2$ the method has converged to a sufficient accuracy. The resulting fit with its asymptotics are plotted in the same figure together with our  numerical results. The corresponding values for the coefficients $a_j$, $b_k$ are listed in the figure caption.

\begin{table}
\centering
\begin{tabular}{ |c|c|c| } 
 \hline
  $\tau$  &  $(f^{+}_{\xi})^{2-z_n}/D_{n}^{+}$  \\
$10^{-10}$ & 2.718 \\ $10^{-9}$ & 
2.718 \\$10^{-8}$ & 
2.718 \\$10^{-7}$ & 
2.718 \\$10^{-6}$ & 
2.638 \\
 \hline
\end{tabular}
\caption{Non-universal amplitudes $(f^{+}_{\xi})^{2-z_n}/D_{n}^{+}$ at different reduced temperatures.}
\label{t_amp}
\end{table}

\section{Conclusion and Outlook}
We have performed a real-time functional renormalization group study of Model G, which is conjectured to be the dynamic universality class of the chiral phase transition of two-flavor QCD in the (chiral) limit of vanishing current quark masses. We have constructed a corresponding path-integral formulation for Model G using the MSR technique, i.e.\ by introducing auxiliary response fields.
We discussed the various symmetries of the MSR action of Model G, which most prominently include a global $O(N)$ symmetry, a temporal gauge and displacement symmetry which reflects the underlying non-reversible mode coupling structure, the discrete symmetry of detailed balance (thermal equilibrium), and a hidden BRST-type symmetry which commonly appears in the MSR approach and expresses the fact that with non-vanishing physical sources, the real-time partition function $Z=1$ is equal to unity.
These symmetries restrict the possible form of the truncation of the effective action.

In the process of formulating a generating functional for Model~G, it turned out that the `traditional' way of introducing source terms as couplings to the elementary MSR response fields on the level of the MSR action is problematic, since for example it does not allow to recover the Boltzmann distribution as a stable equilibrium distribution, and correspondingly one does not recover the correct fluctuation-dissipation relations. Moreover, the underlying Poisson-bracket structure of the reversible mode couplings, which we expressed as an extended displacement symmetry of the MSR action, is violated if the sources are coupled in this traditional way.  We found that the correct way to couple physical source terms is to add them on the level of the thermodynamic free energy, which indeed solves the above two problems, and leaves all relevant symmetries during the FRG flow intact. Because of the structure of the reversible mode couplings, in this approach the sources couple to \emph{composite} response fields on the level of the MSR action. It turns out that the flows of the static couplings generally remain precisely the same as in the dimensionally reduced Euclidean theory which describes the static critical properties. This warrants that the presence of dynamics in Model~G does not change the static critical behavior.

The introduction of composite fields in the real-time FRG also makes the study of FRG flow equations much less cumbersome. We proposed a truncation of the effective average action by promoting all couplings in the bare action to be running, and by introducing an arbitrarily momentum-dependent kinetic coefficient for the conserved charges. Using this truncation, we have studied the flow of the static and dynamic couplings in the $O(4)$ Model~G. Due to the displacement symmetry, we found that the reversible mode coupling coefficients $g_k=g$ are scale-independent, which we showed using a Ward identity originating from the aforementioned displacement symmetry. In future work, we plan to extend this expansion scheme to higher orders in a systematic derivative expansion of the effective average action. However, whereas a derivative expansion of the static free energy poses no additional problems and works just as in the Euclidean case, the time-dependent displacement symmetry implies that time derivatives must come in the form of \emph{covariant} derivatives (see Eq.~\eqref{eq:covarTimeDerivs}), which makes the generalization to higher-order time derivatives more subtle.

As a sanity check, we first focused on the well-known static sector, which is purely described by an $O(4)$ symmetric Landau-Ginzburg-Wilson free energy. As one main feature of our approach, the corresponding flow equations are unaltered by the presence of dynamics. Using the arguably simplest truncation of writing the free energy in local-potential approximation (LPA) combined with a fourth-order Taylor expansion, the static critical exponent $\nu$ is only slightly better than the mean-field result. As in the results section above, we emphasize again that in the present work we are mainly interested in how a real-time FRG flow for systems with reversible mode couplings as Model~G can be formulated \emph{in principle}, and thus consider this result for $\nu$ as sufficient for our purposes. Moreover, because the free energy decouples from the dynamics within our approach, this result can be improved precisely in the same way as in the Euclidean case by including higher-order terms in the truncation of the free energy, see e.g.~\cite{DePolsi:2020pjk,DePolsi:2021cmi}, and thus poses no conceptual difficulty here.

In the dynamic sector, we have solved the flow equations for the kinetic coefficients of the order parameter and the conserved charge densities numerically, and we found that the dynamic critical exponents indeed come out as $z_\phi = z_n = d/2$, as predicted by Rajagopal and Wilczek \cite{Rajagopal:1992qz}. From the numerical results of the FRG flow we have then systematically extracted a dimensionless dynamic scaling function $\mathcal{L}(x)$ describing the universal momentum and temperature dependence of the diffusion coefficient of the conserved charge densities as a function of the dimensionless scaling variable $x = p\xi(\tau)$. In particular, the result is consistent with the analytical limits of large and small scaling variables $x\gg 1$ and $x\ll 1$.

In future work, we plan to generalize this study to $O(N)$ for values of $N \neq 4$. In particular, this includes the large-$N$ limit of Model~G, which would be interesting to study on its own. 
We also plan to study a flow with an improved truncation of the scale-dependent effective action, such as a derivative expansion with higher orders in the covariant time derivative included.
Moreover, the present study was limited to vanishing external fields $H = 0$, which translates to vanishing current quark masses (i.e., the chiral limit) in QCD. Hence, it would be also worthwhile to include an explicit symmetry breaking by keeping a non-vanishing external field $H \neq 0$ in Model~G, which would resemble finite current quark masses in QCD. One can also extend real-time FRG to include fermions, which could then be used to formulate e.g.~the Quark-Meson or Parity-Doublet model on the Schwinger-Keldysh contour and study their non-perturbative real-time dynamics with the FRG.

\section*{Acknowledgements}

We thank Stefan Floerchinger, Adrien Florio, Eduardo Grossi, Markus Huber, Frederic Klette, Patrick Niekamp, Jan Pawlowski, Fabian Rennecke, Leon Sieke, Alexander Soloviev, and Derek Teaney for insightful discussions. This work was supported by by the Deutsche Forschungsgemeinschaft (DFG, German Research Foundation) through the CRC-TR 211 ‘Strong-interaction matter under extreme conditions’-project number 315477589 – TRR 211. JVR is supported by the Studienstiftung des deutschen Volkes. The authors gratefully acknowledge the computing time provided to them on the high-performance computers Noctua 1 and 2 at the NHR Center PC2. These are funded by the German Federal Ministry of Education and Research and the state governments participating on the basis of the resolutions of the GWK for the national high-performance computing at universities (www.nhr-verein.de/unsere-partner).

\appendix

\section{Details on the MSR path-integral formulation of Model G}
\label{sec:detailsOnMSRPathIntegral}

In this Appendix, we discuss some subtleties of the MSR path-integral formulation of Model~G which we did not fully cover in the main text: In subsection \ref{sct:contFormOfIto} we discuss how the Ito discretization can be properly formulated in the continuum (which corresponds, as we shall see, to a regularization of expressions as $\theta(0)$, i.e.~the value of the Heaviside step function at zero). In subsection \ref{sct:jacobianAndGhosts} we show that the presence of the Jacobian determinant in our generating functional \eqref{eq:genFncAfterFieldTrafo} after the non-linear field transformation can be understood as the determinant that is always (but usually implicitly) present in the MSR path integral, and that it can be represented on a diagrammatical level as additional interactions with Grassmann-valued ghost fields that ensure the normalization condition $Z=1$. Finally, we show in subsection \ref{traditionalapproach} that if the physical sources would be coupled in the `traditional' way to the elementary response fields $\tilde{\psi}$ and $\tilde{n}$ instead of the composite response fields $\tilde{\Phi}$ and $\tilde{N}$, the Boltzmann distribution would no longer be the stationary state of the system.

\subsection{Continuum formulation of the Ito discretization}
\label{sct:contFormOfIto}

In perturbation theory, there can be acausal diagrams where a retarded or advanced propagator closes into a loop.
Such diagrams should naively vanish due to causality.
This is true in the Ito discretization since there the response fields appear always at greater times in the interaction terms than the classical fields (a quartic interaction term in a $\lambda \phi^4$ theory, for instance, would be of the form $\tilde{\Phi}_{i+1} \phi_{i} \phi_{i} \phi_{i}$, with the indices $i$, $i-1$ labeling time slices in this subsection).
As such, the acausal loop will be proportional to expressions as $\langle \tilde{\Phi}_{i+1} \phi_{i} \rangle$.
These expressions generally vanish due to causality, since $\langle \tilde{\Phi}_{i} \phi_{j} \rangle \sim G^{A}_{ij}$ is \emph{advanced} (i.e.\ it vanishes for $i>j$), and  $i$ is always \emph{before} $i+1$.
However, in a naive continuum limit the difference between $i+1$ and $i$ disappears.
Instead, in the continuum limit the retarded/advanced propagators are proportional to the Heaviside step function, $G^R(t,t') = G^A(t',t) \sim \theta(t-t')$.
Hence, the ambiguity can be pushed into a specification of $\theta(0)$ in the propagators (since it closes in a loop at the vertex).
Indeed, a possible continuum formulation of the Ito discretization can be obtained by requiring the value $G_{0}^R(t,t) = G_{0}^A(t,t) \sim \theta(0) \equiv 0$ of the bare retarded/advanced propagators at equal times.
With this regularization, acausal diagrams which involve closed loops of retarded/advanced propagators vanish.
However, this regularization becomes slightly subtle to achieve in frequency space since there we would naively have (after Fourier transform),
\begin{equation}
    G_{0}^{R/A}(t,t) = \int_{-\infty}^{\infty} \frac{d\omega}{2\pi} \, G_{0}^{R/A}(\omega) = - \int_{-\infty}^{\infty} \frac{d\omega}{2\pi} \, \frac{\Gamma_0}{\pm i\omega - \Gamma_0 m^2} = \frac{\Gamma_0}{2} \label{eq:nonZeroIntegralAtCoincidingTimes}
\end{equation}
(here for simplicity in $d=0$ spatial dimensions). This is because the integrand does not fall fast enough for $\omega \to \infty$ when applying the residue theorem.
Obtaining a proper continuum limit of the Ito discretization can be achieved by shifting the time argument of the response fields in the definition of the propagators by an infinitesimal amount $\varepsilon$ into the future \cite{Canet:2011wf},
\begin{subequations}
\begin{align}
    G_{\varepsilon}^R(t,t') &= i\langle \phi(t) \tilde{\Phi}(t'+\varepsilon) \rangle \sim \theta(t-(t'+\varepsilon)) \\
    G_{\varepsilon}^A(t,t') &= i\langle \tilde{\Phi}(t+\varepsilon) \phi(t')  \rangle \sim \theta(t'-(t+\varepsilon))
\end{align} \label{eq:itoEpsilonPrescription}%
\end{subequations}
and understanding all diagrams in the sense that the limit $\varepsilon \to 0^+$ is taken after evaluating the integrals.
Then we indeed have $G_{\varepsilon}^R(t,t) = G_{\varepsilon}^A(t,t) \sim \theta(-\varepsilon) = 0$ for every $\varepsilon > 0$, as desired at equal times. After Fourier transform (assuming time-translation invariance) this amounts to the replacement 
\begin{align}
    G_{\varepsilon}^{R/A}(\omega) &= e^{\pm i\omega \varepsilon} G_{0}^{R/A}(\omega)
\end{align}
in frequency space.
The exponential factors $e^{\pm i\omega \varepsilon}$ ensure that the contribution from the large semicircle  in the upper (lower) half-planes in integrals like in \eqref{eq:nonZeroIntegralAtCoincidingTimes} vanishes.
Thus the whole integral is zero since the integrand is analytic in the upper (lower) half-plane and vanishes fast enough for $\omega \to \infty$.
As such, closed loops over retarded/advanced propagators vanish for every $\varepsilon > 0$, as desired.

We can thus set the Jacobian to $\mathcal{J}[\phi,n] = 1$, but have to keep in mind that this goes hand in hand with the $\varepsilon$-prescription \eqref{eq:itoEpsilonPrescription}.
Alternative discretizations like the Stratonovich convention can avoid such an $\varepsilon$-prescription (since they correspond to the regularization $\theta(0) = 1/2$ which coincides with the result \eqref{eq:nonZeroIntegralAtCoincidingTimes} from the `naive' continuum limit), but there the Jacobian is in general non-zero \cite{Bausch1976,DeDominicis:1977fw}.
Usually, the latter is rewritten as an integral over anti-commuting fields (`ghosts') which otherwise have the same quantum numbers as the original fields.
These additional ghost contributions precisely cancel any non-vanishing unphysical contributions from acausal diagrams \cite{Gao:2018bxz}.   

\subsection{Jacobians and ghosts}
\label{sct:jacobianAndGhosts}

The presence of a Jacobian determinant as in \eqref{eq:genFncAfterFieldTrafo} is not unusual in field-theoretic treatments of classical-statistical systems and can be intuitively understood as follows. The generating functional as formulated in \eqref{eq:genFncAfterFieldTrafo} can be alternatively derived without ever referring to the non-linear field transformation \eqref{eq:responseFieldTrafo}, i.e.\ by just appealing to the standard MSR technique. To see this, first combine the two equations of motion \eqref{eq:eoms} into one using vector notation, and rewrite the right-hand side as a field-dependent matrix acting on the (functional) gradient of the free energy (plus a vector containing the contribution from the noises),
\begingroup
\setlength\arraycolsep{6pt}
\begin{align}
    \frac{\partial}{\partial t}
    \begin{pmatrix}\phi_{a} \\ n_{bc}
    \end{pmatrix} &= \label{eq:eomsMatVec} \\ \nonumber
    &\hspace{-1.0cm} 
    - \begin{pmatrix}
        \delta_{ad} \Gamma_{0}  & -\orange{\tfrac{g}{2}}\{\phi_{a},n_{ef}\}  \\
        -g\{n_{bc},\phi_{d}\} & - \tfrac{1}{2}(\delta_{be}\delta_{cf}-\delta_{bf}\delta_{ce}) \gamma \vec{\nabla}^2  - \orange{\tfrac{g}{2}}\{n_{bc},n_{ef}\}
    \end{pmatrix}
    \begin{pmatrix}
        \frac{\delta F}{\delta \phi_{d}} \\
        \frac{\delta F}{\delta n_{ef}} \\
    \end{pmatrix} + 
    \begin{pmatrix}
        \theta_a \\
        \vec{\nabla} \!\cdot\! \vec{\zeta}_{bc}
    \end{pmatrix} 
\end{align}
\endgroup
In front of the functional gradient of the free energy now stands a (field-dependent) matrix. This matrix is, in fact, the transpose $J^T(\psi)$ of $J(\psi)$ in our non-linear transformation \eqref{eq:responseFieldTrafoInMatrixNotation} of the response fields.
The field-dependent matrix $J(\psi)$ and its transpose are invertible for arbitrary field configurations $\psi=\psi(x)$ since the corresponding determinant $\det J(\psi) \neq 0$ does not vanish.\footnote{By the inverse function theorem, its inverse can be expressed by the Jacobian matrix of the inverse of the field transformation \eqref{eq:responseFieldTrafo}, \[ J^{-1}(x,x';\psi) = \frac{\delta \tilde{\psi}(x)}{\delta \tilde{\Psi}(x')}\,. \]} Thus we can multiply by ${J^{-1}}^T(\psi)$ on both sides of \eqref{eq:eomsMatVec} (also using that matrix inversion and transposition commute, $(J^{-1})^T = (J^{T})^{-1}$), and rewrite the equations of motion in symbolic superfield notation as
\begin{equation}
    \int_{x'} {J^{-1}_{ij}}^T(x,x';\psi) \, \frac{\partial \psi_j(t',\vec{x}')}{\partial t'} = -\frac{\delta F}{\delta \psi_i(\vec{x})} \bigg\rvert_{t} + \int_{x'} {J^{-1}_{ij}}^T(x,x';\psi) \, \xi_{j}(x') \label{eq:eomsMultipliedByInvJ}
\end{equation}
with superfield indices $i$ and $j$, and $\xi = (\xi_i) = (\theta_a,\vec{\nabla} \!\cdot\! \vec{\zeta}_{bc})$ denoting the noise superfield. We can use this (equivalent) form of the equations of motion now as our starting point for the MSR path integral. Importantly, after the overall multiplication with the field-dependent factor ${J^{-1}}^T(\psi)$, the noise becomes multiplicative. Correspondingly, we have to specify the discretization of the stochastic equations to make them unambiguous. Moreover, we have to include a field-dependent Jacobian as in \eqref{path_integral} to ensure the overall normalization of the path integral.  In Ito discretization, the time derivative in \eqref{eq:eomsMultipliedByInvJ} becomes a finite backward difference (with finite time step $\Delta t$)
\begin{equation}
    \frac{\partial \psi_i(t,\vec{x})}{\partial t} \;\to\; \frac{\Delta^R \psi_i(t,\vec{x})}{\Delta t} \equiv \frac{\psi_i(t,\vec{x}) - \psi_i(t-\Delta t,\vec{x})}{\Delta t} \,. \label{eq:discTimeDeriv}
\end{equation}
The Jacobian matrix $J_{ij}(x,x';\psi)$ is local in time, so we can also write
\begin{equation}
    {J_{ij}^{-1}}^T(x,x';\psi) \equiv {J_{ij}^{-1}}^T(\vec{x},\vec{x}';\psi,t) \,\delta(t-t')
\end{equation}
for its transposed inverse. We can then express the Ito-discretized equations of motion as
\begin{align}
    \int_{\vec{x}'} {J^{-1}_{ij}}^T(\vec{x},\vec{x}';\psi,t-\Delta t) \, \frac{\Delta^R \psi_j(t,\vec{x}')}{\Delta t}  &= \\ \nonumber
    &\hspace{-3.5cm} -\frac{\delta F}{\delta \psi_i(\vec{x})} \bigg\rvert_{t-\Delta t} + \int_{\vec{x}'} {J^{-1}_{ij}}^T(\vec{x},\vec{x}';\psi,t-\Delta t) \, \xi_{j}(t-\Delta t,\vec{x}') \,, 
\end{align}
which involve the fields only at times $t$ and $t-\Delta t$.
Note that, as an essential property of Ito discretization, only the first term in the discretized time derivative is evaluated at time $t$. All other quantities are evaluated at the time before, $t-\Delta t$.
We define an operator $G$ as a shorthand notation for these Ito-discretized equations of motions,
\begin{align}
    G_{i}(t,\vec{x}) &\equiv \int_{\vec{x}'} {J^{-1}_{ij}}^T(\vec{x},\vec{x}';\psi,t-\Delta t) \, \frac{\Delta^R \psi_j(t,\vec{x}')}{\Delta t} \\ \nonumber
    &\hspace{1.0cm}  + \frac{\delta F}{\delta \psi_i(\vec{x})} \bigg\rvert_{t-\Delta t} - \int_{\vec{x}'} {J^{-1}_{ij}}^T(\vec{x},\vec{x}';\psi,t-\Delta t) \, \xi_{j}(t-\Delta t,\vec{x}') \,. 
\end{align}
The discretized equations of motion are satisfied if $G = 0$.
Its Jacobian matrix can be straightforwardly computed,
\begin{equation}
    \frac{\delta G_{i}(t,\vec{x})}{\delta \psi_{j}(t',\vec{x}')} = {J^{-1}_{ij}}^T(\vec{x},\vec{x}';\psi,t-\Delta t) \, \delta(t-t') + \left( \cdots \right) \, \delta(t-t'-\Delta t) \label{eq:discJacobMat}
\end{equation}
and is a lower triangular block matrix in the discretized time domain. Its determinant can thus be computed as an (infinite) product of the determinants of the blocks at discretized times $t_n=n\Delta t$ along the main diagonal (where the determinants of the blocks are understood with respect to both internal indices $i,j$ and spatial `indices' $\vec{x},\vec{x}'$),
\begin{equation}
    \mathcal{J}'[\psi] = \det \frac{\delta G_{i}(t,\vec{x})}{\delta \psi_{j}(t',\vec{x}')} = \prod_{t_n} \frac{1}{\Delta t} \,\det\, {J^{-1}_{ij}}^T(\vec{x},\vec{x}';\psi,t_n-\Delta t) \,. \label{eq:detOfTransformedEOMs}
\end{equation}
This is the same Jacobian as in  \eqref{eq:jacobianOfNonlinearFieldTrafo}. However, in \eqref{eq:jacobianOfNonlinearFieldTrafo} it appeared due to the non-linear field transformation of the response field. Remarkably, the determinant does not depend on the noise, even though in the alternative formulation \eqref{eq:eomsMultipliedByInvJ} of the equations of motion the noise is multiplicative. This is due to the choice of Ito discretization.

By constructing the generating functional using the standard MSR technique, which we have reviewed in Sec.~\ref{sct:MSRpi}, we obtain
\begin{align}
    Z[J,\tilde{J}] &= \int \mathcal{D}\psi\,\mathcal{D}\tilde{\Psi}\, \mathcal{J}'[\psi] \, \exp\bigg\{ iS'[\psi,\tilde{\Psi}] + \label{eq:genFuncAfterTrafoOfEOMs} i\int_x \big( \tilde{J}_i \psi_i + \tilde{\Psi}_i J_i \big)  \bigg\} 
\end{align}
with
\begin{equation}
    S'[\psi,\tilde{\Psi}] = - \tilde{\Psi}^T {J^{-1}}^T(\psi) \frac{\partial \psi}{\partial t} + iT \tilde{\Psi}^T {J^{-1}}^T(\psi) \tilde{\Psi} - \tilde{\Psi}^T \frac{\delta F}{\delta\psi} \label{eq:S'} 
\end{equation}
in compact superfield notation.

As a side remark, note that when one derives $S'[\psi,\tilde{\Psi}]$ by performing the field transformation in the action $S[\psi,\tilde{\psi}]$ for the elementary response fields, the quadratic term in the response fields looks like
\begin{equation}
    S[\psi,\tilde{\psi}] = iT \tilde{\psi}^T \hat{\gamma} \tilde{\psi} + \cdots
\end{equation}
(with the superfield matrix $\hat{\gamma} = \mathrm{diag}(\Gamma_{0}, -\gamma\vec{\nabla}^2)$ of kinetic coefficients), which translates to
\begin{equation}
    S'[\psi,\tilde{\Psi}] = iT \tilde{\Psi}^T {J^{-1}}^T(\psi) \hat{\gamma} J^{-1}(\psi) \tilde{\Psi} + \cdots \label{eq:quadrTermInS'}
\end{equation}
in the transformed action.
This is, in fact, the same quadratic term as in \eqref{eq:S'}, which can be seen by using that the symmetric contraction of the antisymmetric Poisson bracket $\{\psi,\psi\}$ with $\tilde{\Psi}^T {J^{-1}}^T(\psi) = \tilde{\psi}^T$ and $J^{-1}(\psi)\tilde{\Psi} = \tilde{\psi}$ vanishes, and so can add a suitable zero to \eqref{eq:quadrTermInS'}, 
\begin{equation}
    iT \tilde{\Psi}^T {J^{-1}}^T(\psi) \hat{\gamma} J^{-1}(\psi) \tilde{\Psi} = iT \tilde{\Psi}^T {J^{-1}}^T(\psi) \underbrace{ \left( \hat{\gamma} + g\{\psi,\psi\} \right) }_{J(\psi)} J^{-1}(\psi) \tilde{\Psi} = iT \tilde{\Psi}^T {J^{-1}}^T(\psi) \tilde{\Psi} \,.
\end{equation}
This shows that the quadratic terms in \eqref{eq:S'} and \eqref{eq:quadrTermInS'} are the same.

The MSR action $S'$ in \eqref{eq:S'} corresponds to the stochastic equations of motion in \eqref{eq:eomsMultipliedByInvJ}.
The Jacobian appearing in \eqref{eq:genFncAfterFieldTrafo} is hence just the standard Jacobian that is always (implicitly) present in the MSR path integral.
Its precise value reflects the underlying discretization of the equations of motion.
Notice that we can not simply set it to one here by clever choice of discretization, because ${J^{-1}}^T(\psi)$ appears as a field-dependent factor in front of the time derivative in the equations of motion \eqref{eq:eomsMultipliedByInvJ}.
The field-dependent function ${J^{-1}}^T(\psi)$ thus appears on the diagonal of the Jacobian matrix \eqref{eq:discJacobMat}.
Hence, it will also enter the determinant \eqref{eq:detOfTransformedEOMs}.

On a practical level, a field-dependent Jacobian in the MSR formalism can be treated using well-established techniques.
For example, by rewriting it using anti-commuting `ghost' degrees of freedom.
There one expresses the determinant as a Gaussian integral over a pair of Grassmann-valued fields $c$ and $\tilde{c}$,
\begin{equation}
    \mathcal{J}'[\psi] = \int \mathcal{D}\tilde{c}\,\mathcal{D}c\,\exp\left\{ -\int_{xx'} \tilde{c}_i(x) \frac{\delta \tilde{\Psi}_i(x)}{\delta \psi_j(x')} c_j(x') \right\} \,,
\end{equation}
where we have absorbed the infinite factor $\prod_{t_n} (1/\Delta t)$ from the determinant in \eqref{eq:detOfTransformedEOMs} into the definition of the path integral.
Rewriting the determinant like this can be interpreted as introducing a ghost part 
\begin{equation}
    S_{gh} = \int_{xx'} i \tilde{c}_i(x)\frac{\delta\tilde{\Psi}_i(x)}{\delta \tilde{\psi}_j(x')} c_j(x') \label{eq:ghostAction}
\end{equation}
to the total action.
We can write down a corresponding generating functional by introducing Grassmann-valued sources $\eta_i$, $\tilde{\eta}_j$ for the ghosts $\tilde{c}_i$, $c_j$,
\begin{align}
    Z[J,\tilde{J},\eta,\tilde{\eta}] &\equiv \int \mathcal{D}[\psi,\tilde{\Psi},\tilde{c},c]\,  \exp\bigg\{ iS' + iS_{gh} + \label{eq:genFuncWithGhosts} i\int_x \big( \tilde{J}_i \psi_i + \tilde{\Psi}_i J_i + \tilde{c}_{i} \eta_i + \tilde{\eta}_i c_{i} \big)  \bigg\}\,.
\end{align}
The corresponding 1PI effective action is obtained by performing the Legendre transform \eqref{eq:effActionDef} also with respect to the ghost sources.

As we have already stated in Eq.~\eqref{eq:BRSTMSR} in the main text, for $\tilde{J}=\eta=\tilde{\eta}=0$ but an arbitrary classical source $J\neq 0$ the total action $S_{J}$ in \eqref{eq:genFuncWithGhosts} becomes BRST exact. This can be seen as follows. First, we evaluate the BRST variation in \eqref{eq:BRSTMSR}, which yields
\begin{align}
    S_J &= Q\left\{ -i\tilde{c}_i \left( {J^{-1}_{ij}}^T(\psi) {\partial_t} \psi_j + \frac{\delta F}{\delta \psi_i} - J_i -iT {J^{-1}_{ij}}^T(\psi) \tilde{\Psi}_j \right) \right\} \\
    &=
    -\tilde{\Psi}_i {J^{-1}_{ij}}^T(\psi) \partial_t^R \psi_j - \tilde{\Psi}_i \frac{\delta F}{\delta \psi_i} + iT \tilde{\Psi}_i {J^{-1}_{ij}}^T(\psi) \tilde{\Psi}_j  + i\tilde{c}_i {J^{-1}_{ij}}^T(\psi) {\partial_t} c_j  + \tilde{\Psi}_i J_i \label{eq:actionSuperfield}  + \cdots \, .
\end{align}
The terms not spelled out explicitly here (but summarized as `$\cdots$') vanish in Ito regularization, since they only contribute to the off-diagonal elements of the lower triangular matrix in \eqref{eq:discJacobMat} and thus do not enter the determinant in \eqref{eq:detOfTransformedEOMs}. Moreover, in Ito regularization only the first `half' of the discretized time derivative $\partial_t c_j \to (c_j(t)-c_j(t-\Delta t))/\Delta t$ effectively enters the determinant \eqref{eq:detOfTransformedEOMs}. Hence, one can omit the second half $c_j(t-\Delta t)/\Delta t$ from \eqref{eq:actionSuperfield}. With a field rescaling $c_i(t)/\Delta t \to c_i(t)$ of the first half, and by absorbing the resulting infinite factor $\prod_{t_n}(1/\Delta t)$ again into the definition of the path integral, one arrives at the properly normalized ghost action \eqref{eq:ghostAction}. Hence, the expression obtained in \eqref{eq:actionSuperfield} indeed matches the action in \eqref{eq:genFuncWithGhosts} (which we have shown here within Ito regularization), so $S_J$  is indeed BRST exact.

The action $S_J$ being BRST symmetric implies the normalization condition \eqref{eq:normCond} of the MSR path integral, which can be proven as in Sec.~1.5 of Ref.~\cite{Crossley:2015evo}: A variation $J \to J + \delta J$ in the generating functional \eqref{eq:genFncAfterFieldTrafo} leads to 
\begin{align}
    \delta Z[J] &= \int \mathcal{D}[\psi,\tilde{\Psi},\tilde{c},c]\,e^{iS_J} i\tilde{\Psi}_i \delta{J}_i = - \int \mathcal{D}[\psi,\tilde{\Psi},\tilde{c},c]\,e^{iS_J} \delta{J}_i Q\tilde{c}_i \;= \nonumber \\
    &= - \int \mathcal{D}[\psi,\tilde{\Psi},\tilde{c},c]\,Q\left\{ e^{iS_J} \delta{J}_i \tilde{c}_i \right\} = 0 \label{normalization}
\end{align}
where in the second equality we used that the composite response fields $i\tilde{\Psi}_i = -Q \tilde{c}_i$ are BRST exact, and in the last equality that the integral over a total differential vanishes.
Together with the normalization $Z[J = 0] = 1$ we thus have $Z[J] = 1$ for all physical source configurations $J$ on the level of the full path integral.

\subsection{Traditional approach of constructing generating functionals}\label{traditionalapproach}

In a traditional approach of constructing a generating functional for correlation functions, one would now go on and introduce source terms as linear couplings to the elementary fields,
\begin{align}
    z[h,\tilde{h},\mathfrak{h},\tilde{\mathfrak{h}}] &= \int \mathcal{D}\phi\,\mathcal{D}\tilde{\phi}\,\mathcal{D}n\,\mathcal{D}\tilde{n}\, \exp\bigg\{ iS \; + \label{eq:traditionalGenFunc} \\ \nonumber &\hspace{1.0cm} i\int_x \big( h_a(x) \tilde{\phi}_a(x) + \tilde{h}_a(x) \phi_a(x) + \orange{\frac{1}{2}}  \mathfrak{h}_{ab}(x) \tilde{n}_{ab}(x) + \orange{\frac{1}{2}}  \tilde{\mathfrak{h}}_{ab}(x) n_{ab}(x) \big) \bigg\}    
\end{align}
However, adding the sources like this has the following disadvantage: Even in the absence of the unphysical response sources $\tilde{h}_{a}(x)$ and $\tilde{\mathfrak{h}}_{ab}(x)$ the stationary state of the system is no longer a Boltzmann distribution:
The equations of motion corresponding to the generating functional \eqref{eq:traditionalGenFunc} (for vanishing response sources) read
\begin{align}
    \frac{\partial \phi_a}{\partial t} &= -\Gamma_{0} \frac{\delta F}{\delta \phi_a}+ \orange{\frac{g}{2}}  \{\phi_a,n_{bc}\}\frac{\delta F}{\delta n_{bc}} +h_a(x) +\theta_a  \\
    \frac{\partial n_{ab}}{\partial t} &= \gamma \vec{\nabla}^2 \frac{\delta F}{\delta n_{ab}}+g\{n_{ab},\phi_c\}\frac{\delta F}{\delta \phi_c}+ \orange{\frac{g}{2}}  \{n_{ab},n_{cd}\}\frac{\delta F}{\delta n_{cd}}+\mathfrak{h}_{ab}(x)+\vec{\nabla} \!\cdot\! \vec{\zeta}_{ab}
\end{align}
which can be reformulated as a shift of the free energy in the dissipative terms,
\begin{align}
    \frac{\partial \phi_a}{\partial t} &= -\Gamma_{0} \frac{\delta}{\delta \phi_a}\left[ F - \int_{x'} h_a'(x') \phi_a(x') \right] + \orange{\frac{g}{2}}  \{\phi_a,n_{bc}\}\frac{\delta F}{\delta n_{bc}} +\theta_a  \\
    \frac{\partial n_{ab}}{\partial t} &= \gamma \vec{\nabla}^2 \frac{\delta }{\delta n_{ab}} \left[ F - \orange{\frac{1}{2}}  \int_{x'} \mathfrak{h}_{ab}'(x') n_{ab}(x') \right] + \orange{\frac{g}{2}}  \{n_{ab},\phi_c\}\frac{\delta F}{\delta \phi_c}+g\{n_{ab},n_{cd}\}\frac{\delta F}{\delta n_{cd}}+\vec{\nabla} \!\cdot\! \vec{\zeta}_{ab}
\end{align}
where we have rescaled the source terms by the kinetic coefficients to put them inside the square brackets, 
\begin{equation}
    \Gamma h_a'(x) = h_a(x) \,, \hspace{0.5cm} 
    -\gamma\vec{\nabla}^2 \mathfrak{h}_{ab}'(x) = \mathfrak{h}_{ab}(x) \,.
\end{equation}
This highlights the inconsistency concerning the equilibrium state that arises when the sources couple to the elementary response fields as in \eqref{eq:traditionalGenFunc}.
On the level of the equations of motion, the source terms correspond to shifts in the free energy, but only in the dissipative terms and not in the Poisson-bracket terms. 
Hence, a Boltzmann distribution $\sim \exp{-\beta(F-h' \phi-\tfrac{1}{2}\mathfrak{h}' n)}$ is no longer a stationary solution to the corresponding Fokker-Planck equation since the mode-coupling terms are no longer divergence-free in presence of the sources, i.e.~\eqref{eq:revModeCouplAreDivFree} no longer holds.
However, when the source terms are added consistently as terms in the free energy \eqref{eq:freeEnergyReplacementWithSources}, the reversible mode couplings generate no probability current, i.e.~\eqref{eq:revModeCouplAreDivFree} is satisfied, and a Boltzmann distribution $\sim \exp{-\beta(F-H \phi-\tfrac{1}{2}\mathcal{H} n)}$ is maintained as a stationary solution.

\section{Generalized fluctuation-dissipation relations}\label{appendix:fdr}

In this section, we first show that the bare MSR action has the discrete symmetry \eqref{eq:thermEqSymm}, which essentially expresses detailed balance in thermal equilibrium.
Since one can show on rather general grounds that the effective action $\Gamma$ as defined in \eqref{eq:effActionDef} admits the same symmetries as the bare action $S$, we can afterward derive generalized fluctuation-dissipation relations for the $n$-point 1PI vertex functions, based on the symmetry \eqref{eq:thermEqSymm}.

We start this section by showing that
the transformation \eqref{eq:thermEqSymm} is a symmetry of the bare MSR action $S$, i.e.~we show that $S[\mathcal{T}_\beta(\psi,\tilde{\Psi})] = S[\psi,\tilde{\Psi}]$. The bare MSR action $S$ is in our compact superfield notation given by
\begin{align}
    S &= -\tilde{\Psi}^T {J^{-1}}^T(\psi) \partial_t \psi + iT \tilde{\Psi}^T  J^{-1}(\psi) \tilde{\Psi} - \tilde{\Psi}^T \frac{\delta F}{\delta\psi} \label{shorthandnotation}
\end{align}
where all symbols are interpreted as vectors/matrices with respect to spatial, temporal, and superfield indices.
We divide our derivation that \eqref{eq:thermEqSymm} is a symmetry of \eqref{shorthandnotation} into two steps.
In the first step (i), we apply the transformation (where $\dot{\psi}_i \equiv \partial_t \psi_i$)
\begin{align}
\begin{split}
    \psi_i(t,\vec{x}) &\to \psi_i(-t,\vec{x}) \,, \\
    \tilde{\Psi}_i(t,\vec{x}) &\to i\beta \dot{\psi}_i(-t,\vec{x}) + \tilde{\Psi}_i(-t,\vec{x}) \,,
\end{split} \tag{i}\label{eq:proofThEqSymmStep1}
\end{align}
and in the second step (ii), we multiply the fields with their time-reversal parities,
\begin{align}
\begin{split}
    \psi_i(t,\vec{x}) &\to \epsilon_i \psi_i(t,\vec{x}) \,, \\
    \tilde{\Psi}_i(t,\vec{x}) &\to \epsilon_i \tilde{\Psi}_i(t,\vec{x}) \,.
\end{split} \tag{ii}\label{eq:proofThEqSymmStep2}
\end{align}
Symbolically, our procedure can be represented by
\begin{equation}
    S \xrightarrow{\eqref{eq:proofThEqSymmStep1}} S_1 \xrightarrow{\eqref{eq:proofThEqSymmStep2}} S_2
\end{equation}
where $S_2$ then equals the transformed MSR action $S_2 = S[\mathcal{T}_{\beta}(\psi,\tilde{\Psi})]$. After both steps, we will indeed obtain back the original action, $S_2=S$, so that $\mathcal{T}_{\beta}$ is indeed a symmetry of~$S$.

\begin{enumerate}
\item 
In step \eqref{eq:proofThEqSymmStep1}
the action transforms as (where we have already replaced $t \to -t$ in the integration) 
\begin{align}
    S \to S_1 = \; &\red{ \tilde{\Psi}^T {J^{-1}}^T(\psi) \dot{\psi} } + \violet{ i\beta \dot{\psi}^T {J^{-1}}^T(\psi) \dot{\psi} } \; + \nonumber \\
    &iT \, \tilde{\Psi}^T {J^{-1}}^T(\psi) \tilde{\Psi} + \underbrace{iT\,i\beta}_{(-)}\,\dot{\psi}^T {J^{-1}}^T(\psi) \tilde{\Psi} \; + \nonumber \\
    &\red{ \underbrace{iT\,i\beta}_{(-)}\,\tilde{\Psi}^T {J^{-1}}^T(\psi) \dot{\psi} } + \violet{ \underbrace{iT\,(i\beta)^2}_{-i\beta} \,\dot{\psi}^T {J^{-1}}^T(\psi) \dot{\psi} } \; + \nonumber \\
    &-\tilde{\Psi}^T \frac{\delta F}{\delta \psi} - \blue{i\beta\,\dot{\psi}^T \frac{\delta F}{\delta \psi} } \,.
\end{align}
Here, the \red{red terms} cancel, the \violet{violet terms} cancel, and the \blue{blue term} is a total time derivative, so it vanishes upon integrating from $t=-\infty$ to $+\infty$.
We thus obtain
\begin{align}
    S_1 &= - \dot{\psi}^T {J^{-1}}^T(\psi) \tilde{\Psi} + iT \, \tilde{\Psi}^T {J^{-1}}^T(\psi) \tilde{\Psi} -\tilde{\Psi}^T \frac{\delta F}{\delta \psi} \,.
\end{align}
The first term can be reformulated via (using that the transpose of a (complex) number becomes itself)
\begin{align}
    - \dot{\psi}^T {J^{-1}}^T(\psi) \tilde{\Psi} &= \left( - \dot{\psi}^T {J^{-1}}^T(\psi) \tilde{\Psi} \right)^T = - \tilde{\Psi}^T J^{-1}(\psi) \dot{\psi}
\end{align}
such that we finally get
\begin{align}
    S_1 &= - \tilde{\Psi}^T J^{-1}(\psi) \dot{\psi} + iT \, \tilde{\Psi}^T {J^{-1}}^T(\psi) \tilde{\Psi} -\tilde{\Psi}^T \frac{\delta F}{\delta \psi} \label{eq:actionAfterThEqSymmTrafo}
\end{align}
as our result of step \eqref{eq:proofThEqSymmStep1}.
Note that in contrast to the original MSR action \eqref{shorthandnotation} there is $J^{-1}(\psi)$ in the first term, instead of its transpose ${J^{-1}}^T(\psi)$.\footnote{In the second term this distinction is irrelevant, because due to the symmetric contraction with $\tilde{\Psi}$ from the left and right only the symmetric part of ${J^{-1}}^T(\psi)$ contributes there.}
If the inverse Jacobian $J^{-1}(\psi)$ would be symmetric (such as in the case of vanishing reversible mode couplings), we would have ${J^{-1}}^T(\psi) = J^{-1}(\psi)$, and step \eqref{eq:proofThEqSymmStep1} alone would already constitute a symmetry of our MSR action. To change `$J^{-1}(\psi)$ back to ${J^{-1}}^T(\psi)$' is the goal of step~\eqref{eq:proofThEqSymmStep2}.

\item
When we multiply the fields with their respective time-reversal parities according to step~\eqref{eq:proofThEqSymmStep2}, the terms in \eqref{eq:actionAfterThEqSymmTrafo} change according to
\begin{subequations}
\begin{align}
    -\tilde{\Psi}^T {J^{-1}}(\psi) \partial_t \psi &\to -\tilde{\Psi}^T \epsilon^T {J^{-1}}(\epsilon \psi) \epsilon \partial_t \psi \\
    iT \, \tilde{\Psi}^T {J^{-1}}^T(\psi) \tilde{\Psi} &\to iT \, \tilde{\Psi}^T \epsilon^T {J^{-1}}^T(\epsilon\psi) \epsilon \tilde{\Psi} \\
    -\tilde{\Psi}^T \frac{\delta F}{\delta \psi}[\psi] &\to -\tilde{\Psi}^T \epsilon^T \frac{\delta F}{\delta \psi}[\epsilon \psi]
\end{align}
\end{subequations}
so we need to have
\begin{align}
    \epsilon^T \frac{\delta F}{\delta \psi}[\epsilon \psi] = \frac{\delta F}{\delta \psi}[\psi] \;\;\text{and}\;\; \epsilon^T {J^{-1}}(\epsilon \psi) \epsilon &= {J^{-1}}^T(\psi)  \label{eq:requiredInvariancesUnderTimeReversal}
\end{align}
to ensure that the MSR action is invariant under the thermal equilibrium symmetry.
The first equality in \eqref{eq:requiredInvariancesUnderTimeReversal} is satisfied if the free energy $F$ is invariant under time-reversal, $F[\varepsilon \psi] = F[\psi]$.
This means that a Taylor expansion of $F$ may only contain operators that have even parity under time reversal.
The second equality in \eqref{eq:requiredInvariancesUnderTimeReversal} can be satisfied by choosing parities $\epsilon_\phi = +1$ and $\epsilon_n = -1$, i.e.~the charge densities change sign under time reversal. This is consistent with their microscopic definition \eqref{current1} in the case of the $O(N)$ model, as well as with their interpretation as the isovector and isoaxial-vector currents \eqref{eq:VACurrents} in the context of QCD. In matrix notation in $\psi=(\phi,n)$ superfield space, we have
\begingroup 
\setlength\arraycolsep{6pt} 
\begin{align}
    \epsilon &= \begin{pmatrix}
        \epsilon_{\phi} &0 \\ 0&\epsilon_{n}
    \end{pmatrix} = \begin{pmatrix}
        +1&0 \\ 0&-1
    \end{pmatrix} \,.
\end{align}
\endgroup
In symbolic shorthand notation, we can expand $J(\psi) = J_0 - \delta J(\psi)$ in a Neumann series in $\delta J(\psi)$ around $\delta J(\psi) = 0$ and apply the transformation individually to each term in the series,
\begin{align}
    \epsilon^T J^{-1}(\epsilon \psi) \epsilon &= \epsilon^T \left( J_0^{-1} + J_0^{-1} \delta J(\epsilon \psi) J_0^{-1} + J_0^{-1} \delta J(\epsilon \psi) J_0^{-1} \delta J(\epsilon \psi) J_0^{-1} + \cdots \right) \epsilon \nonumber \\
    &= \epsilon^T J_0^{-1} \epsilon + \epsilon^T J_0^{-1} (\epsilon \epsilon^T) \, \delta J(\epsilon \psi)\, (\epsilon \epsilon^T) J_0^{-1} \epsilon + \nonumber \\ 
    &\hspace{1.0cm} \epsilon^T J_0^{-1} (\epsilon \epsilon^T) \, \delta J(\epsilon \psi)\, (\epsilon \epsilon^T) J_0^{-1} (\epsilon \epsilon^T) \, \delta J(\epsilon \psi) \, (\epsilon \epsilon^T) J_0^{-1} \epsilon + \cdots \nonumber \\
    &= J_0^{-1} + J_0^{-1} (\epsilon^T \delta J(\epsilon \psi) \epsilon) J_0^{-1} + \nonumber \\ 
    &\hspace{1.0cm} J_0^{-1} (\epsilon^T \delta J(\epsilon \psi) \epsilon) J_0^{-1} (\epsilon^T \delta J(\epsilon \psi) \epsilon) J_0^{-1} + \cdots \label{eq:neumannSeriesAfterTimeParity}
\end{align}
where we used that $\epsilon^T J_0^{-1} \epsilon = J_0^{-1}$.
We calculate
\begingroup
\setlength\arraycolsep{6pt}
\begin{align}
    \epsilon^T \delta J(\epsilon \psi) \epsilon &= -g \begin{pmatrix}
        +1&0\\0&-1
    \end{pmatrix}
    \begin{pmatrix}
        0&\{\phi,n\} \\ \{n,\phi\}&-\{n,n\}
    \end{pmatrix}
    \begin{pmatrix}
        +1&0\\0&-1
    \end{pmatrix} = \nonumber \\
    &= -g
    \begin{pmatrix}
        0&-\{\phi,n\}\\-\{n,\phi\}&-\{n,n\}
    \end{pmatrix} = -\delta J(\psi) = \delta J^T(\psi) \,.
\end{align}
\endgroup
We can now resum the series in \eqref{eq:neumannSeriesAfterTimeParity} again and find
\begin{align}
    \epsilon^T J^{-1}(\epsilon \psi) \epsilon &= {J^{-1}}^T(\psi)
\end{align}
as anticipated from step~\eqref{eq:proofThEqSymmStep1}, such that we finally obtain
\begin{equation}
    S_2 = - \tilde{\Psi}^T {J^{-1}}^T(\psi) \dot{\psi} + iT \, \tilde{\Psi}^T {J^{-1}}^T(\psi) \tilde{\Psi} -\tilde{\Psi}^T \frac{\delta F}{\delta \psi} \label{eq:actionAfterThEqSymmTrafo2}
\end{equation}
as our result of the step~\eqref{eq:proofThEqSymmStep2}.
The transformed action $S_2$ in \eqref{eq:actionAfterThEqSymmTrafo2} is the same as the original action $S$ in \eqref{shorthandnotation}. This shows that the MSR action \eqref{shorthandnotation} is indeed invariant under the discrete transformation \eqref{eq:thermEqSymm} and concludes our proof.
\end{enumerate}

Using the symmetry of thermal equilibrium one can derive the general fluctuation-dissipation relations. The recipe starts from writing the effective action in a vertex expansion around the vanishing field expectation value, where for simplicity (and in compliance with our truncation from Sec.~\ref{sec:truncation}) we have only included vertices up to second order in the response fields $\tilde{\Psi}$ (with superfield indices $a_1,\ldots,a_n$ here),
\begin{align}
    \Gamma = \sum_{n=2}^{\infty} \frac{1}{n!} \int_{p_1\ldots p_n} \bigg[ n\,&\Gamma^{\tilde{\Psi} \psi \cdots \psi}_{a_1 a_2 \cdots a_n}(p_1,p_2,\ldots,p_n) \tilde{\Psi}_{a_1}(-p_1) \psi_{a_2}(-p_2) \ldots \psi_{a_n}(-p_n) \; + \label{eq:nthOrderTermInVertExp} \\ \nonumber  
    &\hspace{-3.5cm} \frac{n(n-1)}{2} \,  \Gamma^{\tilde{\Psi} \tilde{\Psi} \psi \cdots \psi}_{a_1 a_2 a_3 \cdots a_n}(p_1,p_2,p_3,\ldots,p_n) \tilde{\Psi}_{a_1}(-p_1) \tilde{\Psi}_{a_2}(-p_2) \psi_{a_3}(-p_3) \ldots \psi_{a_n}(-p_n) + \mathcal{O}(\tilde{\Psi}^3) \bigg] 
\end{align}
Applying the thermal equilibrium transformation defined above, and requiring that the effective action is symmetric under such a transformation, one can derive the general form of the fluctuation-dissipation relation by comparing terms to terms and requiring them to be identical. The final result of the general FDR is given by
\begin{align}
   \Gamma^{\tilde{\Psi} \psi \cdots \psi}_{a_1 a_2 \cdots a_n}(p_1,\ldots,p_n) - \hat{\epsilon}\, \Gamma^{\tilde{\Psi} \psi \cdots \psi}_{a_1 a_2 \cdots a_n}(-p_1,\ldots,-p_n) &= \label{eq:conditionClassVert}  \\  \nonumber
   -\beta \big[
    &p_2^0 \Gamma_{a_1 a_2 a_3 \cdots a_n}^{\tilde{\Psi}\tilde{\Psi}\psi \cdots \psi}(p_1,p_2,p_3,\ldots,p_n) \;+ \\ \nonumber
    &p_3^0 \Gamma_{a_1 a_2 a_3 \cdots a_n}^{\tilde{\Psi}\psi\tilde{\Psi} \cdots \psi}(p_1,p_2,p_3,\ldots,p_n) \;+ \\ \nonumber
    &\cdots\;+ \\ \nonumber
    &p_n^0 \Gamma_{a_1 a_2 \cdots a_{n-1} a_n}^{\tilde{\Psi}\psi \cdots \psi\tilde{\Psi}}(p_1,p_2,\ldots,p_{n-1},p_n) \big] \,,
\end{align}
and
\begin{align}
    \Gamma^{\tilde{\Psi} \tilde{\Psi} \psi \cdots \psi}_{a_1 a_2 a_3 \cdots a_n}(p_1,p_2,p_3,\ldots,p_n) &= \hat{\epsilon} \, \Gamma^{\tilde{\Psi} \tilde{\Psi} \psi \cdots \psi}_{a_1 a_2 a_3 \cdots a_n}(-p_1,-p_2,-p_3,\ldots,-p_n) \,, \label{eq:conditionAnomVert}
\end{align}
where we introduced the shorthand notation $\hat{\epsilon} \equiv \epsilon_{a_1} \cdots \epsilon_{a_n}$ for the product of the time-reversal parities of the involved fields.

\section{Details on the FRG for systems with reversible mode couplings}
\label{sec:detailsOnFRG}

In this Appendix, we discuss more details about our formulation of the FRG for dynamic systems with reversible mode couplings.
In subsection~\ref{derivationfloweq} we first reiterate the derivation of the flow equation of the effective average action in the spirit of the well-known derivation by Wetterich \cite{Wetterich:1992yh}.
A central aspect of our approach is that the regulator couples to the \emph{composite} response fields $\tilde{\Psi} = J(\psi)\tilde{\psi}$, so it is a priori not entirely obvious that the effective average action $\Gamma_k$ actually converges to the bare MSR action $S$ in the UV limit $k \to \Lambda$.
Therefore, we show in subsection~\ref{convergenceofeffectiveaction} that the effective average action $\Gamma_k$ indeed converges to the bare MSR action $S$, up to an additional logarithm of a field-dependent (Jacobian) determinant, which we identify with the one in \eqref{eq:genFncAfterFieldTrafo}.
In subsection \ref{sct:ghostRegs} we discuss how the regulators for the ghosts should be chosen such that the BRST transformation \eqref{eq:BRSTTrafo} is a symmetry of the effective average action at all FRG scales.
In subsection \ref{sct:vertexFunctions} we list the various 1PI vertex functions extracted from our specific truncation \eqref{effAvgActionCompactNewFields} of the effective average action, which are needed for the flow of the kinetic coefficients in Sec.~\ref{sct:flowOfKineticCoeffs}.
In subsection \ref{sct:CorrespondenceRealTimeAndEuclFlows} we explicitly show that the flow of the FRG-scale-dependent free energy, extracted via \eqref{eq:defOfFk} from the effective average MSR action, itself satisfies the closed $d$-dimensional Euclidean flow equation \eqref{eq:staticFlowEq}.

\subsection{Derivation of the flow equation}\label{derivationfloweq}

For a concise derivation of the flow equation, we first define the effective action $\tilde{\Gamma}_k$ as the (unmodified) Legendre transform (with the sources $J=(H,\mathcal{H})$ and $\tilde{J}=(\tilde{H},\tilde{\mathcal{H}})$ in superfield space)
\begin{equation}
    \tilde{\Gamma}_{k}[\psi,\tilde{\Psi}]=W_k[J,\tilde{J}]- \int_x J_i(x)\tilde{\Psi}_i(x)-\int_x \tilde{J}_i(x)\psi_i(x) \label{eq:unmodifiedLegendreTrafo}
\end{equation}
such that the effective \emph{average} action $\Gamma_k$ is given by an additional subtraction of the regulator term, $\Gamma_k = \tilde{\Gamma}_k - \Delta S_k$.
By differentiating the Legendre transform \eqref{eq:unmodifiedLegendreTrafo} with respect to the FRG scale~$k$, we calculate
\begin{align}
    \partial_k \tilde{\Gamma}_{k|\tilde{\Psi}, \psi}&=\partial_k W_{k|J,\tilde{J}}+ \int_x \frac{\delta W_k}{\delta J_i(x)}\partial_k J_i(x)_{|\tilde{\Psi}, \psi}+\int_x \frac{\delta W_k}{\delta \tilde{J}_i(x)}\partial_k \tilde{J}_i(x)_{|\tilde{\Psi}, \psi} \nonumber \\ 
    &\hspace{1.0cm} - \int_x \tilde{\Psi}_i(x)\partial_k J_i(x)_{|\tilde{\Psi}, \psi}-\int_x \psi_i(x)\partial_k \tilde{J}_i(x)_{|\tilde{\Psi}, \psi}\\& =\partial_k W_{k|J,\tilde{J}} \label{1}
\end{align}
where we used that the field expectation values are given by
\begin{align}
    \psi_i(x) = \frac{\delta W_k}{\delta \tilde{J}_i(x)} \;\;\text{and}\;\; \tilde{\Psi}_i(x) = \frac{\delta W_k}{\delta J_i(x)} \,.
\end{align}
Next, we want to show that
\begin{equation}
    \partial_k W_k[J,\tilde{J}]=\partial_k \langle \Delta S_k\rangle \label{2}
\end{equation}
which is straightforward since
\begin{align}
    \partial_kW_k[J,\tilde{J}]&=-i\partial_k \log\int \mathcal{D}\psi\,\mathcal{D}\tilde{\Psi}\, \exp(iS+i\Delta S_k + i\int_x J_i(x)\tilde{\Psi}_i(x)+i\int_x \tilde{J}_i(x)\psi_i(x)) \nonumber \\
    &=\frac{1}{Z_k}\int \mathcal{D}\psi\, \mathcal{D}\tilde{\Psi}\, \exp(iS+i\Delta S_k + i\int_x J_i(x)\tilde{\Psi}_i(x)+i\int_x\tilde{J}_i(x)\psi_i(x))\partial_k \Delta S_k\nonumber \\
    &=\partial_k \langle \Delta S_k\rangle \,.
\end{align}
Using the superfield notation, the regulator can be written as
\begin{equation}
    \Delta S_k= -\frac{1}{2}\int_{xy} \left( \tilde{\Psi}_i(x)R_{ij,k}(x,y)\psi_j(y) + \psi_i(x)R_{ij}(x,y)\tilde{\Psi}_j(y) \right)
\end{equation}
Combining Eq.~\eqref{1} and Eq.~\eqref{2}, we find
\begin{equation}
    \partial_k \tilde{\Gamma}_k = - \frac{1}{2} \int_{xy} (\partial_k R_{ij,k}(x,y)) \left( \langle \tilde{\Psi}_i(x)\psi_j(y)\rangle + \langle \psi_i(x)\tilde{\Psi}_j(y)\rangle \right) \,.
\end{equation}
Using the definitions 
\begin{subequations}
\begin{align}
    -iG^R_{ij,k}(x,y) &= \langle \psi_i(x)\tilde{\Psi}_j(y)\rangle-\langle \psi_i(x)\rangle\langle \tilde{\Psi}_j(y)\rangle \,, \\
    -iG^A_{ij,k}(x,y) &= \langle \tilde{\Psi}_i(x)\psi_j(y)\rangle-\langle \tilde{\Psi}_i(x)\rangle\langle \psi_j(y)\rangle \,,
\end{align} \label{eq:GRandGADefs}%
\end{subequations}
of the connected retarded and advanced 2-point correlation functions,
one can write
\begin{align}
    \partial_k \tilde{\Gamma}_k&= - \frac{1}{2} \int_{xy} (\partial_k R_{ij,k}(x,y)) \big(-iG^A_{ij,k}(x,y) -iG^R_{ij,k}(x,y) \\ \nonumber
    &\hspace{4.0cm} +\langle \tilde{\Psi}_i(x)\rangle\langle\psi_j(y)\rangle+\langle \psi_i(x)\rangle\langle\tilde{\Psi}_j(y)\rangle \big) \,.
\end{align}
Using $\Gamma_k = \tilde{\Gamma}_k-\Delta S_k$, we obtain
\begin{equation}
\begin{split}
    \partial_k \Gamma_k&=\frac{i}{2}(\partial_k R_{ij,k}(x,y)) \left( G^R_{ij,k}(x,y) + G^A_{ij,k}(x,y) \right) \label{eq:flowEqWithGRandGA}
\end{split}
\end{equation}
for the flow of the effective average action.
Moreover, using the symmetry relation
\begin{equation}
    G^R_{ij,k}(x,y) = G^A_{ji,k}(y,x)
\end{equation}
which can be seen directly from their definition \eqref{eq:GRandGADefs},
and by combining the retarded, advanced, statistical, and anomalous propagators into a $2\times 2$-matrix in the MSR $(\psi,\tilde{\Psi})$ space according to (where the subscript $_c$ denotes the connected part of the correlation function)
\begin{equation}
    G_k[\psi,\tilde{\Psi}] =
    \begin{pmatrix}
        i\langle\psi_i(x) \psi_j(y)\rangle_c & i\langle\psi_i(x) \tilde{\Psi}_j(y)\rangle_c \\
        i\langle\tilde{\Psi}_i(x) \psi_j(y)\rangle_c & i\langle\tilde{\Psi}_i(x) \tilde{\Psi}_j(y)\rangle_c
    \end{pmatrix} =
    \begin{pmatrix}
        iF_{ij,k}(x,y) & G^R_{ij,k}(x,y) \\
        G^A_{ij,k}(x,y) & i\widetilde{F}_{ij,k}(x,y)
    \end{pmatrix} \,,
\end{equation}
and analogously the regulator as a $2\times 2$ matrix in the same MSR space,
\begin{equation}
    \hat{R}_k = \begin{pmatrix}
        0 & R_{ij,k}(x,y) \\
        R_{ij,k}(x,y) & 0 \,,
    \end{pmatrix}
\end{equation}
we can compactly express the flow equation \eqref{eq:flowEqWithGRandGA} as
\begin{align}
    \partial_k \Gamma_k &=\frac{i}{2} \tr \left\{ \partial_k R_k \circ G^R_k + \partial_k R_k \circ G^A_k \right\} = \frac{i}{2} \tr \left\{ \partial_k \hat{R}_k \circ G_k \right\}  \label{eq:treeLevelFlowEqResult}
\end{align}
with the trace
\begin{equation}
    \tr \left\{ A \circ B \right\} = \int_{xy} A_{ij}(x,y) B_{ji}(y,x) 
\end{equation}
acting in superfield and coordinate space.
This finalizes the derivation of the tree-level flow equation~\eqref{eq:treeLevelFlowEq}. 

\subsection{Convergence of the MSR effective average action in the UV limit $k\to\Lambda$}
\label{convergenceofeffectiveaction}
In this subsection we show that the effective average action $\Gamma_k$ indeed converges to the bare MSR action $S$ (up to a Jacobian determinant), even if the regulator is coupled to the composite response fields $\tilde{\Psi}$ instead of the usual elementary response fields $\tilde{\psi}$. This is not entirely trivial, since one has to make sure that all fluctuations are properly suppressed in the limit $k\to\Lambda$ \cite{Pawlowski:2005xe}. We start from the background field identity,
\begin{align}
    \Gamma_k[\psi,\tilde{\Psi}] = S'[\psi,\tilde{\Psi}] - i \log \Delta Z_k
\end{align}
with $S'[\psi,\tilde{\Psi}] \equiv S[\psi,\tilde{\psi}]$ and
\begin{align}
    \Delta Z_k = \int \mathcal{D}\psi'\, \mathcal{D}\tilde{\psi}'\,\exp\bigg\{ &i(S[\psi',\tilde{\psi}'] - S[\psi,\tilde{\psi}]) \label{eq:DelZk1} \\ \nonumber 
    -\, &i\left( \frac{\delta \Gamma_k}{\delta \psi},\psi'-\psi \right) - i\left( \frac{\delta \Gamma_k}{\delta \tilde{\Psi}},\tilde{\Psi}'-\tilde{\Psi} \right)  \\ \nonumber 
    -\, &i\left( \frac{\delta \Delta S_k}{\delta \psi},\psi'-\psi \right) - i\left( \frac{\delta \Delta S_k}{\delta \tilde{\Psi}},\tilde{\Psi}'-\tilde{\Psi} \right) \\ \nonumber 
    +\, &i(\Delta S_k[\psi',\tilde{\Psi}'] - \Delta S_k[\psi,\tilde{\Psi}])\bigg\} 
\end{align}
where $(\cdot,\cdot)$ here denotes the inner product
\begin{align}
    (A,B) \equiv \int_x A_i(x)B_i(x) \,.
\end{align}
We now do a variable substitution in the path integral to the new (composite) response fields, $\tilde{\psi}' \to \tilde{\Psi}'$,
\begin{align}
    \mathcal{D} \tilde{\psi}' = \bigg| \det \frac{\delta \tilde{\psi}'}{\delta \tilde{\Psi}'} \bigg| \, \mathcal{D}\tilde{\Psi}'
\end{align}
and perform a subsequent shift in the path integral, $\psi' \to \psi+\psi'$, $\tilde{\Psi}' \to \tilde{\Psi} + \tilde{\Psi}'$, such that afterwards $\psi'$ and $\tilde{\Psi}'$ describe the fluctuations around the expectation values $\psi$ and $\tilde{\Psi}$,
\begin{align}
    \Delta Z_k = \int \mathcal{D}\psi'\,\mathcal{D}\tilde{\Psi}'\, &\left| \det \frac{\delta \tilde{\psi}'}{\delta \tilde{\Psi}'}(\psi+\psi',\tilde{\Psi}+\tilde{\Psi}') \right| \; \times  \label{eq:DelZk2} \\  \nonumber
    \exp\bigg\{ &i(S'[\psi+\psi',\tilde{\Psi}+\tilde{\Psi}'] - S'[\psi,\tilde{\Psi}])  \\   \nonumber
    &- i\left( \frac{\delta \Gamma_k}{\delta \psi},\psi' \right) - i\left( \frac{\delta \Gamma_k}{\delta \tilde{\Psi}},\tilde{\Psi}' \right) + i\Delta S_k[\psi',\tilde{\Psi}'] \bigg\}
\end{align}
where we used that the regulator is quadratic in the (composite) fields.
We can rewrite \eqref{eq:DelZk2} as
\begin{align}
    \Delta Z_k = \exp\bigg\{ i &\left(S'\bigg[\psi-i\frac{\delta}{\delta \tilde{j}},\tilde{\Psi}-i\frac{\delta}{\delta j}\bigg] - S'[\psi,\tilde{\Psi}] \right) \\ \nonumber 
    - &\left( \frac{\delta \Gamma_k}{\delta \psi}, \frac{\delta}{\delta \tilde{j}} \right) - \left( \frac{\delta \Gamma_k}{\delta \tilde{\Psi}}, \frac{\delta}{\delta j} \right) \bigg\}\, \bigg| \det \frac{\delta \tilde{\psi}}{\delta \tilde{\Psi}} \bigg(\psi-i\frac{\delta}{\delta \tilde{j}},\tilde{\Psi}-i\frac{\delta}{\delta j}\bigg) \bigg|  \\  \nonumber
    &\hspace{0.5cm} \int \mathcal{D}(\psi',\tilde{\Psi}')~\exp\bigg\{ i\Delta S_k[\psi',\tilde{\Psi}'] + i(\tilde{j},\psi') + i(j,\tilde{\Psi}') \bigg\} \bigg\rvert_{j=\tilde{j}=0} \,. \label{eq:DelZk3}
\end{align}
Due to the properties of the regulator term $\Delta S_k$, one can now straightforwardly verify that in the UV limit $k\to \Lambda$ the remaining integral becomes independent of sources $j,\tilde{j}$ in the vicinity of $j=\tilde{j}=0$ where the functional derivatives are evaluated.
By taking the limit $k\to\Lambda$ of \eqref{eq:DelZk3} we finally arrive at
\begin{align}
    \Gamma_k[\psi,\tilde{\Psi}] \xrightarrow{k \to \Lambda} S'[\psi,\tilde{\Psi}] - i \log \bigg| \det \frac{\delta \tilde{\psi}}{\delta \tilde{\Psi}} \bigg| + \text{const.}
\end{align}
which coincides with \eqref{effAvgActionUVConvergence}.
(Because the transformation $\tilde{\Psi} = J(\psi)\tilde{\psi}$ is linear in the usual response fields $\tilde{\psi}$, the remaining determinant actually only depends on the classical fields $\psi$, not on the response fields $\tilde{\Psi}$.)

This result can be compared with 
Eq.~\eqref{eq:genFncAfterFieldTrafo}, where the non-linear field transformation is performed on the level of the generating functional.
In this case, the regulator term is actually quadratic in the \emph{transformed} fields $(\psi,\tilde{\Psi})$, such that the effective average action converges to the `bare' MSR action $\Gamma_k \to S' - i\log \mathcal{J}'[\psi]$, but the `bare' MSR action then contains the logarithm of the Jacobian determinant anyway.
This highlights the equivalence of the two approaches of either $(i)$ keeping the `standard' response fields $\tilde{\psi}$ as the fields in the path integral, and coupling the regulator to the composite response fields $\tilde{\Psi} = J(\psi)\tilde{\psi}$ which we followed in the present subsection, or $(ii)$ performing the non-linear field transformation on the level of the path integral, such that the regulator is quadratic in $(\psi,\tilde{\Psi})$.
In both cases, one obtains an additional logarithm of a Jacobian determinant in the action, which can be practically handled by rewriting the determinant as an integral over Grassmann-valued ghost fields, as we have already discussed in more detail in Appendix~\ref{sct:jacobianAndGhosts} above.

\subsection{Regulators for ghosts} \label{sct:ghostRegs}
We need to ensure that the normalization condition $Z_k[J]\equiv Z_k[J,\tilde{J}=0]=1$ is satisfied at all FRG scales $k$. A sufficient condition for this is that the BRST transformation~\eqref{eq:BRSTTrafo} is a symmetry of the scale-dependent effective average action $\Gamma_k$ at all FRG scales $k$. This can be achieved by choosing $\Delta S_k$ to be BRST exact,
\begin{align}
    \Delta S_k &= -\tilde{\Psi}_i R_{ij,k} \psi_j + i \tilde{c}_i R_{ij,k} c_j = Q\left( -i\tilde{c}_i R_{ij,k} \psi_j \right) \,,
\end{align}
which implies $Q( \Delta S_k ) = 0$ due to the nilpotency $Q^2=0$ of the BRST transformation.
In particular, introducing a regulator for the ghost fields ensures that the flow of acausal two-point functions stays zero. Due to causality, two-point functions such as $\Gamma_k^{\psi\psi}$ should be zero at all FRG scales. However, for a general truncation of $\Gamma_k$, there might appear non-vanishing diagrams on the right-hand sides of the flow equations for these acausal two-point functions. This will not cause any problem because one can show that diagrams containing ghost fields in the loop will cancel exactly with the appearing acausal diagrams, which ensures that the flow of these acausal two-point functions vanishes. The relation between the ghost fields and causality is also discussed for example in Refs.~\cite{Canet:2011wf,Marguet:2021gab,Crossley:2015evo,Glorioso:2017fpd,Gao:2018bxz}.
In fact, the ghosts \emph{solely} appear here to cancel contributions from acausal diagrams. 
As such, one can simply omit all ghost diagrams and all acausal diagrams, since the two precisely cancel.  This is analogous to the cancellation scheme of acausal diagrams known from perturbation theory \cite{Canet:2011wf}, but here formulated for the FRG flow.

\subsection{Vertex functions}
\label{sct:vertexFunctions}

In this subsection we list the explicit expressions for the 1PI vertex functions that occur in the diagrams in Eqs.~\eqref{eq:feynDiagPhiPhi1} -- \eqref{eq:feynDiagNN4}, extracted from our truncation of the effective average action \eqref{effAvgActionCompactNewFields} by expanding the Jacobian \eqref{JExpansion} in a Neumann series and subsequently evaluating suitable functional derivatives on the expanded terms.
We have verified our analytical calculation explicitly using \emph{DoFun}~\cite{Huber:2019dkb} in special cases.

\noindent
The relevant `classical' vertices (i.e.~those with one response field) are given by:
\begin{align}
    \Gamma_k^{\phi_a \phi_b \tilde{N}_{cd}}(p,q,r) &=g_k \frac{iZ_{\phi,k}^\omega (q^0-p^0)}{\gamma_{\phi,k}\,\gamma_{n,k}(\vec r)} ~  (\delta_{ac}\delta_{bd} - \delta_{ad}\delta_{bc}) ~ (2\pi)^D \delta(p+q+r)
\end{align}
\begin{align}
    \Gamma_k^{\phi_a \tilde{\Phi}_b n_{cd}}(p,q,r) &= -g_k \frac{iZ_{n,k}^\omega r^0}{\gamma_{\phi,k}\,\gamma_{n,k}(\vec r)} ~ (\delta_{ac}\delta_{bd} - \delta_{ad}\delta_{bc}) ~ (2\pi)^D \delta(p+q+r)
\end{align}
\begin{align}
    \Gamma_k^{n_{ab} n_{cd} \tilde{N}_{ef}}(p,q,r) =   \frac{g_k}{\,\gamma_{n,k}(\vec r)} \bigg[ &\frac{iZ_{n,k}^\omega p^0}{\gamma_{n,k}(\vec p)} (  \delta_{ae}\delta_{bc}\delta_{df} - \delta_{af}\delta_{bc}\delta_{de} - \delta_{ac}\delta_{be}\delta_{df} + \delta_{ac}\delta_{bf}\delta_{de} \nonumber \\
    &\phantom{\frac{iZ_{n,k}^\omega p^0}{\gamma_{n,k}(\vec p)} (} \hspace{-0.5cm} - \delta_{ae}\delta_{bd}\delta_{cf} + \delta_{af}\delta_{bd}\delta_{ce} + \delta_{ad}\delta_{be}\delta_{cf} - \delta_{ad}\delta_{bf}\delta_{ce} ) \nonumber \\
    +&\frac{iZ_{n,k}^\omega q^0}{\gamma_{n,k}(\vec q)} (  \delta_{ad}\delta_{bf}\delta_{ce} - \delta_{ad}\delta_{be}\delta_{cf} - \delta_{ac}\delta_{bf}\delta_{de} + \delta_{ac}\delta_{be}\delta_{df} \nonumber \\
    &\phantom{\frac{iZ_{n,k}^\omega q^0}{\gamma_{n,k}(\vec q)} (} \hspace{-0.5cm} - \delta_{bd}\delta_{af}\delta_{ce} + \delta_{bd}\delta_{ae}\delta_{cf} + \delta_{bc}\delta_{af}\delta_{de} -\delta_{bc}\delta_{ae}\delta_{df} ) \bigg] \times \nonumber \\
    &(2\pi)^D \delta(p+q+r)
\end{align}

\noindent
The relevant `anomalous' vertices (i.e.~those with two response fields) are given by:
\begin{align}
    \Gamma_k^{\phi_a \tilde{\Phi}_b \tilde{N}_{cd}}(p,q,r) &= 0 \hspace{0.5cm} \text{(by symmetry)}
\end{align}
\begin{align}
    \Gamma_k^{n_{ab} \tilde{N}_{cd} \tilde{N}_{ef}}(p,q,r) &= 0 \hspace{0.5cm} \text{(by symmetry)} 
\end{align}
\begin{align}
    &\Gamma_k^{\phi_a \phi_b \tilde{\Phi}_c \tilde{\Phi}_d}(p,q,r,s) = (2Z_{\phi,k}^{\omega}-Z_{n,k}^{\omega})  \frac{2g_k^2 iT}{\gamma_{\phi,k}^2} \times \nonumber \\
    &\hspace{0.8cm} \bigg[ \delta_{ac}\delta_{bd} \frac{1}{\gamma_{n,k}(\vec p+\vec r)} + \delta_{ad}\delta_{bc} \frac{1}{\gamma_{n,k}(\vec p+\vec s)}
     - \delta_{ab}\delta_{cd}\left( \frac{1}{\gamma_{n,k}(\vec p+\vec r)} + \frac{1}{\gamma_{n,k}(\vec p+\vec s)}  \right) \bigg] \times \nonumber \\
     &\hspace{1.0cm} (2\pi)^D \delta(p+q+r+s)  
\end{align}
\begin{align}
    \Gamma_k^{\phi_a \phi_b \tilde{N}_{cd} \tilde{N}_{ef}}(p,q,r,s) &=  (2Z_{n,k}^{\omega}-Z_{\phi,k}^{\omega})  \frac{2 g_k^2 Z_{n,k}^\omega iT}{\gamma_{\phi,k}\,\gamma_{n,k}(\vec r)\,\gamma_{n,k}(\vec s)} \times \nonumber \\  
    &\hspace{0.8cm} \Big[  + \delta_{ad}\delta_{be}\delta_{cf}+\delta_{bd}\delta_{ae}\delta_{cf} - \delta_{ad}\delta_{ce}\delta_{bf} - \delta_{bd}\delta_{ce}\delta_{af} \nonumber \\  
    &\hspace{1.0cm} - \delta_{ac}\delta_{be}\delta_{df} - \delta_{bc}\delta_{ae}\delta_{df} + \delta_{ac}\delta_{de}\delta_{bf} + \delta_{bc}\delta_{de}\delta_{af} \Big] \times  \nonumber \\
    &\hspace{1.0cm} (2\pi)^D \delta(p+q+r+s)
\end{align}
\begin{align}
    \Gamma_k^{n_{gh} n_{ij} \tilde{N}_{kl} \tilde{N}_{mn}}(p,q,r,s) &= \frac{2 g_k^2 Z_{n,k}^\omega iT}{\gamma_{n,k}(\vec s) \, \gamma_{n,k}(\vec r)} \times \nonumber \\
    \hspace{0.8cm}\bigg[  \frac{1}{\gamma_{n,k}(\vec p+\vec r)} \big(
    +\; &\delta _{g n} \delta _{h l} \delta _{i m} \delta _{j k}-\delta _{g n} \delta _{h k} \delta _{i m} \delta _{j l}-\delta _{g
   n} \delta _{h l} \delta _{i k} \delta _{j m}+\delta _{g n} \delta _{h k} \delta _{i l} \delta _{j m} \nonumber \\[-0.7em]
   -\; &\delta _{g l} \delta
   _{h n} \delta _{i m} \delta _{j k}+\delta _{g k} \delta _{h n} \delta _{i m} \delta _{j l}+\delta _{g l} \delta _{h n}
   \delta _{i k} \delta _{j m}-\delta _{g k} \delta _{h n} \delta _{i l} \delta _{j m}  \nonumber \\
   -\; &\delta _{g m} \delta _{h l} \delta
   _{i n} \delta _{j k}+\delta _{g m} \delta _{h k} \delta _{i n} \delta _{j l}+\delta _{g l} \delta _{h m} \delta _{i n}
   \delta _{j k}-\delta _{g k} \delta _{h m} \delta _{i n} \delta _{j l}  \nonumber \\ 
   -\; &\delta _{g l} \delta _{h j} \delta _{i n} \delta
   _{k m}+\delta _{g j} \delta _{h l} \delta _{i n} \delta _{k m}+\delta _{g k} \delta _{h j} \delta _{i n} \delta _{l
   m}-\delta _{g j} \delta _{h k} \delta _{i n} \delta _{l m}  \nonumber \\ 
   +\;&\delta _{g m} \delta _{h l} \delta _{i k} \delta _{j n}-\delta
   _{g m} \delta _{h k} \delta _{i l} \delta _{j n}-\delta _{g l} \delta _{h m} \delta _{i k} \delta _{j n}+\delta _{g k}
   \delta _{h m} \delta _{i l} \delta _{j n}  \nonumber \\ 
   +\;&\delta _{g l} \delta _{h i} \delta _{j n} \delta _{k m}-\delta _{g i} \delta
   _{h l} \delta _{j n} \delta _{k m}-\delta _{g k} \delta _{h i} \delta _{j n} \delta _{l m}+\delta _{g i} \delta _{h k}
   \delta _{j n} \delta _{l m}  \nonumber \\ 
   +\;&\delta _{g l} \delta _{h j} \delta _{i m} \delta _{k n}-\delta _{g j} \delta _{h l} \delta
   _{i m} \delta _{k n}-\delta _{g l} \delta _{h i} \delta _{j m} \delta _{k n}+\delta _{g i} \delta _{h l} \delta _{j m}
   \delta _{k n}  \nonumber \\
   -\;&\delta _{g k} \delta _{h j} \delta _{i m} \delta _{l n}+\delta _{g j} \delta _{h k} \delta _{i m} \delta
   _{l n}+\delta _{g k} \delta _{h i} \delta _{j m} \delta _{l n}-\delta _{g i} \delta _{h k} \delta _{j m} \delta _{l n} \big) \; + \nonumber \\
     \frac{1}{\gamma_{n,k}(\vec q+\vec r)} \big( 
    +\;&\delta _{g n} \delta _{h l} \delta _{i m} \delta _{j k}-\delta _{g n} \delta _{h k} \delta _{i m} \delta _{j l}-\delta _{g
   n} \delta _{h l} \delta _{i k} \delta _{j m}+\delta _{g n} \delta _{h k} \delta _{i l} \delta _{j m}   \nonumber \\[-0.7em]
   -\;&\delta _{g n} \delta
   _{h j} \delta _{i l} \delta _{k m}+\delta _{g n} \delta _{h i} \delta _{j l} \delta _{k m}+\delta _{g n} \delta _{h j}
   \delta _{i k} \delta _{l m}-\delta _{g n} \delta _{h i} \delta _{j k} \delta _{l m}   \nonumber \\  
   -\;&\delta _{g l} \delta _{h n} \delta
   _{i m} \delta _{j k}+\delta _{g k} \delta _{h n} \delta _{i m} \delta _{j l}+\delta _{g l} \delta _{h n} \delta _{i k}
   \delta _{j m}-\delta _{g k} \delta _{h n} \delta _{i l} \delta _{j m}   \nonumber \\  
   +\;&\delta _{g j} \delta _{h n} \delta _{i l} \delta
   _{k m}-\delta _{g i} \delta _{h n} \delta _{j l} \delta _{k m}-\delta _{g j} \delta _{h n} \delta _{i k} \delta _{l
   m}+\delta _{g i} \delta _{h n} \delta _{j k} \delta _{l m}   \nonumber \\ 
   -\;&\delta _{g m} \delta _{h l} \delta _{i n} \delta _{j k}+\delta
   _{g m} \delta _{h k} \delta _{i n} \delta _{j l}+\delta _{g l} \delta _{h m} \delta _{i n} \delta _{j k}-\delta _{g k}
   \delta _{h m} \delta _{i n} \delta _{j l}   \nonumber \\ 
   +\;&\delta _{g m} \delta _{h l} \delta _{i k} \delta _{j n}-\delta _{g m} \delta
   _{h k} \delta _{i l} \delta _{j n}-\delta _{g l} \delta _{h m} \delta _{i k} \delta _{j n}+\delta _{g k} \delta _{h m}
   \delta _{i l} \delta _{j n}   \nonumber \\ 
   +\;&\delta _{g m} \delta _{h j} \delta _{i l} \delta _{k n}-\delta _{g m} \delta _{h i} \delta
   _{j l} \delta _{k n}-\delta _{g j} \delta _{h m} \delta _{i l} \delta _{k n}+\delta _{g i} \delta _{h m} \delta _{j l}
   \delta _{k n}   \nonumber \\ 
   -\;&\delta _{g m} \delta _{h j} \delta _{i k} \delta _{l n}+\delta _{g m} \delta _{h i} \delta _{j k} \delta
   _{l n}+\delta _{g j} \delta _{h m} \delta _{i k} \delta _{l n}-\delta _{g i} \delta _{h m} \delta _{j k} \delta _{l n}
    \big)  \bigg] \times \nonumber \\
    &(2\pi)^D \delta(p+q+r+s)
\end{align}

\subsection{Proof of correspondence between real-time and Euclidean flows}
\label{sct:CorrespondenceRealTimeAndEuclFlows}
The goal of this subsection is to show that the flow of the free energy $F_k$, extracted via \eqref{eq:defOfFk} from the effective average MSR action $\Gamma_k$, is governed by the closed (Euclidean) flow equation \eqref{eq:staticFlowEq}.

Having the identification \eqref{eq:defOfFk} in mind, the general flow equation for the first functional derivative of the free energy with general non-local 3-point vertices is given by (with $a,b,c,\ldots$ generally denoting indices in superfield space during this section)
\begin{align}
    \partial_k \frac{\delta \Gamma_k}{\delta \tilde{\Psi}_c(x)} \bigg\rvert_{\tilde{\Psi}=0}  &= 
	-\frac{i}{2} \int_{yz} \bigg(
	\Gamma_{acb,k}^{\psi \tilde{\Psi} \psi}(y,x,z) \, B_{ba,k}^F(z,y) \; + \label{dGamdPhiQ} \\ \nonumber
	&\hspace{1em} \Gamma_{acb,k}^{\tilde{\Psi}\tilde{\Psi}\psi}(y,x,z) \, B_{ba,k}^R(z,y) +
	\Gamma_{acb,k}^{\psi\tilde{\Psi}\tilde{\Psi}}(y,x,z) \, B_{ba,k}^A(z,y) \bigg) \, .
\end{align}
On the right-hand side, the propagators and $n$-point function are evaluated for a classical field configuration $\psi = \psi(x) = \psi(\vec{x})$ independent of time, and vanishing response fields $\tilde{\Psi} = 0$.
We can do a Fourier transform of the 3-point function and the propagators such that \eqref{dGamdPhiQ} can be expressed as
\begin{align}
	\label{dGamdPhiQFourierTransform}
	\partial_k \frac{\delta \Gamma_k}{\delta \tilde{\Psi}_c(x)} \bigg\rvert_{\tilde{\Psi}=0} &=
	-\frac{i}{2} \int_{pq} e^{-iq \cdot x} \bigg(
	\Gamma_{acb,k}^{\psi\tilde{\Psi}\psi}(p,q,-p) \, B_{ba,k}^F(p) \; + \\ \nonumber
	&\hspace{4em} \Gamma_{acb,k}^{\tilde{\Psi}\tilde{\Psi}\psi}(p,q,-p) \, B_{ba,k}^R(p) +
	\Gamma_{acb,k}^{\psi\tilde{\Psi}\tilde{\Psi}}(p,q,-p) \, B_{ba,k}^A(p) \bigg) \, .
\end{align}
Since the system is in thermal equilibrium, we now employ fluctuation-dissipation relations for the propagators and the three-point functions.
For the $B$-propagators \eqref{eq:BProps}, the FDR can be straightforwardly derived from~\eqref{eq:FDRForPropagators} and reads
\begin{equation}
    B_{ab,k}^F(p) = F(p^0)  \left( B_{ab,k}^R(p) - B_{ab,k}^A(p) \right)\,, \label{eq:FDRForBProps}
\end{equation}
where $F(p^0) \equiv T/p^0$ denotes the Rayleigh-Jeans distribution.
For three-point functions, the FDR is given by 
\begin{align}
	\Gamma_{abc,k}^{\psi\tilde{\Psi}\tilde{\Psi}} &= F(p_2^0) \frac{1+\epsilon_a \epsilon_b \epsilon_c}{2}\left( {\Gamma_{abc,k}^{\tilde{\Psi}\psi\psi}}^* - \Gamma_{abc,k}^{\psi\psi\tilde{\Psi}} \right) + F(p_3^0) \frac{1+\epsilon_a \epsilon_b \epsilon_c}{2}\left( {\Gamma_{abc,k}^{\tilde{\Psi}\psi\psi}}^* - \Gamma_{abc,k}^{\psi\tilde{\Psi}\psi} \right) \, , \label{eq:FDR3PtFnc1} \\
	\Gamma_{abc,k}^{\tilde{\Psi}\tilde{\Psi}\psi} &= F(p_1^0) \frac{1+\epsilon_a \epsilon_b \epsilon_c}{2}\left( {\Gamma_{abc,k}^{\psi\psi\tilde{\Psi}}}^* - \Gamma_{abc,k}^{\psi\tilde{\Psi}\psi} \right) + F(p_2^0) \frac{1+\epsilon_a \epsilon_b \epsilon_c}{2}\left( {\Gamma_{abc,k}^{\psi\psi\tilde{\Psi}}}^* - \Gamma_{abc,k}^{\tilde{\Psi}\psi\psi} \right) \, , \label{eq:FDR3PtFnc2}
\end{align}
and which can be shown to be equivalent to the three-point function case of Eq.~\eqref{eq:conditionClassVert} and Eq.~\eqref{eq:conditionAnomVert}.
Inserting the FDR's \eqref{eq:FDR3PtFnc1} and \eqref{eq:FDR3PtFnc2} for the 3-point function and the FDR \eqref{eq:FDRForBProps} for the $B$-propagators 
into \eqref{dGamdPhiQFourierTransform}, we find
\begin{align}
	\partial_k \frac{\delta \Gamma_k}{\delta \tilde{\Psi}_c(x)} \bigg\rvert_{\tilde{\Psi}=0} &=
	-\frac{i}{2} \int_{pq}  e^{-iq \cdot x} \times \label{eq:flowOfPhiTildeDerivativeOfGammaWithFDRs} \\ \nonumber \sum_{ab} \bigg\{
	&F(p^0)\frac{1+\epsilon_a\epsilon_b\epsilon_c}{2} \, \Gamma_{acb,k}^{\psi\tilde{\Psi}\psi}(p,q,-p) \, \left( \red{ B_{ba,k}^R(p) } \blue{-  B_{ba,k}^A(p) } \right) \; + \\ \nonumber
    &F(p^0)\frac{1-\epsilon_a\epsilon_b\epsilon_c}{2} \, \Gamma_{acb,k}^{\psi\tilde{\Psi}\psi}(p,q,-p) \, \left(  B_{ba,k}^R(p)-  B_{ba,k}^A(p) \right) \; + \\ \nonumber
	&F(p^0)\frac{1+\epsilon_a\epsilon_b\epsilon_c}{2} \left( {\Gamma_{acb,k}^{\psi\psi\tilde{\Psi}}}^*(p,q,-p)\red{  - \Gamma_{acb,k}^{\psi\tilde{\Psi}\psi}(p,q,-p) } \right) \, B_{ba,k}^R(p) \; + \\ \nonumber
	&\violet{ F(q^0) \frac{1+\epsilon_a\epsilon_b\epsilon_c}{2}\left( {\Gamma_{acb,k}^{\psi\psi\tilde{\Psi}}}^*(p,q,-p) - \Gamma_{acb,k}^{\tilde{\Psi}\psi\psi}(p,q,-p) \right) \, B_{ba,k}^R(p) } \; + \\ \nonumber
	&\violet{ F(q^0)\frac{1+\epsilon_a\epsilon_b\epsilon_c}{2} \left( {\Gamma_{acb,k}^{\tilde{\Psi}\psi\psi}}^*(p,q,-p) - \Gamma_{acb,k}^{\psi\psi\tilde{\Psi}}(p,q,-p) \right) \, B_{ba,k}^A(p) } \; + \\ \nonumber
	&F(-p^0)\frac{1+\epsilon_a\epsilon_b\epsilon_c}{2} \left( {\Gamma_{acb,k}^{\tilde{\Psi}\psi\psi}}^*(p,q,-p) \blue{ - \Gamma_{acb,k}^{\psi\tilde{\Psi}\psi}(p,q,-p) } \right) \, B_{ba,k}^A(p)
	\bigg\} \, .
\end{align}
The two blue and the two red terms cancel each other, respectively.
The violet terms are retarded/advanced functions of $p$ and thus their $p^0$-integral vanishes.

To proceed further, we now need another ingredient for systems with reversible mode couplings, namely that propagators $G_{ab}^R(p)$ of two fields $\psi_a$ and $\psi_b$ vanish at zero frequency if the fields have different time-reversal parities, i.e.~$\epsilon_a \epsilon_b = -1$. To show this, we use that the propagators satisfy the Onsager symmetry relation \cite{Glorioso:2017fpd} 
\begin{align}
    G^R_{ab}(p)=\epsilon_a \epsilon_b G^R_{ba}(p) \,. \label{eq:OnsagerRel}
\end{align}
At zero frequency, the retarded and advanced propagator both have to be real, since $(G^{R/A}(\omega))^*=G^{R/A}(-\omega)$ and therefore $(G^{R/A}(0))^*=G^{R/A}(0)$. Using the FDR
$ G^R_{ab}(0)-G^A_{ab}(0)=\frac{\omega}{T}iF_{ab}(0)$, and knowing that $iF_{ab}(0)$ has to be imaginary, we obtain 
\begin{align}
   G^R_{ab}(0)=G^A_{ab}(0)=(G^R_{ba}(0))^*=G^R_{ba}(0) \,.
\end{align}
Combined with the Onsager relation \eqref{eq:OnsagerRel} at zero frequency, we get
\begin{align}
    G^R_{ab}(0)=G^A_{ab}(0)=G^R_{ba}(0)=G^A_{ba}(0)=0\,, \label{eq:mixedPropagatorsVanishAt0}
\end{align}
as desired.

We now discuss the cases $\epsilon_c=-1$ and $\epsilon_c=+1$ in \eqref{eq:flowOfPhiTildeDerivativeOfGammaWithFDRs} separately.

First, we discuss $\epsilon_c=-1$. In this case, the remaining terms with $(1+\epsilon_a\epsilon_b\epsilon_c)/2$ will only contribute when $\epsilon_a=-\epsilon_b$, which corresponds to mixed propagator in the flow equation. The $p^0$ integral can be evaluated using the residue theorem, where we only pick up the pole at $p^0=0$ from the classical distribution function $F(p^0) = T/p^0$.\footnote{In a derivative expansion of the effective average action based on timeline derivatives $\partial_t$, we know that the 3-point functions, being the third functional derivative of $\Gamma_k$, will be polynomial in their frequency arguments, and hence admit no additional poles.} Hence, upon integration over $p^0$, the mixed propagator will be evaluated at zero frequency, where they vanish according to Eq.~\eqref{eq:mixedPropagatorsVanishAt0}. This means that contributions from terms with the factor $(1+\epsilon_a\epsilon_b\epsilon_c)/2$ all vanish, in the case $\epsilon_c = -1$ here. The remaining terms with the factor $(1-\epsilon_a\epsilon_b\epsilon_c)/2$ will only contribute when $\epsilon_a=\epsilon_b$ for $\epsilon_c=-1$. After integrating over $p^0$, the three point vertices will be evaluated at $\Gamma_{acb,k}^{\psi\tilde{\Psi}\psi}(0,q,0)$, which will also be zero in our truncation. All together, the flow of $\frac{\delta \Gamma_k}{\delta \tilde{\Psi}_c(x)} $ in the case of $\epsilon_c=-1$ will be zero in our truncation, which indeed matches the Euclidean flow of $\frac{\delta F_k}{\delta n_{ab}(x)} $, since in the effective potential $n$ field only couple is quadratically and will not be renormalized.

Next, we discuss $\epsilon_c=+1$. For the term with $\frac{1-\epsilon_a\epsilon_b\epsilon_c}{2}$, it only contributes when there are mixed propagators. After performing the $p^0$ integration, the mixed propagator coming from the $\frac{1-\epsilon_a\epsilon_b\epsilon_c}{2}$ term will be evaluated at zero frequency, which will then also vanish.

Hence, we will now focus on the case $\epsilon_c=+1$, where $\tilde{\Psi}_c$ corresponds to a component of the order-parameter response field $\tilde{\Phi}$, and
\begin{align}
	\partial_k \frac{\delta \Gamma_k}{\delta \tilde{\Psi}_c(x)} \bigg\rvert_{\tilde{\Psi}=0} &=
	-\frac{i}{2} \int_{pq}  e^{-iq \cdot x} \times \\ \nonumber \sum_{ab} &F(p^0) \frac{1+\epsilon_a\epsilon_b}{2} \bigg\{
	{\Gamma_{acb,k}^{\psi\psi\tilde{\Psi}}}^*(p,q,-p) \, B_{ba,k}^R(p) - {\Gamma_{ac b,k}^{\tilde{\Psi}\psi\psi}}^*(p,q,-p) \, B_{ba,k}^A(p)\bigg\}
\end{align}
remains.
We can also drop the redundant factor $(1+\epsilon_a\epsilon_b)/2$. This is because, in the case of $\epsilon_a=\epsilon_b$, such a factor is equivalent to one, while in the mixed case of $\epsilon_a=-\epsilon_b$ the integrand vanishes anyway: Upon integration of $p^0$, the mixed propagators are evaluated at $p^0=0$, where again according to Eq.~\eqref{eq:mixedPropagatorsVanishAt0} they vanish anyway. Hence, the factor $(1+\epsilon_a\epsilon_b)/2$ is redundant here and can be safely omitted.

We get rid of the complex conjugation in the 3-point function by equivalently flipping all momentum arguments $p\to -p$, $q\to-q$ (since the $n$-point vertices are real-valued functions in the time domain),
\begin{align}
	\partial_k \frac{\delta \Gamma_k}{\delta \tilde{\Psi}_c(x)} \bigg\rvert_{\tilde{\Psi}=0} &=
	- \frac{i}{2} \int_{pq}  e^{-iq \cdot x} \times \\ \nonumber \sum_{ab} & F(p^0)  \bigg\{  {\Gamma_{acb,k}^{\psi\psi\tilde{\Psi}}}(-p,-q,p) \, B_{ba,k}^R(p)- {\Gamma_{acb,k}^{\tilde{\Psi}\psi\psi}}(-p,-q,p) \, B_{ba,k}^A(p) \bigg\} \, ,
\end{align}
Using the symmetry
\begin{equation}
    B^{R}_{ba,k}(-p) = B^{A}_{ab,k}(p) = \epsilon_a \epsilon_b B^{A}_{ba,k}(p)
\end{equation}
where the second equality again follows from the Onsager relation \eqref{eq:OnsagerRel},
we can flip the integration variables back, $p\to-p$ and $q\to-q$ (together with the antisymmetry $F(-p^0) = -F(p^0)$ of the distribution function),  and obtain
\begin{align}
    \partial_k \frac{\delta \Gamma_k}{\delta \tilde{\Psi}_c(x)} \bigg\rvert_{\tilde{\Psi}=0} &=
	- \frac{i}{2} \int_{pq}  e^{+iq \cdot x} \times \\ \nonumber \sum_{ab} & F(p^0) \epsilon_a \epsilon_b \bigg\{ {\Gamma_{acb,k}^{\tilde{\Psi}\psi\psi}}(p,q,-p) \, B_{ba,k}^R(p) -  {\Gamma_{acb,k}^{\psi\psi\tilde{\Psi}}}(p,q,-p) \, B_{ba,k}^A(p)  \bigg\} \, .
\end{align}
We have argued above that the integrand is only non-vanishing in case $\epsilon_a \epsilon_b = +1$ (and vanishes anyway in the mixed case $\epsilon_a \epsilon_b = -1$), so we can safely omit the factor $\epsilon_a \epsilon_b$ from the expression,
\begin{align}
    \partial_k \frac{\delta \Gamma_k}{\delta \tilde{\Psi}_c(x)} \bigg\rvert_{\tilde{\Psi}=0} &=
	- \frac{i}{2} \int_{pq}  e^{+iq \cdot x} \times \\ \nonumber \sum_{ab} & F(p^0) \bigg\{ {\Gamma_{acb,k}^{\tilde{\Psi}\psi\psi}}(p,q,-p) \, B_{ba,k}^R(p) -  {\Gamma_{acb,k}^{\psi\psi\tilde{\Psi}}}(p,q,-p) \, B_{ba,k}^A(p)  \bigg\} \, ,
\end{align}
We can now insert the definitions \eqref{eq:BPropsRet} and \eqref{eq:BPropsAdv} of $B_{ba,k}^R(p)$ and $B_{ba,k}^A(p)$, respectively, and reorder the factors in the appearing product,
\begin{align}
	\partial_k \frac{\delta \Gamma_k}{\delta \tilde{\Psi}_c(x)} \bigg\rvert_{\tilde{\Psi}=0} 
	= \frac{i}{2} \int_{pq}  e^{+iq \cdot x} \times & \\ \nonumber  \sum_{ab} \bigg\{
	 &G_{ea,k}^R(p) {\Gamma_{acb,k}^{\tilde{\Psi}\psi\psi}}(p,q,-p) \, G_{bd,k}^R(p)  F(p^0)\partial_k R_{de,k}(p)  \\ \nonumber 
	-\,&G_{ea,k}^A(p) { \Gamma_{acb,k}^{\psi\psi\tilde{\Psi}}}(p,q,-p) \, G_{bd,k}^A(p) F(p^0) \partial_k R_{de,k}(p) \bigg\}
\end{align}
which we again transform back into coordinate space,
\begin{align}
	\partial_k \frac{\delta \Gamma_k}{\delta \tilde{\Psi}_c(x)} = & \label{dGamdqNotYetIntegrated} \\ \nonumber
\frac{i}{2}  \int_{uvwyz} \big\{
	 &G_{ea,k}^R(v,y) {\Gamma_{acb,k}^{\tilde{\Psi}\psi\psi}}(y,-x,z) \, G_{bd,k}^R(z,w)  F(w,u)\partial_k R_{de,k}(u,v)  \\ \nonumber 
	-\, &G_{ea,k}^A(v,y) { \Gamma_{acb}^{\psi\psi\tilde{\Psi}}}(y,-x,z) \, G_{bd,k}^A(z,w) F(w,u) \partial_k R_{de,k}(u,v) \big\}
\end{align}
with the classical distribution function
\begin{equation}
    F(w,u) =  -\frac{iT}{2} \,\mathrm{sgn}(w^0-u^0)\, \delta(\vec{w}-\vec{u})
\end{equation}
in coordinate space.

To proceed further with \eqref{dGamdqNotYetIntegrated}, recall that the retarded and advanced propagators are in the presence of an arbitrary field configuration $\psi=\psi(x)$ and $\tilde{\Psi}=\tilde{\Psi}(x)$.
 generally given by (see e.g.~Refs.~\cite{Berges:2012ty,Huelsmann:2020xcy})
\begin{subequations}
\begin{align}
	G_k^R &= -\left[ \left( \Gamma_k^{\tilde{\Psi}\psi} - R_k \right) - \Gamma_k^{\tilde{\Psi}\tilde{\Psi}} \circ \left( \Gamma_k^{\psi\tilde{\Psi}} - R_k \right)^{-1} \circ \Gamma_k^{\psi\psi} \right]^{-1} \, , \label{retPropGenFRG} \\
	G_k^A &= -\left[ \left( \Gamma_k^{\psi\tilde{\Psi}} - R_k \right) - \Gamma_k^{\psi\psi} \circ \left( \Gamma_k^{\tilde{\Psi}\psi} - R_k \right)^{-1} \circ \Gamma_k^{\tilde{\Psi}\tilde{\Psi}} \right]^{-1} \, . \label{advPropGenFRG}
\end{align} \label{propGenFRG}%
\end{subequations}
In the relevant case of a vanishing response field $\tilde{\Psi} = 0$ but an arbitrary and possibly spacetime-dependent classical field $\psi = \psi(x)$ here, the unphysical 2-point function $\Gamma_k^{\psi\psi}[\tilde{\Psi}=0] = 0$ vanishes identically\footnote{This is a consequence of the BRST symmetry discussed in Appendix~\ref{sct:jacobianAndGhosts}.} such that the propagators in \eqref{propGenFRG} become in this case
\begin{subequations}
\begin{align}
    G_{ed,k}^R[\tilde{\Psi}=0](v,w) &= -\left[ \left(\Gamma_k^{\tilde{\Psi}\psi}[\tilde{\Psi}=0] - R_k \right)^{-1} \right]_{ed}(v,w) \, , \label{retPropGenFRGPhiQZero} \\
    G_{ed,k}^A[\tilde{\Psi}=0](v,w) &= -\left[ \left( \Gamma_k^{\psi\tilde{\Psi}}[\tilde{\Psi}=0] - R_k \right)^{-1} \right]_{ed}(v,w) \, . \label{advPropGenFRGPhiQZero}
\end{align} \label{propGenFRGPhiQZero}%
\end{subequations}
A functional derivative of \eqref{propGenFRGPhiQZero}  with respect to $\psi_c(x)$ yields the identities
\begin{subequations}
\begin{align}
    \frac{\delta G_{ed,k}^R[\tilde{\Psi}=0](v,w)}{\delta \psi_c(x)} &= \int_{yz} G_{ea,k}^R(v,y) \Gamma_{acb,k}^{\tilde{\Psi}\psi\psi}(y,x,z) G_{bd,k}^R(z,w) \, , \\
    \frac{\delta G_{ed,k}^A[\tilde{\Psi}=0](v,w)}{\delta \psi_c(x)} &= \int_{yz} G_{ea,k}^A(v,y) \Gamma_{acb,k}^{\psi\psi\tilde{\Psi}}(y,x,z) G_{bd,k}^A(z,w) \, ,
\end{align}
\end{subequations}
where we have used the functional chain rule.
By inserting these derivatives into Eq.~\eqref{dGamdqNotYetIntegrated} we obtain
\begin{align}
	\partial_k \frac{\delta \Gamma_k}{\delta \tilde{\Psi}_c(x)} \bigg\rvert_{\tilde{\Psi}=0} = \frac{i}{2}\int_{uvw} \bigg\{&
	 \frac{\delta}{\delta \psi_c(x)} G_{ed,k}^R(v,w)  F(w,u)\partial_k R_{de,k}(u,v) \\ \nonumber
  - \, &\frac{\delta}{\delta \psi_c(x)} G_{ed,k}^A(v,w) F(w,u) \partial_k R_{de,k}(u,v) \bigg\} \, .
\end{align}
Since both the regulator $R_k(u,v)$ and the distribution function $F(w,u)$ are not field dependent, we can pull the functional $\psi_c(x)$-derivative outside the integral,
\begin{align}
	\partial_k \frac{\delta \Gamma_k}{\delta \tilde{\Psi}_c(x)} \bigg\rvert_{\tilde{\Psi}=0} = \frac{i}{2}\frac{\delta}{\delta \psi_c(x)}\int_{uvw} \bigg\{
	  &G_{ed,k}^R(v,w)  F(w,u)\partial_k R_{de,k}(u,v) \\ \nonumber 
   - \, &G_{ed,k}^A(v,w) F(w,u) \partial_k R_{de,k}(u,v) \bigg\} \, .
\end{align}
and find, after Fourier transform (where the convolutions become simple products),
\begin{align}
	\partial_k \frac{\delta \Gamma_k}{\delta \tilde{\Psi}_c(x)} \bigg\rvert_{\tilde{\Psi}=0} &= \frac{i}{2} \, \frac{\delta }{\delta \psi_c(x)} \int_{p}  F(p^0)\left\{
	 G_{ed,k}^R(p)-G_{ed,k}^A(p)  \right\} \partial_k R_{de,k}(p) \, .\label{dGamdPhiQAsPhiCDeriv}
\end{align}
To bring this finally into the form of a flow equation for the free energy $F_k$, we first combine \eqref{dGamdPhiQAsPhiCDeriv} and the definition \eqref{eq:defOfFk} of the free energy $F_k$ from the effective MSR action $\Gamma_k$,
\begin{align}
    \partial_k \frac{\delta F_k}{\delta \psi_c(\vec{x})} = -\frac{i}{2} \frac{\delta}{\delta \psi_c(x)} \int_{p}  F(p^0)\bigg\{
	 G_{ed,k}^R(p) -G_{ed,k}^A(p) \bigg\}  \partial_k R_{de,k}(p)
\end{align}
formally integrate both sides with respect to $\psi_c$,
and conclude that the flow equation of the free energy $F_k$ is given by
\begin{align}
	\partial_k F_k = -\frac{i}{2} \int_{p}  F(p^0)\bigg\{
	 G_{ed,k}^R(p) -G_{ed,k}^A(p)  \bigg\} \partial_k R_{de,k}(p)
\end{align}
(up to possible constant terms that only depend on the FRG scale~$k$).

We can solve the remaining frequency integral over $p^0$ by closing the contour either in the upper half-plane (in case of $G^R$), or in the lower half-plane (in case of $G^A$). In both cases we only pick up (half) the pole at $p^0$ from the distribution function $F(p^0)=T/p^0$, such that we get
\begin{align}
	\partial_k F_k = \frac{T}{4} \int_{\vec{p}} \bigg\{
	 G_{ed,k}^R(0,\vec{p}) +G_{ed,k}^A(0,\vec{p})  \bigg\} \partial_k R_{de,k}(\vec{p}) \label{eq:dkFkWithGRandGA}
\end{align}
By symmetry we have $G_{ed,k}^R(0,\vec{p})=G_{ed,k}^A(0,\vec{p})$ at zero frequency. Moreover, in this zero-frequency limit the retarded/advanced propagators simply become the static susceptibilities \eqref{eq:staticSuscs} 
(which follows from the identification \eqref{eq:defOfFk}),
\begin{equation}
    G_{ed,k}^R(0,\vec{p}) = -\left[ \left(\Gamma_k^{\tilde{\Psi}\psi}[\tilde{\Psi}=0] - R_k \right)^{-1} \right]_{ed}(0,\vec{p}) = \left[ \left( F_k^{(2)}[\psi] + R_k \right)^{-1} \right]_{ed}(0,\vec{p}) = \chi_{ed,k}(\vec{p}) 
\end{equation}
the flow \eqref{eq:dkFkWithGRandGA} of $F_k$ can be written in closed form,
\begin{equation}
    \partial_k F_k = \frac{T}{2}\,\tr\left\{ \partial_k R_k \circ \left( F_k^{(2)}+R_k \right)^{-1} \right\}
\end{equation}
which proves the final flow equation~\eqref{eq:staticFlowEq} of the free energy.

\section{Details on the extraction of the dynamic critical exponent}\label{extraction_app}
\begin{figure}[t]
    \centering
    \begin{subfigure}{0.48\textwidth}
    \includegraphics[width=\linewidth]{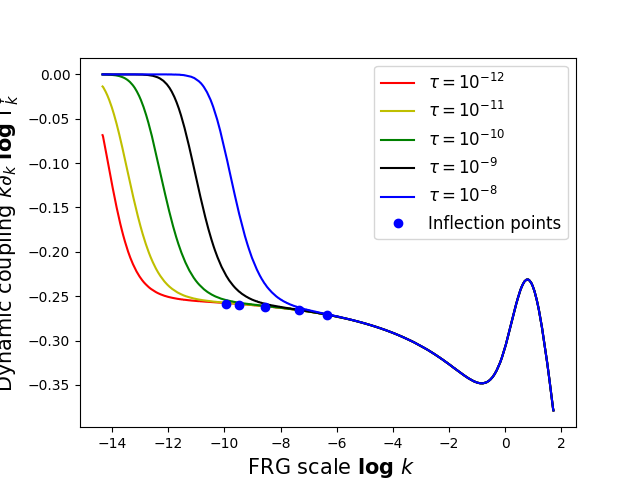}
    \end{subfigure}
    \begin{subfigure}{0.48\textwidth}
    \includegraphics[width=\linewidth]{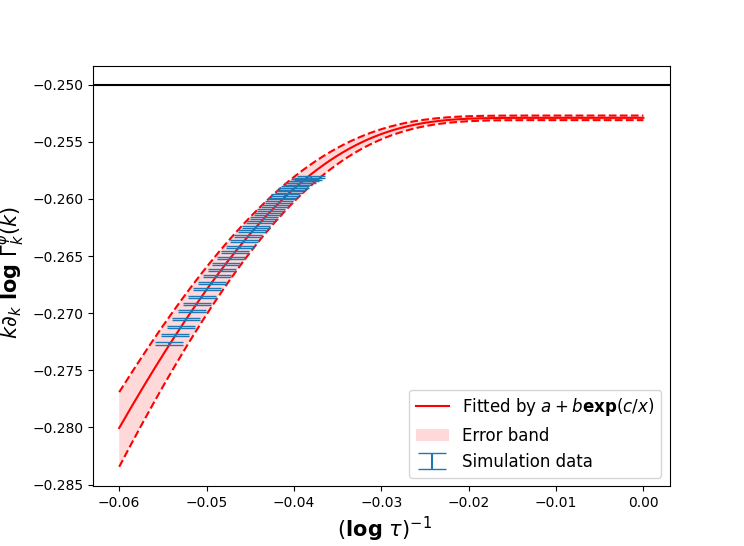}
    \end{subfigure}
    \caption{Extraction of $z^\phi$ in the spatial dimension $d=3.5$, the error band is determined by the error of the parameters from the fit function.}
            \label{fig:determine_z_phi}
\end{figure}

\begin{figure}[t]
    \centering
    \begin{subfigure}{0.48\textwidth}
    \includegraphics[width=\linewidth]{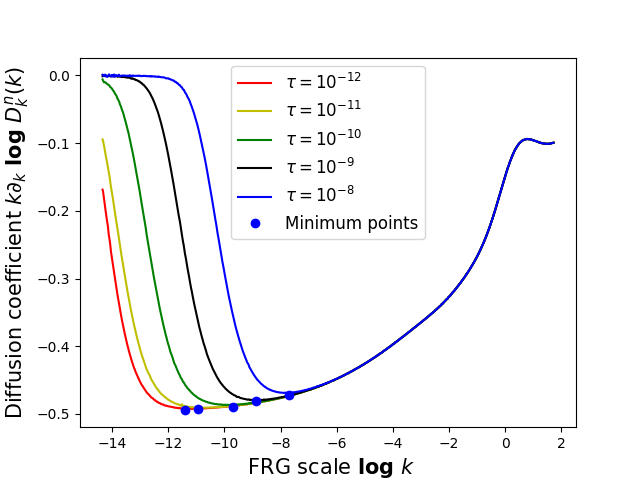}
    \end{subfigure}
    \begin{subfigure}{0.48\textwidth}
    \includegraphics[width=\linewidth]{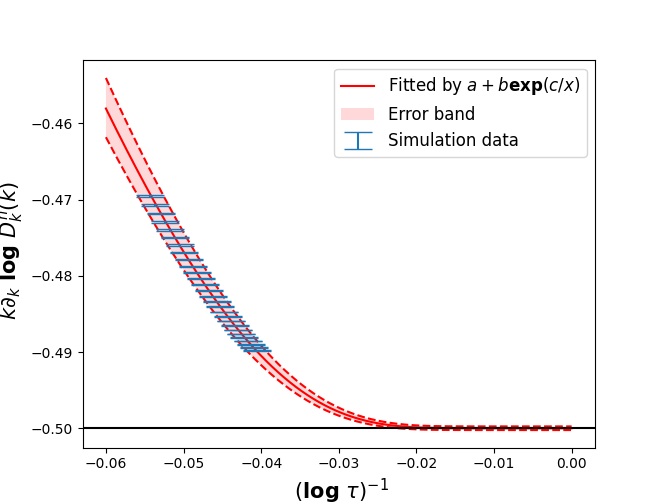}
    \end{subfigure}
    \caption{Extraction of $z^n$ in the spatial dimension $d=3$ using the minimum point of $k\partial_k \textbf{log} \gamma^n_k(k)/k^2$ versus $\textbf{log}k$, the error band is determined by the error of the parameters from the fit function.}
            \label{fig:extraction_zn_1}
\end{figure}

\begin{figure}[t]
    \centering
    \begin{subfigure}{0.48\textwidth}
    \includegraphics[width=\linewidth]{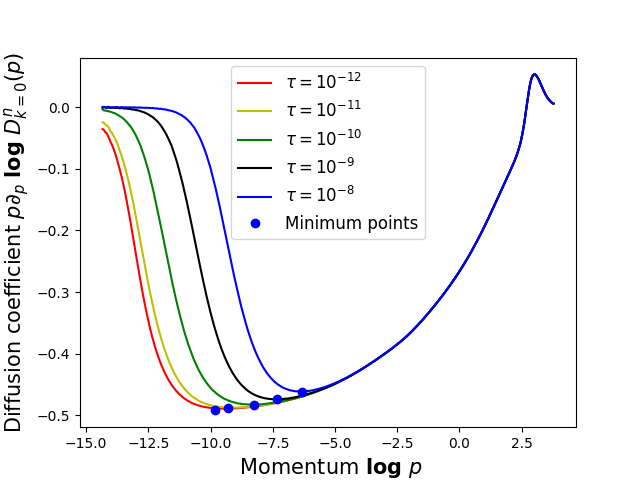}
    \end{subfigure}
    \begin{subfigure}{0.48\textwidth}
    \includegraphics[width=\linewidth]{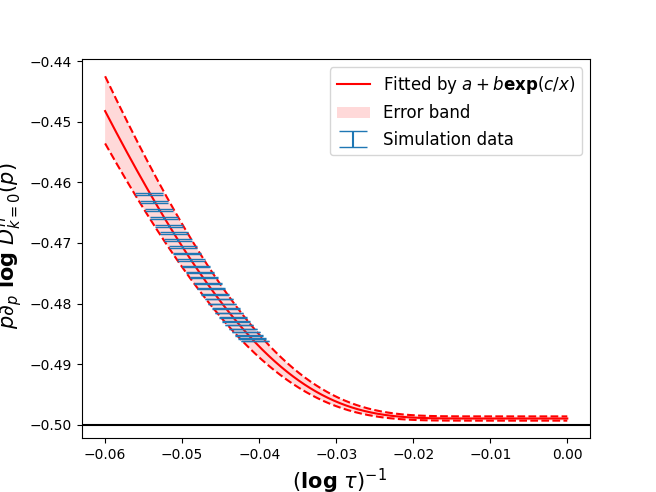}
    \end{subfigure}
    \caption{Extraction of $z^n$ in the spatial dimension $d=3$ using the minimum point of $p\partial_p \textbf{log} \gamma^n_{k=0}(p)/p^2$ versus $\textbf{log}p$, the error band is determined by the error of the parameters from the fit function.}
            \label{extraction_zn_2}
\end{figure}

To extract the dynamic critical exponent $z^\phi$, one first determine the inflection points of $k\partial_k \log \Gamma^\phi_k$ w.r.t $\log k$ under different reduced temperature. As an example, we plot these inflection points for a few different reduced temperatures in spatial dimension $d=3.5$ in Fig.~\ref{fig:determine_z_phi}. One can see that these inflection points approach $-0.25$ when approaching the critical temperature. After determining the inflection points, one plot these points w.r.t $(\log\tau)^{-1}$. In order to determine the value of $z^\phi$, one need to extrapolate these points to $(\log\tau)^{-1}=0$, which is that when the temperature is equivalent to the critical temperature. In order to perform such an extrapolation, one fit these points with an exponential function which is given by $f(x)=a+b\exp(c/x)$. Then we take the value of $f(0)=a$ as the extracted value of the dynamic critical exponent $z^\phi$, and the error of the extraction is given by $\sqrt{\sigma_{a}^2+\Delta_{\textbf{sys}}^2}$, in which $\sigma_a$ is the uncertainty of the coefficient $a$ from fitting the inflection points with $f(x)$, and $\Delta_{\textbf{sys}}$ is the systematic error from the truncation. When fitting the inflection points, we want to make sure that the fit gives the chi-square per degree of freedom $\chi^2\sim 1$. Since we know that for very small reduced temperature, the simulation will lose accuracy due to that the lowest momentum bound in our simulation is not small enough to resolve the infrared physics, one can exclude data points for very small $T_r$, until that the data points and the fit satisfies $\chi^2\sim 1$(but not smaller that 1). In spatial dimension $d=3$ and $d=2.5$, the extractions of $z^\phi$ are similar.
To extract $z^n$, one can start from determining the minimum point of $k\partial_k \log \gamma^n_k(k)/k^2$ w.r.t $\log k$. These points are plotted taking a few different reduced temperature as an example in Fig.~\ref{fig:extraction_zn_1} in spatial dimension $d=3$. Then, similarly, one plot these minimum points w.r.t $(\log\tau)^{-1}$, and extrapolate them to $(\log\tau)^{-1}=0$ using the same method we described above.
Similarly, using exactly the same method, one can equivalently extract the dynamic critical exponent $z^n$ using the minimum points of $p\partial_p \log\gamma^n_{k=0}(p)/p^2$ w.r.t $\log p$, as in Fig.~\ref{extraction_zn_2}.

\begin{table}
\centering
\begin{tabular}{ |c|c|c| } 
 \hline
  d  &  $a$ & $\sqrt{\sigma_a^2+\Delta_{\text{sys}}^2}$ \\
3.5 & -0.253 & 0.003\\
3.0 & -0.501 & 0.001\\ 2.5 & 
-0.750 & 0.000\\
 \hline
\end{tabular}
\caption{$a$ and $\sqrt{\sigma_a^2+\Delta_{\text{sys}}^2}$ from fitting the inflection points of $k\partial_k \log \Gamma^\phi_k$ w.r.t $\log k$ with $f(x)= a + b*\exp(c/x)$, which gives the result and error of $z^\phi -2$.}
\label{t3}
\end{table}

\begin{table}
\centering
\begin{tabular}{ |c|c|c| } 
 \hline
  d  &  $a$ & $\sqrt{\sigma_a^2+\Delta_{\text{sys}}^2}$ \\
3.5 & -0.246 & 0.005 \\ 3.0 & 
-0.500 & 0.000\\ 2.5 & 
-0.749 & 0.001\\
 \hline
\end{tabular}
\caption{$a$ and $\sqrt{\sigma_a^2+\Delta_{\text{sys}}^2}$ from fitting the minimum points of $k\partial_k \log \gamma^n_k(k)/k^2$ w.r.t $\log k$ with $f(x)= a + b*\exp(c/x)$, which gives the result and error of $z^n -2$.}
\label{t2}
\end{table}

\begin{table}
\centering
\begin{tabular}{ |c|c|c| } 
 \hline
  d  &  $a$ & $\sqrt{\sigma_a^2+\Delta_{\text{sys}}^2}$ \\
3.5 & -0.260 & 0.010\\ 3.0 & 
-0.499 & 0.001\\ 2.5 & 
-0.749 & 0.001\\
 \hline
\end{tabular}
\caption{$a$ and $\sqrt{\sigma_a^2+\Delta_{\text{sys}}^2}$ from fitting the minimum points of $p\partial_p \log \gamma^n_{k=0}(p)/p^2$ w.r.t $\log p$ with $f(x)= a + b*\exp(c/x)$, which gives the result and error of $z^n -2$.}
\label{t1}
\end{table}

The value of $a$ and $\sigma_a$ obtained from the obove three different ways are listed in Table. \ref{t3}, Table. \ref{t2}. and Table. \ref{t1}.

\bibliographystyle{JHEP}
\bibliography{refs}
\end{document}